\documentclass[11pt,oneside,reqno,american]{amsart}
\usepackage[T1]{fontenc}
\usepackage[latin9]{inputenc}
\usepackage{verbatim}
\usepackage{units}
\usepackage{amsbsy}
\usepackage{amstext}
\usepackage{amsthm}
\usepackage{amssymb}
\usepackage{undertilde}
\usepackage{graphicx}
\usepackage{xargs}[2008/03/08]

\makeatletter
\numberwithin{equation}{section}
\theoremstyle{plain}
\newtheorem{thm}{\protect\theoremname}[section]
\theoremstyle{remark}
\newtheorem{rem}[thm]{\protect\remarkname}
\theoremstyle{plain}
\newtheorem{cor}[thm]{\protect\corollaryname}
\theoremstyle{remark}
\newtheorem*{acknowledgement*}{\protect\acknowledgementname}

\usepackage{mathtools}
\usepackage{pdfsync} 
\usepackage{hyperref}
\usepackage[all]{xy}
\newcommand{\ie}{\textit{i.e.}}
\newcommand{\eg}{\textit{e.g.}}
\usepackage[lining,tabular]{fbb} 
\usepackage[libertine,bigdelims]{newtxmath}
\usepackage[cal=boondoxo,bb=boondox,frak=boondox]{mathalfa}

   \usepackage{bm}
\usepackage{bm}
\renewcommand{\mathbf}{\bm}
 \theoremstyle{plain}
  
\usepackage{xcolor} 
\definecolor{brown(traditional)}{rgb}{0.59, 0.29, 0.0}
\definecolor{blue(ryb)}{rgb}{0.01, 0.28, 1.0}
\definecolor{red}{rgb}{1.0, 0.0, 0.0}
\definecolor{magenta}{rgb}{1.0, 0.0, 1.0}
\definecolor{mahogany}{rgb}{0.75, 0.25, 0.0}
\definecolor{lavenderpurple}{rgb}{0.59, 0.48, 0.71}
\definecolor{olive}{rgb}{0.5, 0.5, 0.0}
\definecolor{brickred}{rgb}{0.8, 0.25, 0.33}
\definecolor{antiquefuchsia}{rgb}{0.57, 0.36, 0.51}
\definecolor{bole}{rgb}{0.47, 0.27, 0.23}
\definecolor{darkolivegreen}{rgb}{0.33, 0.42, 0.18}
\definecolor{deepjunglegreen}{rgb}{0.0, 0.29, 0.29}

\makeatother

\usepackage{babel}
\providecommand{\acknowledgementname}{Acknowledgement}
\providecommand{\corollaryname}{Corollary}
\providecommand{\remarkname}{Remark}
\providecommand{\theoremname}{Theorem}

\begin{document}

\global\long\def\ga{\alpha}%
\global\long\def\gb{\beta}%
\global\long\def\ggm{\gamma}%
\global\long\def\go{\omega}%
\global\long\def\gs{\sigma}%
\global\long\def\gd{\delta}%
\global\long\def\gD{\Delta}%
\global\long\def\vph{\varphi}%
\global\long\def\gf{\varphi}%
\global\long\def\gk{\kappa}%
\global\long\def\gl{\lambda}%
\global\long\def\gz{\zeta}%
\global\long\def\gh{\eta}%
\global\long\def\gy{\upsilon}%
\global\long\def\gth{\theta}%

\global\long\def\eps{\varepsilon}%
\global\long\def\epss#1#2{\varepsilon_{#2}^{#1}}%
\global\long\def\ep#1{\eps_{#1}}%

\global\long\def\wh#1{\widehat{#1}}%
\global\long\def\hi{\hat{\imath}}%
\global\long\def\hj{\hat{\jmath}}%
\global\long\def\hk{\hat{k}}%

\global\long\def\spec#1{\textsf{#1}}%

\global\long\def\ui{\wh{\boldsymbol{\imath}}}%
\global\long\def\uj{\wh{\boldsymbol{\jmath}}}%
\global\long\def\uk{\widehat{\boldsymbol{k}}}%

\global\long\def\uI{\widehat{\mathbf{I}}}%
\global\long\def\uJ{\widehat{\mathbf{J}}}%
\global\long\def\uK{\widehat{\mathbf{K}}}%

\global\long\def\bs#1{\boldsymbol{#1}}%
\global\long\def\vect#1{\mathbf{#1}}%
\global\long\def\bi#1{\textbf{\emph{#1}}}%

\global\long\def\uv#1{\widehat{\boldsymbol{#1}}}%
\global\long\def\cross{\times}%

\global\long\def\ddt{\frac{\dee}{\dee t}}%
\global\long\def\dbyd#1{\frac{\dee}{\dee#1}}%
\global\long\def\dby#1#2{\frac{\partial#1}{\partial#2}}%

\global\long\def\vct#1{\mathbf{#1}}%
\global\long\def\v#1{\vct{#1}}%

\global\long\def\ga{\alpha}%
\global\long\def\gb{\beta}%
\global\long\def\ggm{\gamma}%
\global\long\def\go{\omega}%
\global\long\def\gs{\sigma}%
\global\long\def\gd{\delta}%
\global\long\def\gD{\Delta}%
\global\long\def\vph{\varphi}%
\global\long\def\gf{\varphi}%
\global\long\def\gk{\kappa}%
\global\long\def\gl{\lambda}%
\global\long\def\gz{\zeta}%
\global\long\def\gh{\eta}%
\global\long\def\gy{\upsilon}%
\global\long\def\gth{\theta}%

\global\long\def\eps{\varepsilon}%
\global\long\def\epss#1#2{\varepsilon_{#2}^{#1}}%
\global\long\def\ep#1{\eps_{#1}}%

\global\long\def\wh#1{\widehat{#1}}%
\global\long\def\hi{\hat{\imath}}%
\global\long\def\hj{\hat{\jmath}}%
\global\long\def\hk{\hat{k}}%

\global\long\def\spec#1{\textsf{#1}}%

\global\long\def\ui{\wh{\boldsymbol{\imath}}}%
\global\long\def\uj{\wh{\boldsymbol{\jmath}}}%
\global\long\def\uk{\widehat{\boldsymbol{k}}}%

\global\long\def\uI{\widehat{\mathbf{I}}}%
\global\long\def\uJ{\widehat{\mathbf{J}}}%
\global\long\def\uK{\widehat{\mathbf{K}}}%

\global\long\def\bs#1{\boldsymbol{#1}}%
\global\long\def\vect#1{\mathbf{#1}}%
\global\long\def\bi#1{\textbf{\emph{#1}}}%

\global\long\def\uv#1{\widehat{\boldsymbol{#1}}}%
\global\long\def\cross{\times}%

\global\long\def\ddt{\frac{\dee}{\dee t}}%
\global\long\def\dbyd#1{\frac{\dee}{\dee#1}}%
\global\long\def\dby#1#2{\frac{\partial#1}{\partial#2}}%

\global\long\def\vct#1{\mathbf{#1}}%
\global\long\def\v#1{\vct{#1}}%

\global\long\def\partialby#1#2{\frac{\partial#1}{\partial x^{#2}}}%
\newcommandx\parder[2][usedefault, addprefix=\global, 1=]{\frac{\partial#2}{\partial#1}}%

\global\long\def\fall{,\quad\text{for all}\quad}%

\global\long\def\reals{\mathbb{R}}%

\global\long\def\rthree{\reals^{3}}%
\global\long\def\rsix{\reals^{6}}%
\global\long\def\rn{\reals^{n}}%
\global\long\def\rt#1{\reals^{#1}}%

\global\long\def\les{\leqslant}%
\global\long\def\ges{\geqslant}%

\global\long\def\dee{\textrm{d}}%
\global\long\def\di{d}%
\global\long\def\dx{\dee\bp}%

\global\long\def\from{\colon}%
\global\long\def\tto{\longrightarrow}%
\global\long\def\lmt{\longmapsto}%
\global\long\def\lhr{\lhook\joinrel\longrightarrow}%

\global\long\def\abs#1{\left|#1\right|}%

\global\long\def\isom{\cong}%

\global\long\def\comp{\circ}%

\global\long\def\cl#1{\overline{#1}}%

\global\long\def\fun{\varphi}%

\global\long\def\interior{\textrm{Int}\,}%
\global\long\def\inter#1{\kern0pt  #1^{\mathrm{o}}}%

\global\long\def\sign{\textrm{sign}\,}%
\global\long\def\sgn#1{(-1)^{#1}}%
\global\long\def\sgnp#1{(-1)^{\abs{#1}}}%

\global\long\def\dimension{\textrm{dim}\,}%

\global\long\def\esssup{\textrm{ess}\,\sup}%

\global\long\def\ess{\textrm{{ess}}}%

\global\long\def\kernel{\mathop{\textrm{\textup{Kernel}}}}%

\global\long\def\support{\mathop{\textrm{\textup{supp}}}}%

\global\long\def\image{\mathop{\textrm{\textup{Image}}}}%

\global\long\def\diver{\mathop{\textrm{\textup{div}}}}%

\global\long\def\sp{\mathop{\textrm{\textup{span}}}}%

\global\long\def\tr{\mathop{\textrm{\textup{tr}}}}%

\global\long\def\resto#1{|_{#1}}%
\global\long\def\incl{\iota}%
\global\long\def\iden{\imath}%
\global\long\def\idnt{\textrm{Id}}%
\global\long\def\rest{\rho}%
\global\long\def\extnd{e_{0}}%

\global\long\def\proj{\textrm{pr}}%

\global\long\def\L#1{L\bigl(#1\bigr)}%
\global\long\def\LS#1{L_{S}\bigl(#1\bigr)}%

\global\long\def\ino#1{\int_{#1}}%

\global\long\def\half{\frac{1}{2}}%
\global\long\def\shalf{{\scriptstyle \half}}%
\global\long\def\third{\frac{1}{3}}%

\global\long\def\empt{\varnothing}%

\global\long\def\paren#1{\left(#1\right)}%
\global\long\def\bigp#1{\bigl(#1\bigr)}%
\global\long\def\biggp#1{\biggl(#1\biggr)}%
\global\long\def\Bigp#1{\Bigl(#1\Bigr)}%

\global\long\def\braces#1{\left\{  #1\right\}  }%
\global\long\def\sqbr#1{\left[#1\right]}%
\global\long\def\anglep#1{\left\langle #1\right\rangle }%

\global\long\def\lsum{{\textstyle \sum}}%

\global\long\def\bigabs#1{\bigl|#1\bigr|}%

\global\long\def\stp{\text{\small\ensuremath{\bigodot}}}%
\global\long\def\tp{\text{\small\ensuremath{\bigotimes}}}%

\global\long\def\mi#1{#1}%
\global\long\def\mii{I}%
\global\long\def\mie#1#2{#1_{1}\cdots#1_{#2}}%

\global\long\def\smi#1{\boldsymbol{#1}}%
\global\long\def\asmi#1{#1}%
\global\long\def\ordr#1{\left\langle #1\right\rangle }%

\global\long\def\symm#1{\paren{#1}}%
\global\long\def\smtr{\mathcal{S}}%

\global\long\def\perm{p}%
\global\long\def\sperm{\mathcal{P}}%

\global\long\def\oneto{1,\dots,}%

\global\long\def\lisub#1#2#3{#1_{1}#2\dots#2#1_{#3}}%

\global\long\def\lisup#1#2#3{#1^{1}#2\dots#2#1^{#3}}%

\global\long\def\lisubb#1#2#3#4{#1_{#2}#3\dots#3#1_{#4}}%

\global\long\def\lisubbc#1#2#3#4{#1_{#2}#3\cdots#3#1_{#4}}%

\global\long\def\lisubbwout#1#2#3#4#5{#1_{#2}#3\dots#3\widehat{#1}_{#5}#3\dots#3#1_{#4}}%

\global\long\def\lisubc#1#2#3{#1_{1}#2\cdots#2#1_{#3}}%

\global\long\def\lisupc#1#2#3{#1^{1}#2\cdots#2#1^{#3}}%

\global\long\def\lisupp#1#2#3#4{#1^{#2}#3\dots#3#1^{#4}}%

\global\long\def\lisuppc#1#2#3#4{#1^{#2}#3\cdots#3#1^{#4}}%

\global\long\def\lisuppwout#1#2#3#4#5#6{#1^{#2}#3#4#3\wh{#1^{#6}}#3#4#3#1^{#5}}%

\global\long\def\lisubbwout#1#2#3#4#5#6{#1_{#2}#3#4#3\wh{#1}_{#6}#3#4#3#1_{#5}}%

\global\long\def\lisubwout#1#2#3#4{#1_{1}#2\dots#2\widehat{#1}_{#4}#2\dots#2#1_{#3}}%

\global\long\def\lisupwout#1#2#3#4{#1^{1}#2\dots#2\widehat{#1^{#4}}#2\dots#2#1^{#3}}%

\global\long\def\lisubwoutc#1#2#3#4{#1_{1}#2\cdots#2\widehat{#1}_{#4}#2\cdots#2#1_{#3}}%

\global\long\def\twp#1#2#3{\dee#1^{#2}\wedge\dee#1^{#3}}%

\global\long\def\thp#1#2#3#4{\dee#1^{#2}\wedge\dee#1^{#3}\wedge\dee#1^{#4}}%

\global\long\def\fop#1#2#3#4#5{\dee#1^{#2}\wedge\dee#1^{#3}\wedge\dee#1^{#4}\wedge\dee#1^{#5}}%

\global\long\def\idots#1{#1\dots#1}%
\global\long\def\icdots#1{#1\cdots#1}%

\global\long\def\norm#1{\|#1\|}%

\global\long\def\nonh{\heartsuit}%

\global\long\def\nhn#1{\norm{#1}^{\nonh}}%

\global\long\def\bigmid{\,\bigl|\,}%

\global\long\def\trps{^{{\scriptscriptstyle \textsf{T}}}}%

\global\long\def\testfuns{\mathcal{D}}%

\global\long\def\ntil#1{\tilde{#1}{}}%

\global\long\def\chart{\varphi}%
\global\long\def\Chart{\Phi}%

\global\long\def\eucl{E}%

\global\long\def\mind{\alpha}%
\global\long\def\vb{W}%
\global\long\def\vbp{\pi}%

\global\long\def\man{\mathcal{M}}%
\global\long\def\odman{\mathcal{N}}%
\global\long\def\subman{\mathcal{A}}%

\global\long\def\vbt{\mathcal{E}}%
\global\long\def\fib{\mathbf{V}}%
\global\long\def\vbts{W}%
\global\long\def\avb{U}%

\global\long\def\chart{\varphi}%
\global\long\def\vbchart{\Phi}%

\global\long\def\jetb#1{J^{#1}}%
\global\long\def\jet#1{j^{1}(#1)}%
\global\long\def\tjet{\tilde{\jmath}}%

\global\long\def\Jet#1{J^{1}(#1)}%

\global\long\def\jetm#1{j_{#1}}%

\global\long\def\coj{\mathfrak{d}}%

\global\long\def\alt{\mathfrak{A}}%

\global\long\def\pou{\eta}%

\global\long\def\ext{{\textstyle \bigwedge}}%
\global\long\def\forms{\Omega}%

\global\long\def\dotwedge{\dot{\mbox{\ensuremath{\wedge}}}}%

\global\long\def\vel{\theta}%

\global\long\def\Jac{\mathcal{J}}%

\global\long\def\contr{\raisebox{0.4pt}{\mbox{\ensuremath{\lrcorner}}}\,}%
\global\long\def\fcor{\llcorner}%
\global\long\def\bcor{\lrcorner}%
\global\long\def\fcontr{\raisebox{0.4pt}{\mbox{\ensuremath{\llcorner}}}\,}%

\global\long\def\lie{\mathcal{L}}%

\global\long\def\ssym#1#2{\ext^{#1}T^{*}#2}%

\global\long\def\sh{^{\sharp}}%

\global\long\def\nfo{\ext^{n}T^{*}\base}%

\global\long\def\spc{\mathcal{S}}%
\global\long\def\sptm{\mathcal{E}}%
\global\long\def\evnt{e}%
\global\long\def\frame{\Phi}%

\global\long\def\timeman{\mathcal{T}}%
\global\long\def\zman{t}%
\global\long\def\dims{n}%
\global\long\def\m{\dims-1}%
\global\long\def\dimw{m}%

\global\long\def\wc{z}%

\global\long\def\fourv#1{\mbox{\ensuremath{\mathfrak{#1}}}}%

\global\long\def\pbform#1{\utilde{#1}}%
\global\long\def\util#1{\raisebox{-5pt}{\ensuremath{{\scriptscriptstyle \sim}}}\!\!\!#1}%

\global\long\def\utilJ{\util J}%

\global\long\def\utilRho{\util{\rho}}%

\global\long\def\body{B}%
\global\long\def\man{\mathcal{M}}%
\global\long\def\var{\mathcal{V}}%
\global\long\def\base{\mathcal{X}}%
\global\long\def\fb{\mathcal{Y}}%
\global\long\def\srfc{\mathcal{Z}}%
\global\long\def\dimb{n}%
\global\long\def\dimf{m}%

\global\long\def\bdry{\partial}%

\global\long\def\gO{\varOmega}%

\global\long\def\reg{\mathcal{R}}%
\global\long\def\bdrr{\bdry\reg}%

\global\long\def\bdom{\bdry\gO}%

\global\long\def\bndo{\partial\gO}%

\global\long\def\pis{x}%
\global\long\def\xo{\pis_{0}}%
\global\long\def\x{x}%

\global\long\def\pib{X}%
\global\long\def\bp{X}%
\global\long\def\ii{i}%
\global\long\def\ia{\alpha}%
\global\long\def\sp{x}%
\global\long\def\fp{y}%

\global\long\def\ib{i}%
\global\long\def\is{\alpha}%

\global\long\def\pbndo{\Gamma}%
\global\long\def\bndoo{\pbndo_{0}}%
 
\global\long\def\bndot{\pbndo_{t}}%

\global\long\def\cloo{\cl{\gO}}%

\global\long\def\nor{\mathbf{n}}%
\global\long\def\Nor{\mathbf{N}}%

\global\long\def\dA{\,\dee A}%

\global\long\def\dV{\,\dee V}%

\global\long\def\eps{\varepsilon}%

\global\long\def\vs{\mathbf{W}}%
\global\long\def\avs{\mathbf{V}}%
\global\long\def\affsp{\mathbf{A}}%
\global\long\def\pt{p}%

\global\long\def\vbase{e}%
\global\long\def\sbase{\mathbf{e}}%
\global\long\def\msbase{\mathfrak{e}}%
\global\long\def\vect{v}%
\global\long\def\dbase{\sbase}%

\global\long\def\vf{w}%

\global\long\def\avf{u}%

\global\long\def\stn{\varepsilon}%

\global\long\def\rig{r}%

\global\long\def\rigs{\mathcal{R}}%

\global\long\def\qrigs{\!/\!\rigs}%

\global\long\def\qd{\!/\,\!\kernel\diffop}%

\global\long\def\dis{\chi}%
\global\long\def\conf{\kappa}%
\global\long\def\csp{\mathcal{Q}}%

\global\long\def\embds{\textrm{Emb}}%

\global\long\def\lc{A}%

\global\long\def\lv{\dot{A}}%
\global\long\def\alv{\dot{B}}%

\global\long\def\fc{F}%

\global\long\def\st{\sigma}%

\global\long\def\bfc{\mathbf{b}}%
\global\long\def\bfcc{b}%

\global\long\def\sfc{\mathbf{t}}%
\global\long\def\sfcc{t}%

\global\long\def\stm{\varsigma}%
\global\long\def\std{S}%
\global\long\def\tst{\sigma}%
\global\long\def\tstd{s}%

\global\long\def\gdiv{\bdry\textrm{iv\,}}%

\global\long\def\smc#1{\mathfrak{#1}}%

\global\long\def\nhs{P}%
\global\long\def\nhsa{P}%
\global\long\def\nhsb{\underline{P}}%

\global\long\def\soc{Z}%

\global\long\def\tran{\mathrm{tr}}%

\global\long\def\slf{R}%

\global\long\def\sts{\varSigma}%
\global\long\def\spstd{\mathfrak{S}}%
\global\long\def\sptst{\mathfrak{T}}%
\global\long\def\spnhs{\mathcal{P}}%
\global\long\def\Ljj{\L{J^{1}(J^{k-1}\vb),\ext^{n}T^{*}\base}}%

\global\long\def\spsb{\text{\Large\ensuremath{\Delta}}}%

\global\long\def\ened{\mathfrak{w}}%
\global\long\def\energy{\mathfrak{W}}%

\global\long\def\ebdfc{T}%
\global\long\def\optimum{\st^{\textrm{opt}}}%
\global\long\def\scf{K}%

\global\long\def\pform{\varsigma}%
\global\long\def\vform{\beta}%
\global\long\def\sform{\tau}%
\global\long\def\flow{J}%
\global\long\def\n{\m}%
\global\long\def\cmap{\mathfrak{t}}%
\global\long\def\vcmap{\varSigma}%

\global\long\def\mvec{\mathfrak{v}}%
\global\long\def\mveco#1{\mathfrak{#1}}%
\global\long\def\smbase{\mathfrak{e}}%
\global\long\def\spx{\simp}%

\global\long\def\hp{H}%
\global\long\def\ohp{h}%

\global\long\def\hps{G_{\dims-1}(T\spc)}%
\global\long\def\ohps{G_{\dims-1}^{\perp}(T\spc)}%
\global\long\def\hpsx{G_{\dims-1}(\tspc)}%
\global\long\def\ohpsx{G_{\dims-1}^{\perp}(\tspc)}%

\global\long\def\fbun{F}%

\global\long\def\flowm{\Phi}%

\global\long\def\tgb{T\spc}%
\global\long\def\ctgb{T^{*}\spc}%
\global\long\def\tspc{T_{\pis}\spc}%
\global\long\def\dspc{T_{\pis}^{*}\spc}%

\global\long\def\fflow{\fourv J}%
\global\long\def\fvform{\mathfrak{b}}%
\global\long\def\fsform{\mathfrak{t}}%
\global\long\def\fpform{\mathfrak{s}}%

\global\long\def\maxw{\mathfrak{g}}%
\global\long\def\frdy{\mathfrak{f}}%
\global\long\def\ptnl{A}%

\global\long\def\sobp#1#2{W_{#2}^{#1}}%

\global\long\def\inner#1#2{\left\langle #1,#2\right\rangle }%

\global\long\def\fields{\sobp pk(\vb)}%

\global\long\def\bodyfields{\sobp p{k_{\partial}}(\vb)}%

\global\long\def\forces{\sobp pk(\vb)^{*}}%

\global\long\def\bfields{\sobp p{k_{\partial}}(\vb\resto{\bndo})}%

\global\long\def\loadp{(\sfc,\bfc)}%

\global\long\def\strains{\lp p(\jetb k(\vb))}%

\global\long\def\stresses{\lp{p'}(\jetb k(\vb)^{*})}%

\global\long\def\diffop{D}%

\global\long\def\strainm{E}%

\global\long\def\incomps{\vbts_{\yieldf}}%

\global\long\def\devs{L^{p'}(\eta_{1}^{*})}%

\global\long\def\incompsns{L^{p}(\eta_{1})}%

\global\long\def\testf{\mathcal{D}}%
\global\long\def\dists{\mathcal{D}'}%

\global\long\def\codiv{\boldsymbol{\partial}}%

\global\long\def\currof#1{\tilde{#1}}%

\global\long\def\chn{c}%
\global\long\def\chnsp{\mathbf{F}}%

\global\long\def\current{T}%
\global\long\def\curr{R}%

\global\long\def\curd{S}%
\global\long\def\curwd#1{\wh{#1}}%
\global\long\def\curnd#1{\wh{#1}}%

\global\long\def\prop{P}%

\global\long\def\aprop{Q}%

\global\long\def\flux{\omega}%
\global\long\def\aflux{S}%

\global\long\def\fform{\tau}%

\global\long\def\dimn{n}%

\global\long\def\sdim{{\dimn-1}}%

\global\long\def\contrf{{\scriptstyle \smallfrown}}%

\global\long\def\prodf{{\scriptstyle \smallsmile}}%

\global\long\def\ptnl{\varphi}%

\global\long\def\form{\omega}%

\global\long\def\dens{\rho}%

\global\long\def\simp{s}%
\global\long\def\ssimp{\Delta}%
\global\long\def\cpx{K}%

\global\long\def\cell{C}%

\global\long\def\chain{B}%

\global\long\def\ach{A}%

\global\long\def\coch{X}%

\global\long\def\scale{s}%

\global\long\def\fnorm#1{\norm{#1}^{\flat}}%

\global\long\def\chains{\mathcal{A}}%

\global\long\def\ivs{\boldsymbol{U}}%

\global\long\def\mvs{\boldsymbol{V}}%

\global\long\def\cvs{\boldsymbol{W}}%

\global\long\def\ndual#1{#1'}%

\global\long\def\nd{'}%

\global\long\def\cee#1{C^{#1}}%

\global\long\def\lone{L^{1}}%

\global\long\def\linf{L^{\infty}}%

\global\long\def\lp#1{L^{#1}}%

\global\long\def\ofbdo{(\bndo)}%

\global\long\def\ofclo{(\cloo)}%

\global\long\def\vono{(\gO,\rthree)}%

\global\long\def\vonbdo{(\bndo,\rthree)}%
\global\long\def\vonbdoo{(\bndoo,\rthree)}%
\global\long\def\vonbdot{(\bndot,\rthree)}%

\global\long\def\vonclo{(\cl{\gO},\rthree)}%

\global\long\def\strono{(\gO,\reals^{6})}%

\global\long\def\sob{W_{1}^{1}}%

\global\long\def\sobb{\sob(\gO,\rthree)}%

\global\long\def\lob{\lone(\gO,\rthree)}%

\global\long\def\lib{\linf(\gO,\reals^{12})}%

\global\long\def\ofO{(\gO)}%

\global\long\def\oneo{{1,\gO}}%
\global\long\def\onebdo{{1,\bndo}}%
\global\long\def\info{{\infty,\gO}}%

\global\long\def\infclo{{\infty,\cloo}}%

\global\long\def\infbdo{{\infty,\bndo}}%

\global\long\def\ld{LD}%

\global\long\def\ldo{\ld\ofO}%
\global\long\def\ldoo{\ldo_{0}}%

\global\long\def\trace{\gamma}%

\global\long\def\pr{\proj_{\rigs}}%

\global\long\def\pq{\proj}%

\global\long\def\qr{\,/\,\reals}%

\global\long\def\aro{S_{1}}%
\global\long\def\art{S_{2}}%

\global\long\def\mo{m_{1}}%
\global\long\def\mt{m_{2}}%

\global\long\def\yieldc{B}%

\global\long\def\yieldf{Y}%

\global\long\def\trpr{\pi_{P}}%

\global\long\def\devpr{\pi_{\devsp}}%

\global\long\def\prsp{P}%

\global\long\def\devsp{D}%

\global\long\def\ynorm#1{\|#1\|_{\yieldf}}%

\global\long\def\colls{\Psi}%

\global\long\def\semib{\mathrm{SB}}%

\global\long\def\tm#1{\overrightarrow{#1}}%
\global\long\def\tmm#1{\underrightarrow{\overrightarrow{#1}}}%

\global\long\def\itm#1{\overleftarrow{#1}}%
\global\long\def\itmm#1{\underleftarrow{\overleftarrow{#1}}}%

\global\long\def\ptrac{\mathcal{P}}%

\global\long\def\nh#1{\hat{#1}}%
\global\long\def\nj{\hat{\jmath}}%
\global\long\def\nJ{\hat{J}}%
\global\long\def\rin#1{\mathfrak{#1}}%
\global\long\def\npi{\hat{\pi}}%
\global\long\def\rp{\rin p}%
\global\long\def\rq{\rin q}%
\global\long\def\rr{\rin r}%

\global\long\def\xty{(\base,\fb)}%
\global\long\def\xts{(\base,\spc)}%
\global\long\def\r{r}%
\global\long\def\ntm{(\reals^{n},\reals^{m})}%

\title[Global Stress Theory]{Notes on Global Stress and Hyper-Stress Theories}
\author{Reuven Segev}
\address{Reuven Segev\\
Department of Mechanical Engineering\\
Ben-Gurion University of the Negev\\
Beer-Sheva, Israel\\
rsegev@bgu.ac.il}
\thanks{\today}
\begin{abstract}
The fundamental ideas and tools of the global geometric formulation
of stress and hyper-stress theory of continuum mechanics are introduced.
The proposed framework is the infinite dimensional counterpart of
statics of systems having finite number of degrees of freedom, as
viewed in the geometric approach to analytical mechanics. For continuum
mechanics, the configuration space is the manifold of embeddings of
a body manifold into the space manifold. Generalized velocity fields
are viewed as elements of the tangent bundle of the configuration
space and forces are continuous linear functionals defined on tangent
vectors, elements of the cotangent bundle. It is shown, in particular,
that a natural choice of topology on the configuration space, implies
that force functionals may be represented by objects that generalize
the stresses of traditional continuum mechanics.

\end{abstract}

\maketitle
\tableofcontents{}%

\section{Introduction}

These notes provide an introduction to the fundamentals of global
analytic continuum mechanics as developed in \cite{epstein_differentiable_1980,segev_differentiable_1981,Box_in_Marsden_Hughes,segev_existence_1986,segev_forces_1986,segev_continuum_2016}.
The terminology ``global analytic'' is used to imply that the formulation
is based on the notion of a configuration space of the mechanical
system as in analytic classical mechanics. As such, this review is
complementary to that of \cite{segev_notes_2013}, which describes
continuum mechanics on differentiable manifolds using a generalization
of the Cauchy approach to flux and stress theory.

The setting for the basics of kinematics and statics is quite simple
and provides an elegant geometric picture of mechanics. Consider the
configuration space $\csp$ containing all admissible configuration
of the system. Then, construct a differentiable manifold structure
on the configuration space, define generalized (or virtual) velocities
as tangent vectors, elements of $T\csp$, and define generalized forces
as linear functions defined on the space of generalized velocities,
elements of $T^{*}\csp$. The result of the action of a generalized
force $\fc$ on a generalized velocity $\vf$ is interpreted as mechanical
power. Thus, such a structure may be used to encompass both classical
mechanics of mass particles and rigid bodies as well as continuum
mechanics. The difference is that the configuration space for continuum
mechanics and other field theories is infinite dimensional.

It is well known that the transition from the mechanics of mass particles
and rigid bodies to continuum mechanics is not straightforward and
requires the introduction of new notions and assumptions. The global
analytic formulation explains this observation as follows. Linear
functions, and forces in particular, are identically continuous when
defined on a finite dimensional space. However, in the infinite dimensional
situation, one has to specify exactly the topology on the infinite
dimensional space of generalized velocities with respect to which
forces should be continuous. Then, the properties of force functionals
are deduced from the continuity requirement through a representation
theorem. In other words, the properties of forces follow directly
from the kinematics of the theory.

For continuum mechanics of a body $\base$ in space $\spc$, the basic
kinematic assumption is traditionally referred to as the axiom of
material impenetrability. A configuration of the body in space is
specified by a mapping $\conf:\base\to\spc$ which is assumed to be
injective and of full rank at each point\textemdash an embedding.
Hence, the configuration space for continuum mechanics should be the
collection of embeddings of the body manifold into the space manifold.
It turns out that the $C^{1}$-topology is the natural one to use
in order to endow the collection of embeddings with a differentiable
structure of a Banach manifold. The $C^{r}$-topologies for $r>1$
are admissible also.

It follows that forces are continuous linear functionals on the space
of vector fields over the body of class $C^{r}$, $r\ge1$, equipped
with the $C^{r}$-topology with a special role for the case $r=1$.
A standard procedure based on the Hahn-Banach theorem leads to a representation
theorem for a force functional in terms of vector valued measures.

The measures representing a force generalize the stress and hyper-stress
objects of continuum mechanics. On the one hand, as expected, a stress
measure is not determined uniquely by a force. This is in accordance
with the inherent static indeterminacy of continuum mechanics and
it follows directly from the representation procedure. While the case
$r=1$ leads to continuum mechanics of order one, the cases $r>1$
are extensions of higher order continuum mechanics. Thus, an existence
theorem for hyper-stresses follows naturally. The relation between
a force and a representing stress object is a generalization of the
principle of virtual work in continuum mechanics and so it is analogous
to the equilibrium equations.

The representation of forces by stress measures is significant for
two reasons. First, the existence of the stress object as well as
the corresponding equilibrium condition are obtained for stress distributions
that may be as singular as Radon measures. In addition, while force
functionals cannot be restricted to subsets of a body, measures may
be restricted to subsets. This reflects a fundamental feature of stress
distributions\textemdash they induce force systems on bodies. It is
emphasized that in no further assumptions of mathematical or physical
nature are made.

The framework described above applies to continuum mechanics on general
differentiable manifolds without any additional structure such as
a Riemannian metric or a connection. The body manifold is assumed
here to be a compact manifold with corners. However, as described
in \cite{Michor2019}, it is now possible to extend the applicability
of this framework to a wider class of geometric object\textemdash Whitney
manifold germs.

\medskip{}

Starting with the introduction of notation used in the manuscript
in Section \ref{sec:Notation-and-Preliminaries}, we continue with
the construction of the manifold structure on the space of embeddings.
Thus, Section \ref{sec:Banachable} describes the Banachable vector
spaces used to construct the infinite dimensional manifold structure
on the configuration space and Section \ref{sec:Man_of_Maps} is concerned
with the Banach manifold structure on the set of $C^{r}$-sections
of a fiber bundle $\xi:\fb\to\base$. This includes, as a special
case, the space $C^{r}(\base,\spc)$ of $C^{r}$-mappings of the body
into space and also provides a natural extension to continuum mechanics
of generalized media. After describing the topology in $C^{r}(\base,\spc)$
in Section \ref{sec:Topology-C^r}, we show in Section \ref{sec:Embeddings},
that the set of embeddings is open in $C^{r}(\base,\spc)$, $r\ge1$.
As such, it is a Banach manifold also and the tangent bundle is inherited
from that of $C^{r}(\base,\spc)$. In Section \ref{sec:The-General-Framework}
we outline the framework for the suggested force and stress theory
as described roughly above. Sections \ref{sec:Duals-to-Sections},
\ref{sec:de_Rham} and \ref{sec:Vector-valued-currents} introduce
relevant spaces of linear functionals on manifolds , and present some
of their properties. These include some standard classes of functionals
such as de Rham currents and Schwartz distributions on manifolds.
The representation theorem of forces by stress measures in considered
in Section \ref{sec:Stresses}. Section \ref{sec:Simple-Forces-St}
discusses the natural situation of simple forces and stress, that
is, the case $r=1$.

Concluding remarks and references to further studies are made in Section
\ref{sec:Concluding-Remarks}.

\section{Notation and Preliminaries\label{sec:Notation-and-Preliminaries}}

\subsection{General notation}

A collection of indices $i_{1}\cdots i_{k}$, $i_{r}=\oneto n$ will
be represented as a multi-index $I$ and we will write $\abs I=k$,
the length of the multi-index. In general, multi-indices will be denoted
by upper-case roman letters and the associated indices will be denoted
by the corresponding lower case letters. Thus, a generic element in
a $k$-multilinear mapping $A\in\otimes^{k}\reals$ is given in terms
of the array $(A_{\mii})$, $\abs I=k$. In what follows, we will
use the summation convention for repeated indices as well as repeated
multi-indices. Whenever the syntax is violated, \eg, when a multi-index
appears more than twice in a term, it is understood that summation
does not apply.

A multi-index $I$ induces a sequence $(\lisub I,n)$ in which $I_{r}$
is the number of times the index $r$ appears in the sequence $\mie ik$.
Thus, $\abs I=\sum_{r}I_{r}$. Multi-indices may be concatenated naturally
such that $\abs{IJ}=\abs I+\abs J$.

In case an array $A$ is symmetric, the independent components of
the array may be listed as $A_{\lisubbc i1{}k}$with $i_{l}\le i_{l+1}$.
A non-decreasing multi-index, that is, amulti-index that satisfies
the condition $i_{l}\le i_{l+1}$, will be denoted by boldface, upper-case
roman characters so that a symmetric tensor $A$ is represented by
the components $(A_{\smi I})$, $\abs{\smi I}=k$. In particular,
for a function $u:\reals^{n}\to\reals$, a particular partial derivative
of order $k$ is written in the form
\begin{equation}
u_{,\smi I}:=\bdry_{\smi I}u:=\frac{\bdry^{\abs{\smi I}}u}{(\bdry\bp^{1})^{I_{1}}\cdots(\bdry\bp^{n})^{I_{n}}},
\end{equation}
where $\smi I$ is a non-decreasing multi-index with $\abs{\smi I}=k$.

The notation $\bdry_{i}=\bdry/\bdry\bp^{i}$, will be used for both
the partial derivatives in $\reals^{n}$ and for the elements of the
basis of the tangent space $T_{\bp}\base$ of a manifold $\base$
at a point $\bp$. The corresponding dual basis for $T_{X}^{*}\base$
will be denoted by $\{\dee\bp^{i}\}$.

Greek letters, $\gl,$ $\mu$, $\nu$, will be used for strictly increasing
multi-indices used in the representation of alternating tensors and
forms, \eg, 
\begin{equation}
\go=\go_{\gl}\dee\bp^{\gl}:=\go_{\mie{\gl}{\abs{\gl}}}\lisuppc{\dee\bp}{\gl_{1}}{\wedge}{\gl_{\abs{\gl}}}.
\end{equation}
Given a strictly increasing multi-index $\gl$ with $\abs{\gl}=p$,
we will denote the strictly increasing $(n-p)$-multi-index that complements
$\gl$ to $\oneto n$ by $\hat{\gl}$. In this context, $\hat{\mu}$,
$\hat{\nu}$, etc. will indicate generic increasing $(n-p)$ multi-indices.
The Levi-Civita symbol will be denoted as $\eps_{\mii}$ or $\eps^{\mii}$,
$\abs{\mii}=n$ so that for example $\dx^{\gl}\wedge\dx^{\hat{\gl}}=\eps^{\gl\hat{\gl}}\dx$,
where we also set 
\begin{equation}
\bdry_{\bp}:=\lisubbc{\bdry}1{\wedge}n,\qquad\dx:=\lisuppc{\dx}1{\wedge}n.
\end{equation}
(Note that we view $\gl$ and $\hat{\gl}$ as two distinct indices
so summation is not implied in a term such as $\dx^{\gl}\wedge\dx^{\hat{\gl}}$.)

The following identifications will be implied for tensor products
of vector spaces and vector bundles
\begin{equation}
V^{*}\otimes U\isom L(V,U),\qquad\text{(V\ensuremath{\otimes U)^{*}\isom V^{*}\otimes U^{*}}.}
\end{equation}
For vector bundles $V$ and $U$ over a manifold $\base$, let $\std$
be a section of $V^{*}\otimes U$ and $\chi$ a section of $V$. The
notation $\std\cdot\chi$ is used for the section of $U$ given by
\begin{equation}
(\std\cdot\chi)(\bp)=\std(\bp)(\chi(\bp)).
\end{equation}

For two manifolds $\base$ and $\fb$, $C^{r}(\base,\fb)$ will denote
the collection of $C^{r}$-mappings from $\base$ to $\fb$. If $\xi:\fb\to\base$
is a fiber bundle, $C^{r}(\xi):=C^{r}(\base,\fb)$ is the space of
$C^{r}$-sections $\base\to\fb$.

\subsection{Manifolds with corners}

Our basic object will be a fiber bundle $\xi:\fb\tto\base$ where
$\base$ is assumed to be an oriented manifold with corners. We recall
(\eg, \cite{Corners_Douady,michor_manifolds_1980,Corners_Melrose,j.m._lee_introduction_2002,Corners_Handbook})
that an $n$-dimensional manifold with corners is a manifold whose
charts assume values in the $n$-quadrant of $\reals^{n}$, that is,
in
\begin{equation}
\overline{\reals}_{+}^{n}:=\{\bp\in\reals^{n}\mid\bp^{i}\ge0,\,i=\oneto n\}.
\end{equation}
In the construction of the manifold structure, it is understood that
a function defined on a quadrant is said to be differentiable if it
is the restriction to the quadrant of a differentiable function defined
on $\reals^{n}$. If $\base$ is an $n$-dimensional manifold with
corners, a subset $\mathcal{Z}\subset\base$ is defined to be a $k$-dimensional,
$k\le n$, submanifold with corners of $\base$ if for any $Z\in\mathcal{Z}$
there is a chart $(U,\vph)$, $Z\in U$, such that $\vph(\mathcal{Z}\cap U)\subset\{\bp\in\overline{\reals}_{+}^{n}\mid\bp^{l}=0,\,k<l\les n\}$.

With an appropriate natural definition of the integral of an $(n-1)$-form
over the boundary of a manifold with corners, Stokes's theorem holds
for manifolds with corners (see \cite[pp. 363--370]{j.m._lee_introduction_2002}).

Relevant to the subject at hand is the following result (see \cite{Corners_Douady,michor_manifolds_1980,Corners_Melrose,Michor2019}).
Every $n$-dimensional manifold with corners $\base$ is a submanifold
with corners of a manifold $\tilde{\base}$ without boundary of the
same dimension. In addition, if $\base$ is compact, it can be embedded
as a submanifold with corners in a compact manifold without boundary
$\tilde{\base}$ of the same dimension \cite[pp. I.24--26]{Corners_Melrose}.
Furthermore, $C^{k}$-forms defined on $\base$, may be extended continuously
and linearly to forms defined on $\tilde{\base}$. Such a manifold
$\tilde{\base}$ is referred to as\emph{ }an\emph{ extension} of $\base$.
Each smooth vector bundle over $\base$ extends to a smooth vector
bundle over $\tilde{\base}$. Each immersion (embedding) of $\base$
into a smooth manifold $\mathcal{Y}$ without boundary is the restriction
of an immersion (embedding) of $\tilde{\base}$ into $\mathcal{Y}$.

It is emphasized that manifold with corners do not model some basic
geometric shapes such as a pyramid with a rectangular base or a cone.
However, much of material presented in this review is valid for a
class of much more general objects,\emph{ Whitney manifold germs }as
presented in \cite{Michor2019}.

\subsection{Bundles, jets and iterated jets\label{subsec:Jets-1}}

We will consider a fiber bundle $\xi:\fb\to\base$, where $\base$
is $n$-dimensional and the typical fiber is $m$-dimensional. The
projection $\xi$ is represented locally by $(\bp^{i},y^{\ga})\mapsto(\bp^{i})$,
$i=\oneto n$, $\ga=\oneto m$. Let 
\begin{equation}
T\xi:T\fb\tto T\base
\end{equation}
be the tangent mapping represented locally by 
\begin{equation}
(\bp^{i},y^{\ga},\dot{\bp}^{j},\dot{y}^{\gb})\lmt(\bp^{i},\dot{\bp}^{j}).
\end{equation}
The \emph{vertical sub-bundle} $V\fb$ of $T\fb$ is the kernel of
$T\xi$. An element $v\in V\fb$ is represented locally as $(\bp^{i},y^{\ga},0,\dot{y}^{\gb})$.
With some abuse of notation, we will write both $\tau:T\fb\to\fb$
and $\tau:V\fb\to\fb$. For $v\in V\fb$ with $\tau(v)=y$ and $\xi(y)=\bp$,
we may view $v$ as an element of $T_{y}(\fb_{\bp})=T_{y}(\xi^{-1}(\bp))$.
In other words, elements of the vertical sub-bundle are tangent vectors
to $\fb$ that are tangent to the fibers.

Let $\conf:\base\to\fb$ be a section and let 
\begin{equation}
\conf^{*}\tau:\conf^{*}V\fb\tto\base
\end{equation}
 be the pullback of the vertical sub-bundle. Then, we may identify
$\conf^{*}V\fb$ with the restriction of the vertical bundle to $\image\conf$.

\subsubsection{Jets}

We will denote by $\xi^{r}:J^{r}(\base,\fb)\to\base$ the corresponding
$r$-jet bundle of $\xi$. When no ambiguity may occur, we will often
use the simpler notation $\xi^{r}:J^{r}\fb\to\base$ and refer to
a section of $\xi^{r}$ as a section of $J^{r}\fb$. One has the additional
natural projections $\xi_{l}^{r}:J^{r}(\base,\fb)\to J^{l}(\base,\fb)$,
$l<r$, and in particular $\xi_{0}^{r}:J^{r}(\base,\fb)\to\fb=J^{0}\fb$
\cite{saunders_geometry_1989}. The jet extension mapping associates
with a $C^{r}$-section, $\conf$, of $\xi$, a continuous section
$j^{r}\conf$ of the jet bundle $\xi^{r}$.

Let $\conf:\base\to\fb$ be a section of $\xi$ which is represented
locally by 
\begin{equation}
\vct X\lmt(\vct X,\vct y=\bs{\conf}(\vct X)),\qquad\text{or,}\qquad\bp^{i}\lmt(\bp^{i},\conf^{\ga}(\bp^{j})),
\end{equation}
$i=\oneto n$, $\ga=\oneto m$. Then, denoting the $k$-th derivative
of $\bs{\conf}$ by $D^{k}$, a local representative of $j^{r}\conf$
is of the form
\begin{equation}
\vct X\lmt(\vct X,\bs{\conf}(\vct X),\dots,D^{r}\bs{\conf}(\vct{\bp})),\qquad\text{or,}\qquad\bp^{i}\lmt(\bp^{i},\conf_{,\smi I}^{\ga}(\bp^{j})),\quad0\le\abs{\smi I}\le r.
\end{equation}
Accordingly, an element $A\in J^{r}(\base,\fb)$ is represented locally
by the coordinates
\begin{equation}
(\bp^{i},A_{\smi I}^{\ga}),\quad0\le\abs{\smi I}\le r.
\end{equation}

\subsubsection{Iterated (non-holonomic) jets}

Completely non-holonomic jets for the fiber bundle $\xi:\fb\to\base$
are defined inductively as follows. Firstly, one defines the fiber
bundles
\begin{equation}
\nJ^{0}(\base,\fb)=\fb,\qquad\nJ^{1}(\base,\fb):=J^{1}(\base,\fb),
\end{equation}
and projections
\begin{equation}
\nh{\xi}^{1}=\xi^{1}:\nJ(\base,\fb)\tto\base,\qquad\nh{\xi}_{0}^{1}=\xi_{0}^{1}:\nJ^{1}(\base,\fb)\tto\fb.
\end{equation}
Then, we define the \emph{iterated $\r$-jet bundle }as
\begin{equation}
\nJ^{\r}\xty:=J^{1}(\base,\nJ^{\r-1}\xty),
\end{equation}
with projection
\begin{equation}
\nh{\xi}^{\r}=\nh{\xi}^{\r-1}\comp\xi_{\r-1}^{1,\r}:\nJ^{\r}\xty\tto\base,\label{eq:der_NHJ-1}
\end{equation}
where,
\begin{equation}
\xi_{\r-1}^{1,\r}:\nJ^{\r}\xty=J^{1}(\base,\nJ^{\r-1}\xty)\tto\nJ^{\r-1}\xty.
\end{equation}
By induction, $\nh{\xi}^{\r}:\nJ^{\r}\xty\to\base$ is a well defined
fiber bundle..

When the projections $\xi_{\r-1}^{1,\r}$ are used inductively $l$-times,
we obtain a projection
\begin{equation}
\nh{\xi}_{\r-l}^{\r}:\nJ^{\r}\xty\tto\nJ^{\r-l}\xty.
\end{equation}

Let $\conf:\base\to\fb$ be a $C^{\r}$-section of $\xi$. The iterated
jet extension mapping
\begin{equation}
\nj^{\r}:C^{\r}(\xi)\tto C^{0}(\nh{\xi}^{\r})
\end{equation}
is naturally defined by 
\begin{equation}
\nj^{1}=j^{1}:C^{l}(\xi)\tto C^{l-1}(\nh{\xi}^{1}),\qquad\text{and}\qquad\nj^{\r}=j^{1}\comp\nj^{\r-1}.
\end{equation}
Note that we use $j^{1}$ here as a generic jet extension mapping,
omitting the indication of the domain.

There is a natural inclusion 
\begin{equation}
\incl^{\r}:J^{\r}\xty\tto\nJ^{\r}\xty,\qquad\text{given by}\qquad j^{\r}\conf(\bp)\lmt\nj^{\r}\conf(\bp).\label{eq:Incl_Jets_in_NH-1}
\end{equation}
Let $\pi:\vb\to\base$ be a vector bundle, then $\nh{\pi}^{1}=\pi^{1}:\nJ^{1}W=J^{1}\vb\to\base$
is a vector bundle. Continuing inductively, 
\begin{equation}
\nh{\pi}^{\r}:\nJ^{\r}\vb\tto\base
\end{equation}
is a vector bundle. In this case, the inclusion $\incl^{r}:J^{r}\pi\to\nh J^{r}\pi$
is linear. Naturally, elements in the image of $\incl^{r}$ are referred
to as \emph{holonomic}.

\subsubsection{Local representation of iterated jets}

The local representatives of iterated jets are also constructed inductively.
Hence, at each step, $G$, to which we refer as\emph{ generation,
}the number of arrays is multiplied. Hence powers of two are naturally
used below. Thus, it is proposed to use multi-indices of the form
$\mii_{\rp}$, where $\rp$, $\rq$, etc. are binary numbers that
enumerate the various arrays included in the representation.  For
example, a typical element of $\nJ^{3}(\base,\fb)$, in the form
\begin{equation}
(\bp^{j};y^{\ga};y_{i_{1}}^{1\gb_{1}};y_{i_{2}}^{2\gb_{2}},y_{i_{3}i_{4}}^{3\gb_{3}};y_{i_{5}}^{4\gb_{4}},y_{i_{6}i_{7}}^{5\gb_{5}},y_{i_{8}i_{9}}^{6\gb_{6}},y_{i_{10}i_{11}i_{12}}^{7\gb_{3}})\label{eq:Example-RepNH-1-1}
\end{equation}
is written as
\begin{equation}
(\bp^{j};y_{0}^{0\gb_{0}};y_{I_{1}}^{1\gb_{1}};y_{\mii_{10}}^{10\gb_{10}},y_{\mii_{11}}^{11\gb_{11}};y_{I_{100}}^{100\gb_{100}},y_{\mii_{101}}^{101\gb_{101}},y_{\mii_{110}}^{110\gb_{110}},y_{\mii_{111}}^{111\gb_{111}}),\label{eq:Example-RepNH-2-1}
\end{equation}
and for short
\begin{equation}
(\bp^{j};y_{\mii_{\rp}}^{\rp\gb_{\rp}}),\quad\text{for all }\rp\text{ with }0\le G_{\rp}\le3.\label{eq:Exmample_RepNH-3-1}
\end{equation}
Here, $G_{p}$ is the generation where the $\rp$-th array appears
and it is given by
\begin{equation}
G_{\rp}=\lfloor\log_{2}\rp\rfloor+1,
\end{equation}
where $\lfloor\log_{2}\rp\rfloor$ denotes the integer part of $\log_{2}\rp$.
In (\ref{eq:Example-RepNH-1-1},\ref{eq:Example-RepNH-2-1}) the generations
are separates by semicolons. As indicated in the example above, with
each $\rp$ we associate a multi-index $\mii_{\rp}=i_{1}\cdots i_{p}$
as follows. For each binary digit 1 in $\rp$ there is an index $i_{l}$,
$l=\oneto p$. Thus, the total number of digits 1 in $\rp$, which
is denoted by $\abs{\rp}$, is the total number of indices, $p$,
in $\mii_{\rp}$. In other words, the length, $|\mii_{\rp}|$, of
the induced multi-index $\mii_{\rp}$ satisfies
\begin{equation}
|\mii_{\rp}|=p=\abs{\rp}.
\end{equation}
Note also that the expression $\gb_{\rp}$, is not a multi-index since
we use upper-case letters to denote multi-indices. Here, the subscript
$\rp$ serves for the enumeration of the $\gb$ indices. If no ambiguity
may arise, we will often make the notation somewhat shorter and write
$y_{\mii_{\rp}}^{\gb_{\rp}}$ for $y_{\mii_{\rp}}^{\rp\gb_{\rp}}$.
 Continuing by induction, let a section $\hat{A}$ of $\nJ^{\r-1}\xty$
be represented locally by $(\bp^{j};y_{\mii_{\rp}}^{\rp\ga_{\rp}}(\bp^{j}))$,
$G_{\rp}\le\r-1$. Then, its $1$-jet extension, a section of $\nJ^{r}(\base,\fb)$,
is of the form
\begin{equation}
(\bp^{j};y_{\mii_{\rp}}^{\rp\ga_{\rp}}(\bp^{J});y_{\mii_{\rp},k_{\rp}}^{\rp\gb_{\rp}}(\bp^{j})),\qquad G_{\rp}\le\r-1,
\end{equation}
or equivalently,
\begin{equation}
(\bp^{j};y_{\mii_{\rp}}^{\rp\ga_{\rp}}(\bp^{j});y_{\mii_{\rp},k_{\rp}}^{1\rp\ga_{1\rp}}(\bp^{j})),\qquad G_{\rp}\le\r-1,
\end{equation}
where $1\rp$ is the binary representation of $2^{\r}+p$. It is
noted that the array $y^{1\rp}$ contains the derivatives of the array
$y^{\rp}$, and that $G_{1\rp}=\r$. Thus indeed, the number of digits
$1$ that appear in $\rq$, \ie, $\abs{\rq}$, determine the length
of the index $\mii_{\rq}$.

It follows that an element of $\nJ^{\r}\xty$ may be represented in
the form
\begin{equation}
(\bp^{j};y_{\mii_{\rp}}^{\rp\ga_{\rp}};y_{\mii_{\rp}k_{\rp}}^{\rp\ga_{1\rp}}),\qquad G_{\rp}\le\r-1
\end{equation}
or
\begin{equation}
(\bp^{j};y_{\mii_{\rq}}^{\rq\ga_{\rq}}),\quad\text{for all }\rq\text{ with }G_{\rq}\le\r.\label{eq:rep_NH_jet-1}
\end{equation}
That is, for each $\rp$ with $G_{\rp}\le r-1$, we have an index
$\rq=\rq(\rp)$ such that $\rq=\rq(\rp)=1\rp$ if $G_{\rq}=r$ and
$\rq=\rq(\rp)=\rp$ if $G_{\rq}<r$. 

A similar line of reasoning leads to the expression for the local
representatives of the iterated jet extension mapping. For a section
$\conf:\base\to\fb$, the iterated jet extension $B=\nj^{\r}\conf$,
a section of $\nh{\xi}^{\r}$, the local representation $(\bp^{j};B_{\mii_{\rq}}^{\rq\ga_{\rq}})$,
$G_{\rq}\le\r$, $|\mii_{\rq}|=\abs{\rq}$, satisfies 
\begin{equation}
B_{\mii_{\rq}}^{\rq\ga_{\rq}}=\conf_{,\mii_{\rq}}^{\ga_{\rq}},\quad|\mii_{\rq}|=\abs{\rq},\text{ indpendenly of the particular value of }\rq.\label{eq:Rep_It_Jet_Ext-1}
\end{equation}
Indeed, if $(\bp^{j},B_{\mii_{\rq}}^{\rq\ga_{\rq}})$, $G_{\rq}\le\r-1$,
with $B_{\mii_{\rq}}^{\rq\ga_{\rq}}=\conf_{,\mii_{\rq}}^{\ga_{\rq}}$
represent $\nj^{\r-1}\conf$, then, $\nj^{\r}\conf$ is represented
locally by 
\begin{equation}
(\bp^{\ib},B_{\mii_{\rq}}^{\rq\ga_{\rq}};B_{\mii_{\rq},j_{\rq}}^{\rq\ga_{1\rq}}),\quad G_{\rq}\le\r-1.
\end{equation}
Thus, by induction, any $\rp=\rp(\rq)$ with $G_{\rp}=\r$ and $G_{\rq}<r$,
may be written as $\rp=2^{\r-1}+\rq$, $\mii_{\rp}=\mii_{\rq}j_{\rq}$,
so that $B_{\mii_{\rp}}^{\rp\ga_{\rp}}=B_{\mii_{\rq},j_{\rq}}^{\rq\ga_{1\rq}}=\conf_{,\mii_{\rp}}^{\ga_{\rp}}$.

Let an element $A\in\nJ^{\r}\xty$ be represented by $(\bp^{j};y_{\mii_{\rp}}^{\rp\ga_{\rp}}),$
$G_{\rp}\le\r$, then $\nh{\xi}_{l}^{\r}(A)$ is represented by $(\bp^{\ib};y_{\mii_{\rp}}^{\rp\ga_{\rp}}),$
$G_{\rp}\le l$.

\subsection{Contraction\label{subsec:Isom-e-1}}

The right and left contractions of a $(p+r)$-form and a $p$-vector
are given respectively by
\begin{equation}
(\theta\fcontr\eta)(\eta')=\gth(\eta'\wedge\eta),\qquad(\eta\contr\gth)(\eta')=\gth(\eta\wedge\eta'),
\end{equation}
for every $r$-vector $\eta'$. We will use the notation
\begin{equation}
\spec C_{\llcorner}:\ext^{p}T\base\otimes\ext^{p+r}T^{*}\base\isom L(\ext^{p}T^{*}\base,\ext^{p+r}T^{*}\base)\tto\ext^{r}T^{*}\base,
\end{equation}
and
\begin{equation}
\spec C_{\lrcorner}:\ext^{p}T\base\otimes\ext^{p+r}T^{*}\base\isom L(\ext^{p}T^{*}\base,\ext^{p+r}T^{*}\base)\tto\ext^{r}T^{*}\base,
\end{equation}
for the mappings satisfying 
\begin{equation}
\spec C_{\llcorner}(\xi\otimes\theta)=\theta\fcontr\xi,\qquad\text{and}\qquad\spec C_{\lrcorner}(\xi\otimes\theta)=\xi\contr\theta,
\end{equation}
respectively. The left and right contraction differ by a factor of
$(-1)^{rp}$.

For the case $r+p=n$, $\dimension(\ext^{p}T\base\otimes\ext^{p+r}T^{*}\base)=\dimension\ext^{r}T^{*}\base$;
as the mappings $\spec C_{\llcorner}$ and $\spec C_{\lrcorner}$
are injective, they are invertible. Specifically, consider the mappings
\begin{equation}
e_{\llcorner}:\ext^{n-p}T^{*}\base\tto\ext^{p}T\base\otimes\ext^{n}T^{*}\base\isom L(\ext^{p}T^{*}\base,\ext^{n}T^{*}\base),
\end{equation}
and 
\begin{equation}
e_{\lrcorner}:\ext^{n-p}T^{*}\base\tto\ext^{p}T\base\otimes\ext^{n}T^{*}\base\isom L(\ext^{p}T^{*}\base,\ext^{n}T^{*}\base),
\end{equation}
given by 
\begin{equation}
e_{\fcor}(\go)(\psi)=\psi\wedge\go,\qquad\text{and}\qquad e_{\bcor}(\go)(\psi)=\go\wedge\psi,
\end{equation}
respectively. One can easily verify that these mappings are isomorphisms,
and in fact, they are the inverses of the the contraction mappings
defined above.

For example,
\begin{equation}
\begin{split}e_{\bcor}(\spec C_{\fcor}(\xi\otimes\theta))(\psi) & =\spec C_{\fcor}(\xi\otimes\theta)\wedge\psi,\\
 & =(\theta\fcontr\xi)\wedge\psi,\\
 & =\psi(\xi)\theta,\\
 & =(\xi\otimes\theta)(\psi),
\end{split}
\end{equation}
where we view $\xi$ s an element of the double dual. Thus, 
\begin{equation}
e_{\bcor}=\spec C_{\fcor}^{-1},\qquad\text{and}\qquad e_{\fcor}=\spec C_{\bcor}^{-1}.
\end{equation}

\section{Banachable Spaces of Sections of Vector Bundles over Compact Manifolds\label{sec:Banachable}}

For a compact manifold $\base$, the infinite dimensional Banach manifold
of mappings to a manifold $\spc$ and the manifold of sections of
the fiber bundle $\xi:\fb\to\base$, are modeled by Banachable spaces
of sections of vector bundles over $\base$, as will be described
in the next section. In this section we describe the Banachable structure
of such a space of differentiable vector bundle sections and make
some related observations. Thus, we consider in this section a vector
bundle $\pi:\vb\to\base$, where $\base$ is a smooth compact $n$-dimensional
manifold with corners and the typical fiber of $\vb$ is an $m$-dimensional
vector space. The space of $C^{\r}$-sections $\vf:\base\to\vb$,
$\r\ge0$, will be denoted by $C^{\r}(\pi)$ or by $C^{r}(\vb)$ if
no confusion may arise. A natural real vector space structure is induced
on $C^{\r}(\pi)$ by setting $(\vf_{1}+\vf_{2})(\bp)=\vf_{1}(\bp)+\vf_{2}(\bp)$
and $(c\vf)(\bp)=c\vf(\bp)$, $c\in\reals$.

\subsection{Precompact atlases\label{subsec:Precompact-atlases}}

Let $K_{a}$, $a=1,\dots,A$, be a finite collection of compact subsets
whose interiors cover $\base$ such that for each $a$, $K_{a}$ is
a subset of a domain of a chart $\vph_{a}:U_{a}\to\reals^{n}$ on
$\base$ and

\begin{equation}
(\vph_{a},\Phi_{a}):\pi^{-1}(U_{a})\tto\reals^{n}\times\reals^{m},\quad v\lmt(\bp^{\ib},v^{\ga})
\end{equation}
is some given vector bundle chart on $\vb$. Such a covering may always
be found by the compactness of $\base$ (using coordinate balls as,
for example, in \cite[p. 16]{j.m._lee_introduction_2002} or \cite[p. 10]{palais_foundations_1968}).
We will refer to such a structure as a \emph{precompact atlas}. The
same terminology will apply for the case of a fiber bundle.

\subsection{The $C^{r}$-topology on $C^{\protect\r}(\pi)$\label{subsec:Cr-Compact}}

For a section $\vf$ of $\pi$ and each $a=\oneto A$, let
\begin{equation}
\vf_{a}:\vph_{a}(K_{a})\tto\reals^{m},
\end{equation}
satisfying
\begin{equation}
\vf_{a}(\vph_{a}(\bp))=\Phi_{a}(\vf(\bp))\fall\bp\in K_{a},
\end{equation}
be a local representative of $\vf$.

Such a choice of a vector bundle atlas and subsets $K_{a}$ makes
it possible to define, for a section $\vf$,
\begin{equation}
\norm{\vf}^{\r}=\sup_{a,\ga,\abs{\smi{\mii}}\le\r}\braces{\sup_{\vct{\bp}\in\vph_{a}(K_{a})}\braces{\abs{(\vf_{a}^{\ga})_{,\smi{\mii}}(\vct{\bp})}}}.\label{eq:Cr-norm}
\end{equation}
Palais \cite[in particular, Chapter 4]{palais_foundations_1968} shows
that $\norm{\cdot}^{r}$ is indeed a norm endowing $C^{r}(\pi)$ with
a Banach space structure. The dependence of this norm on the particular
choice of atlas and sets $K_{a}$, makes the resulting space Banachable,
rather than a Banach space. Other choices will correspond to different
norms. However, norms induced by different choices will induce equivalent
topological vector space structures on $C^{\r}(\pi)$.

\subsection{The jet extension mapping\label{subsec:jet-extension}}

Next, one observes that the foregoing may be applied, in particular,
to the vector space $C^{0}(\pi^{\r})=C^{0}(J^{\r}\vb)$ of continuous
sections of the $r$-jet bundle $\pi^{\r}:J^{\r}\vb\to\base$ of $\pi$.
As a continuous section $B$ of $\pi^{\r}$ is locally of the form
\begin{equation}
(\bp^{\ib})\lmt(\bp^{\ib},B_{\smi I}^{\ga}(\bp^{\ib})),\qquad\abs{\smi I}\le\r,
\end{equation}
the analogous expression for the norm induced by a choice of a precompact
vector bundle atlas is
\begin{equation}
\norm B^{0}=\sup_{a,\ga,\abs{\smi{\mii}}\le\r}\braces{\sup_{\vct{\bp}\in\vph_{a}(K_{a})}\braces{\abs{B_{a\smi I}^{\ga}(\vct{\bp})}}}.
\end{equation}

Once, the topologies of $C^{\r}(\pi)$ and $C^{0}(\pi^{\r})$ have
been defined, one may consider the jet extension mapping
\begin{equation}
j^{r}:C^{\r}(\pi)\tto C^{0}(\pi^{\r}).
\end{equation}
For a section $\vf\in C^{\r}(\pi)$, with local representatives $\vf_{a}^{\ga}$,
$j^{\r}\vf$ is represented by a section $B\in C^{0}(\pi^{k})$, the
local representatives of which satisfy, 
\begin{equation}
B_{a\smi I}^{\ga}=\vf_{a,\smi I}^{\ga}.\label{eq:Cond-comp}
\end{equation}
Clearly, the mapping $j^{\r}$ is injective and linear. Furthermore,
it follows that 
\begin{equation}
\norm{j^{\r}\vf}^{0}=\sup_{a,\ga,\abs{\smi{\mii}}\le\r}\braces{\sup_{\vct{\bp}\in\vph_{a}(K_{a})}\braces{\abs{\vf_{a,\smi I}^{\ga}(\vct{\bp})}}}.
\end{equation}
(Note that since we take the supremum of all partial derivatives,
we could replace the non-decreasing multi-index $\smi{\mii}$ by a
regular multi-index $\mii$.) Thus, in view of Equation (\ref{eq:Cr-norm}),
\begin{equation}
\norm{j^{\r}\vf}^{0}=\norm{\vf}^{\r}
\end{equation}
and we conclude that $j^{\r}$ is a linear embedding of $C^{\r}(\pi)$
into $C^{0}(\pi^{\r})$. Evidently, $j^{\r}$ is not surjective as
a section $A$ of $\pi^{k}$ need not be compatible, \ie, it need
not satisfy (\ref{eq:Cond-comp}), for some section $\vf$ of $\pi$.
As a result of the above observations, $j^{\r}$ has a continuous
right inverse
\begin{equation}
(j^{\r})^{-1}:\image j^{\r}\subset C^{0}(\pi^{\r})\tto C^{\r}(\pi).\label{eq:Inv_Jet_Ext}
\end{equation}

\subsection{The iterated jet extension mapping\label{subsec:iterated-jet-ext}}

In analogy, we now consider the iterated jet extension mapping
\begin{equation}
\nj^{\r}:C^{\r}(\pi)\tto C^{0}(\nh{\pi}^{\r}).
\end{equation}
Specializing Equation (\ref{eq:rep_NH_jet-1}) for the case of the
non-holonomic $\r$-jet bundle 
\begin{equation}
\nh{\pi}^{\r}:\nJ^{\r}\vb\tto\base,
\end{equation}
a section $B$ of $\nh{\pi}^{\r}$ is represented locally in the form
\begin{equation}
\bp^{i}\lmt(\bp^{i},B_{a\mii_{\rq}}^{\rq\ga_{\rq}}(\bp^{i})),\quad\text{for all }\rq\text{ with }G_{\rq}\le\r.
\end{equation}
Thus, the induced norm on $C^{0}(\nh{\pi}^{\r})$ is given by
\begin{equation}
\norm B^{0}=\sup\braces{\bigl|B_{a\mii_{\rq}}^{\rq\ga_{\rq}}(\bp^{i})\bigr|},\label{eq:Norm_Sec_It_Jet}
\end{equation}
where the supremum is taken over all $\vct{\bp}\in\vph_{a}(K_{a})$,
$a=\oneto A$, $\ga_{\rq}=\oneto m$, $\mii_{\rq}$ with $\abs{\mii}=\abs{\rq}$,
and $\rq$ with $G_{\rq}\le\r$.

Specializing (\ref{eq:Rep_It_Jet_Ext-1}) for the case of a vector
bundle, it follows that if the section $B$ of $\nh{\pi}^{\r}$, satisfies
$B=\nj^{\r}\vf$, its local representatives satisfy
\begin{equation}
B_{a\mii_{\rq}}^{\rq\ga_{\rq}}=\vf_{a,\mii_{\rq}}^{\ga_{\rq}}\quad|\mii_{\rq}|=\abs{\rq},\text{ indpendenly of the particular value of }\rq.
\end{equation}
It follows that in 
\begin{equation}
\norm{\nj^{\r}\vf}^{0}=\sup\braces{\bigl|\vf_{a,\mii_{\rq}}^{\ga_{\rq}}(\bp^{\ib})\bigr|}
\end{equation}
(where the supremum is taken over all $\bp^{\ib}\in\vph_{a}(K_{a})$,
$a=\oneto A$, $\ga=\oneto m$, $\mii_{\rq}$ with $\abs{\mii}=\abs{\rq}$,
and $\rq$ with $G_{\rq}\le\r$), it is sufficient to take simply
all derivatives $\vf_{a,\mii}^{\ga}(\bp^{\ib})$, for $\abs I\le r$.
Hence,
\begin{equation}
\norm{\nj^{\r}\vf}^{0}=\sup\braces{\bigl|\vf_{a,\mii}^{\ga_{\rq}}(\bp^{\ib})\bigr|}
\end{equation}
where the supremum is taken over all $\bp^{\ib}\in\vph_{a}(K_{a})$,
$a=\oneto A$, $\ga=\oneto m$, and $\mii$ with $\abs{\mii}\le\r.$
It is therefore concluded that
\begin{equation}
\norm{\nj^{\r}\vf}^{0}=\norm{j^{\r}\vf}^{0}=\norm{\vf}^{\r}.\label{eq:Isometries}
\end{equation}
In other words, one has a sequence of linear embeddings

\begin{equation}
\label{eq:IterJetExt}
\begin{xy}
(40,25)*+{C^0(\nh\pi^\r)} ="TS";
(-10,25)*+{C^\r(\pi)} = "TQ";
(15,25)*+{C^0(\pi^\r)} ="jTS";
{\ar@{->}@/^{2pc}/^{\nj^\r} "TQ"; "TS"};
{\ar@{->}^{j^\r} "TQ"; "jTS"};
{\ar@{->}^{C^0(\incl^\r)} "jTS"; "TS"};
\end{xy}
\end{equation}where, $\incl^{r}:J^{r}\vb\to\hat{J}^{r}\vb$ is the natural inclusion
(\ref{eq:Incl_Jets_in_NH-1}) and $C^{0}(\incl^{r})$ defined as $C^{0}(\incl^{r})(A):=\incl^{r}\comp A$,
for every continuous section $A$ of $J^{\r}\vb$, is the inclusion
of sections. These embeddings are not surjective. In particular, sections
of $\nJ^{\r}\vb$ need not have the symmetry properties that hold
for sections of $J^{\r}\vb$.

\section{The Construction of Charts for the Manifold of Sections\label{sec:Man_of_Maps}}

In this section, we outline the construction of charts for the Banach
manifold structure on the collection of sections $C^{\r}(\xi)$ as
in \cite{palais_foundations_1968}. (See a detailed presentation of
the subject in this volume \cite{Michor2019}.)

Let $\conf$ be a $C^{\r}$-section of $\xi$. Similarly to the construction
of tubular neighborhoods, the basic idea is to identify points in
a neighborhood of $\image\conf$ with vectors at $\image\conf$ which
are tangent to the fibers. This is achieved by defining a second order
differential equation, so that a neighboring point $y$ in the same
fiber as $\conf(\bp)$ is represented through the solution $c(t)$
of the differential equations with the initial condition $v\in T_{\conf(\bp)}(\fb_{\bp})$
by $y=c(t=1)$. In other words, $y$ is the image of $v$ under the
exponential mapping.

To ensure that the image of the exponential mapping is located on
the same fiber, $\fb_{\bp}$, the spray inducing the second order
differential equation is a vector field
\begin{equation}
\go:V\fb\tto T(V\fb)
\end{equation}
which is again tangent to the fiber in the sense that for
\begin{equation}
T\tau_{\fb}:T(V\fb)\tto T\fb,\quad\text{one has, }\quad T\tau_{\fb}\comp\go\in V\fb.
\end{equation}
This condition, together with the analog of the standard condition
for a second order differential equation, namely, 
\begin{equation}
T\tau_{\fb}\comp\go(v)=v\fall v\in V\fb,
\end{equation}
imply that $\go$ is represented locally in the form
\begin{equation}
(\bp^{\ib},y^{\ga},0,\dot{y}^{\gb})\lmt(\bp^{\ib},y^{\ga},0,\dot{y}^{\gb},0,\dot{y}^{\ggm},0,\tilde{\go}^{\ga}(\bp^{\ib},y^{\ga},\dot{y}^{\gb})).
\end{equation}
Finally, $\go$ is a \emph{bundle spray }so that
\begin{equation}
\tilde{\go}^{\ga}(\bp^{\ib},y^{\ga},a_{0}\dot{y}^{\gb})=a_{0}^{2}\tilde{\go}^{\ga}(\bp^{\ib},y^{\ga},\dot{y}^{\gb}).
\end{equation}
Bundle sprays can always be defined on compact manifolds using partitions
of unity and the induced exponential mappings have the required properties.

The resulting structure makes it possible to identify an open neighborhood
$U$\textemdash \emph{a vector bundle neighborhood\textemdash }of
$\image\conf$ in $\fb$ with 
\begin{equation}
V\fb\resto{\image\conf}\simeq\conf^{*}V\fb.
\end{equation}
(We note that a rescaling is needed if $U$ is to be identified with
the whole of $\conf^{*}V\fb$. Otherwise, only an open neighborhood
of the zero section of $\conf^{*}V\fb$ will be used to parametrize
$U$.)

Once the identification of $U$ with $\conf^{*}V\fb$ is available,
the collection of sections $C^{\r}(\base,U)$ may be identified with
$C^{\r}(\conf^{*}V\fb)$, $\conf\in C^{\r}(\xi)$. Thus, a chart into
a Banachable space is constructed, where $\conf$ is identified with
the zero section.

\begin{figure}
\includegraphics[scale=0.65]{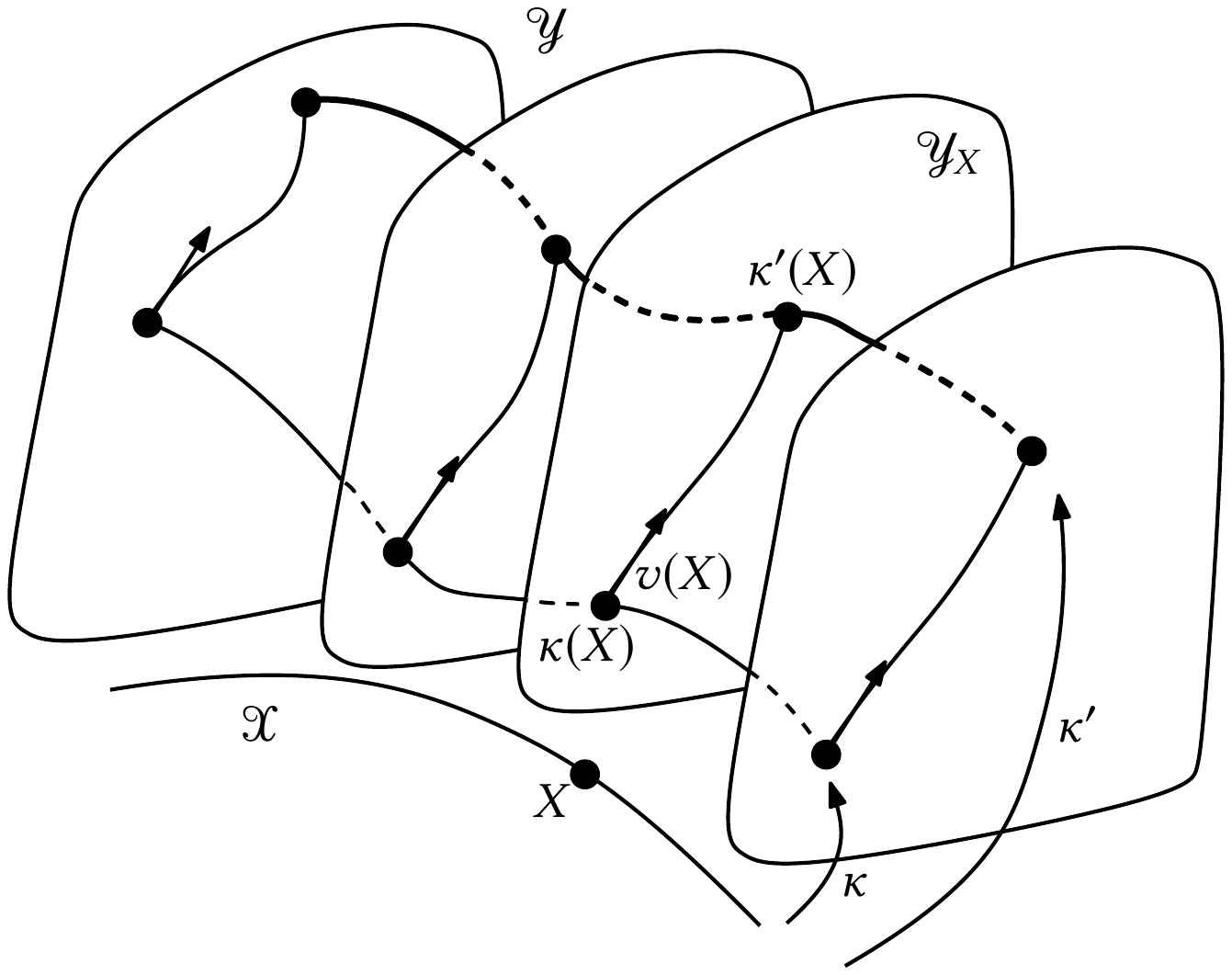}

\caption{Constructing the manifold of sections, a rough illustration.}
\end{figure}

The construction of charts on the manifold of sections, implies that
curves in $C^{r}(\xi)$ in a neighborhood of $\conf$ are represented
locally by curves in the Banachable space $C^{r}(\conf^{*}V\fb)$.
Thus, tangent vectors $\vf\in T_{\conf}C^{r}(\xi)$ may be identified
with elements of $C^{r}(\conf^{*}V\fb)$. We therefore make the identification
\begin{equation}
T_{\conf}C^{r}(\xi)=C^{r}(\conf^{*}V\fb).\label{eq:TanBund_Man_of_Sections}
\end{equation}

\section{The $C^{\protect\r}$-Topology on the Space of Sections of a Fiber
Bundle\label{sec:Topology-C^r}}

The topology on the space of sections of fiber bundles is conveniently
described in terms of filters of neighborhoods (\eg,  \cite{treves_topological_1967}).

\subsection{Local representatives of sections\label{subsec:rep-sections}}

We consider a fiber bundle $\xi:\fb\to\base$, where $\base$ is assumed
to be a compact manifold with corners and the typical fiber is a manifold
$\spc$ without a boundary. Let $\braces{(U_{a},\vph_{a},\Phi_{a})}$,
$a=\oneto A$, and $K_{a}\subset U_{a}$, be a precompact (as in Section
\ref{subsec:Precompact-atlases}) fiber bundle trivialization on $\fb$.
That is, the interiors of $K_{a}$ cover $\base$, and $(\vph_{a},\Phi_{a}):\xi^{-1}(U_{a})\to\reals^{n}\times\spc$.
 Let $\braces{(V_{b},\psi_{b})}$, $b=\oneto B$, be an atlas on
$\spc$ so that $\{V_{b}\}$ cover $\spc$.

Consider a $C^{k}$-section $\conf:\base\to\fb$. For any $a=\oneto A$,
we can set
\begin{equation}
\tilde{\conf}_{a}:U_{a}\tto\spc,\quad\text{by }\tilde{\conf}_{a}:=\Phi_{a}\comp\conf\resto{U_{a}}.
\end{equation}
Let 
\begin{equation}
U_{ab}:=U_{a}\cap\tilde{\conf}_{a}^{-1}(V_{b}),
\end{equation}
so that $\conf(U_{ab})\subset V_{b}$, and so, the local representatives
of $\conf$ are
\begin{equation}
\conf_{ab}:=\psi_{b}\comp\tilde{\conf}_{a}\comp\vph_{a}^{-1}\resto{\vph_{a}(U_{ab})}:\vph_{a}(U_{ab})\tto\psi_{b}(V_{b})\subset\reals^{m}.
\end{equation}
Thus, re-enumerating the subsets $\{U_{ab}\}$ and $\{V_{b}\}$ we
may assume that we have a precompact trivialization $\braces{(U_{a},\vph_{a},\Phi_{a})}$,
$a=\oneto A$, $K_{a}\subset U_{a}$, on $\fb$, and an atlas $\braces{(V_{a},\psi_{a})}$
on $\spc$ such that $\tilde{\conf}_{a}(U_{a})\subset V_{a}$. The
local representatives of $\conf$ relative to these atlases are
\begin{equation}
\conf_{a}:=\psi_{a}\comp\tilde{\conf}_{a}\comp\vph_{a}^{-1}\resto{\vph_{a}(U_{a})}:\vph_{a}(U_{a})\tto\psi_{a}(V_{a})\subset\reals^{m}.
\end{equation}

\subsection{Neighborhoods for $C^{r}(\xi)$ and the $C^{r}$-topology\label{subsec:neighborhoods-Cr}}

Let $\conf\in C^{r}(\xi)$ be given. Consider sets of sections of
the form $U_{\conf,\eps}$ induced by the collection of local representatives
as above and numbers $\eps>0$ in the form
\begin{equation}
U_{\conf,\eps}=\braces{\conf'\in C^{\r}(\xi)\mid\sup\braces{\abs{((\conf'_{a})^{\ga}-\conf_{a}^{\ga})_{,\mii}(\vct{\bp})}}<\eps},\label{eq:Defn_Open_N_Cr}
\end{equation}
where the supremum is taken over all
\[
\vct{\bp}\in\vph_{a}(K_{a}),\,\ga=\oneto m,\,\abs{\mii}\le r,\,a=\oneto A.
\]
The $C^{\r}$-topology on $C^{\r}(\xi)$ uses all such sets as a basis
of neighborhoods. Using the transformation rules for the various variables,
it may be shown that other choices of precompact trivialization and
atlas will lead to equivalent topologies. It is noted that we use
here the compactness of $\base$ which implies that the weak and strong
$C^{r}$-topologies (see \cite[p. 35]{m._hirsch_differential_1976})
become identical.
\begin{rem}
\label{rem:Subbasis}The collection of neighborhoods $\{U_{\conf,\eps}\}$
for the various values of $\eps$ generate a basis of neighborhoods
for the topology of $C^{r}(\xi)$. If one keeps the value of $a=\oneto A$,
fixed, then the collection of sections
\[
U_{\conf,\eps_{a},a}=\braces{\conf'\in C^{\r}(\xi)\mid\sup\braces{\abs{((\conf'_{a})^{\ga}-\conf_{a}^{\ga})_{,\mii}(\vct{\bp})}}<\eps_{a}},
\]
where the supremum is taken over all
\[
\vct{\bp}\in\vph_{a}(K_{a}),\,\ga=\oneto m,\,\abs{\mii}\le r,
\]
is a neighborhood as it contains the open neighborhood $U_{\conf,\eps_{a}}$.
In fact, since $\base$ is assumed to be compact, the collection of
sets of the form $\{U_{\conf,\eps_{a},a}\}$ is a sub-basis of neighborhoods
of $\conf$ for the topology on $C^{r}(\xi)$.
\end{rem}

\subsection{Open neighborhoods for $C^{\protect\r}(\xi)$ using vector bundle
neighborhoods}

In order to specialize the preceding constructions for the case where
a vector bundle neighborhood is used, we first consider local representations
of sections.

Let $\conf\in C^{\r}(\xi)$ be a section and let
\begin{equation}
\conf^{*}\tau_{\fb}:\conf^{*}V\fb\tto\base
\end{equation}
be the vector bundle identified with an open subbundle $U$ of $\fb$.
(We will use the two aspects of the vector bundle neighborhood, interchangeably.)
Since the typical fiber of $\conf^{*}V\fb$ is $\reals^{m}$, one
may choose a precompact vector bundle atlas $\braces{(U_{a},\vph_{a},\Phi_{a})}$,
$K_{a}\subset U_{a}$, $a=\oneto A$, on $\conf^{*}V\fb$, such that
\begin{equation}
(\vph_{a},\Phi_{a}):(\conf^{*}\tau_{\fb})^{-1}(U_{a})\tto\reals^{n}\times\reals^{m}.
\end{equation}
Thus, if we identify all open subsets $V_{a}$ in Section \ref{subsec:rep-sections}
above with the typical fiber $\reals^{m}$, the representatives of
a section are of the form 
\begin{equation}
\conf_{a}:=\Phi_{a}\comp\conf\resto{U_{a}}\comp\vph_{a}^{-1}\resto{\vph_{a}(U_{a})}:\vph_{a}(U_{a})\tto\reals^{m}.
\end{equation}

A basic neighborhood of $\conf$ is given by Equation (\ref{eq:Defn_Open_N_Cr}).
However, from the point of view of a vector bundle neighborhood, $\conf$
is represented by the zero section and each $\conf'$ is viewed as
a section of the vector bundle $\conf^{*}\tau_{\fb}$, which we may
denote by $\vf'$. Thus
\begin{equation}
U_{\conf,\eps}\simeq\braces{\vf'\in C^{\r}(\conf^{*}\tau_{\fb})\mid\sup\braces{\abs{(\vf'_{a})_{,\mii}^{\ga}(\bp^{\ib})}}<\eps}.
\end{equation}
In other words, using the structure of a vector bundle neighborhood
we have
\begin{equation}
U_{\conf,\eps}\simeq\braces{\vf'\in C^{\r}(\conf^{*}\tau_{\fb})\mid\norm{w'}^{r}<\eps},
\end{equation}
so that $U_{\conf,\eps}$ is identified with a ball of radius $\eps$
in $C^{\r}(\conf^{*}\tau_{\fb})$ at the zero section.

It is concluded that the charts on $C^{\r}(\xi)$ induced by the vector
bundle neighborhood are compatible with the $C^{\r}$-topology on
$C^{\r}(\xi)$.

\section{The Space of Embeddings\label{sec:Embeddings}}

The kinematic aspect of the Lagrangian formulation of continuum mechanics
is founded on the notion of a configuration, an embeddings of a body
manifold $\base$ into the space manifold $\spc$. The restriction
of configurations to be embeddings, rather than generic $C^{r}$-mappings
of the body into space follows from the traditional principle of material
impenetrability which requires that configurations be injective and
that infinitesimal volume elements are not mapped into elements of
zero volume.

It is noted that any configuration $\conf:\base\to\spc$ may be viewed
as a section of the trivial fiber bundle $\xi:\base\to\fb=\base\times\spc$.
Thus, the constructions described above apply immediately to configurations
in continuum mechanics. In this particular case, we will write $C^{\r}\xts$
for the collection of all $C^{r}$-mappings. Our objective in this
section is to describe how the set of embeddings $\embds^{\r}(\base,\spc)$
constitutes an open subset of $C^{\r}\xts$ for $\r\ge1$. In particular,
it will follow that that at each configuration $\conf$, $T_{\conf}\embds^{\r}\xts=T_{\conf}C^{r}\xts$.
Since the $C^{\r}$-topologies, for $\r>1$, are finer than the $C^{1}$-topology,
it is sufficient to prove that $\embds^{1}\xts$ is open in $C^{1}\xts$.
This brings to light the special role that the case $\r=1$ plays
in continuum mechanics.

\subsection{The case of a trivial fiber bundle\textemdash manifolds of mappings}

It is observed that the definitions of Sections \ref{subsec:rep-sections}
and \ref{subsec:neighborhoods-Cr} hold with natural simplifications
for the case of the trivial bundle. Thus, we use a precompact atlas
$\braces{(U_{a},\vph_{a})}$, $a=\oneto A$, and $K_{a}\subset U_{a}$,
in $\base$ (the interiors, $\inter{K_{a}}$, cover $\base$). Given
$\conf\in C^{1}\xts$, we can find an atlas $\braces{(V_{a},\psi_{a})}$
on $\spc$ such that $\conf(U_{a})\subset V_{a}$. The local representatives
of $\conf$ are of the form 
\begin{equation}
\conf_{a}=\psi_{a}\comp\conf\resto{U_{a}}\comp\vph_{a}^{-1}:\vph_{a}(U_{a})\tto\psi_{a}(V_{a})\subset\reals^{m}.
\end{equation}
For the case $\r=1$, Equation (\ref{eq:Defn_Open_N_Cr}) reduces
to
\begin{equation}
U_{\conf,\eps}=\braces{\conf'\in C^{1}\xts\,\bigl|\,\sup\braces{\bigl|(\conf'_{a})^{\ga}(\vct{\bp})-\conf_{a}^{\ga}(\vct{\bp})\bigr|,\bigl|(\conf'_{a})_{,j}^{\ga}(\vct{\bp})-\conf_{a,j}^{\ga}(\vct{\bp})\bigr|}<\eps},\label{eq:Nbhds-Cr(X,S)}
\end{equation}
where the supremum is taken over all
\[
\vct{\bp}\in\vph_{a}(K_{a}),\,\ga=\oneto m,\,a=\oneto A.
\]

\begin{rem}
\label{rem:Subbasis_Man_of_Maps}It is noted that in analogy with
Remark \ref{rem:Subbasis}, for a fixed $a=\oneto A$, a subset of
mappings of the form
\begin{equation}
U_{\conf,\eps,a}=\braces{\conf'\in C^{1}\xts\bigmid\sup\braces{\bigl|(\conf'_{a})^{\ga}(\vct{\bp})-\conf_{a}^{\ga}(\vct{\bp})\bigr|,\bigl|(\conf'_{a})_{,j}^{\ga}(\vct{\bp})-\conf_{a,j}^{\ga}(\vct{\bp})\bigr|}<\eps},\label{eq:Nbhds-subbasis}
\end{equation}
where the supremum is taken over all
\[
\vct{\bp}\in\vph_{a}(K_{a}),\,j=1,\dots,n,\,\ga=\oneto m,
\]
is a neighborhood of $\conf$ as it contains a neighborhood as defined
above. The collection of such sets for various values of $a$ and
$\eps$ form a subbasis of neighborhoods for the topology on $C^{1}(\base,\spc)$.
\end{rem}

\subsection{The space of immersions\label{subsec:Immersions}}

Let $\conf\in C^{1}\xts$ be an immersion, so that $T_{\bp}\conf:T_{\bp}\base\to T_{\conf(\bp)}\spc$
is injective for every $\bp\in\base$. We show that there is a neighborhood
$U_{\conf}\subset C^{1}\xts$ of $\conf$ such that all $\conf'\in U_{\conf}$
are immersions.

Note first, that since the evaluation of determinants of $n\times n$
matrices is a continuous mapping, the collection of $m\times n$ matrices
for which all $n\times n$ minors vanish is a closed set. Hence, the
collection $L_{\text{In}}(\reals^{n},\reals^{m})$ of all injective
$m\times n$ matrices is open in $L(\reals^{n},\reals^{m})$. Let
$\conf$ be an immersion with representatives $\conf_{a}$ as above.
For each $a$, the derivative mapping 
\begin{equation}
D\conf_{a}:\vph_{a}(U_{a})\tto L(\reals^{n},\reals^{m}),\quad\vct{\bp}\lmt D\conf_{a}(\vct{\bp}),
\end{equation}
is continuous, hence, $D\conf_{a}(K_{a})$ is a compact set of injective
linear mappings. Choosing any norm in $L(\reals^{n},\reals^{m})$,
one can cover $D\conf_{a}(K_{a})$ by a finite number of open balls
all containing only injective mappings. In particular, setting 
\begin{equation}
\norm T=\max_{i,\ga}\braces{\abs{T_{i}^{\ga}}},\qquad T\in L\ntm,
\end{equation}
let $\eps_{a}$ be the least radius of balls in this covering. Thus,
we are guaranteed that any linear mapping $T$, such that $\norm{T-D\conf_{a}(\vct{\bp})}<\eps_{a}$
for some $\vct{\bp}\in\vph_{a}(K_{a})$, is injective. Specifically,
for any $\conf'\in C^{1}\xts$, if 
\begin{equation}
\sup_{\vct{\bp}\in\vph_{a}(K_{a})}\bigl|(\conf'_{a})_{,j}^{\ga}(\vct{\bp})-\conf_{a,j}^{\ga}(\vct{\bp})\bigr|\les\eps_{a},
\end{equation}
$D\conf_{a}'$ is injective everywhere in $\vph_{a}(K_{a})$. Letting
$\eps=\min_{a}\eps_{a}$, any configuration in $U_{\conf,\eps}$ as
in (\ref{eq:Nbhds-Cr(X,S)}) is an immersion.

\subsection{Open neighborhoods of local embeddings}

Let $\conf\in C^{1}(\base,\spc)$ and $\bp\in\base.$ It is shown
below that if $T_{\bp}\conf$ is injective, then there is a neighborhood
of mappings $U_{\conf,\bp}$ of $\conf$ such that every $\conf'\in U_{\conf,\bp}$
is injective in some fixed neighborhood of $\bp$. Specifically, there
is a neighborhood $W_{\bp}$ of $\bp$, and a neighborhood $U_{\conf,\bp}$
of $\conf$ such that for each $\conf'\in U_{\conf,\bp}$, $\conf'\resto{W_{\bp}}$
is injective.

Let $(U,\vph)$ and $(V,\psi)$ be coordinate neighborhoods of $\bp$
and $\conf(\bp)$, respectively, such that $\conf(U)\subset V$ .
Let $\vct{\bp}$ and $\vct{\conf}$ be the local representative of
$\bp$ and $\conf$ relative to these charts. Thus, we are guaranteed
that
\begin{equation}
M:=\inf_{\abs{\vct v}=1}\abs{D\vct{\conf}(\vct{\bp})(\vct v)}>0.
\end{equation}
By a standard corollary of the inverse function theorem, due to the
injectivity of $T_{\bp}\conf$, we can choose $U$ to be small enough
so that the restrictions of $\conf$ and $\bs{\conf}$ to $U$ and
its image under $\vph$, respectively, are injective. Next, let $W_{\bp}$
be a neighborhood of $\bp$ such that $\vph(W_{\bp})$ is convex and
its closure, $\overline{W}_{\bp}$, is a compact subset of $U$. Thus,
define the neighborhood $U_{\conf,\bp}\subset C^{1}\xts$ whose elements,
$\conf'$, satisfy the conditions
\begin{equation}
\conf'(\overline{W}_{\bp})\subset V,\text{ \ensuremath{\quad}and}\quad\abs{D\vct{\conf}'(\vct{\bp}')-D\vct{\conf}(\vct{\bp})}<\frac{M}{2}\fall\vct{\bp}'\in\vph(\overline{W}_{\bp}).
\end{equation}
By the definition of neighborhoods in $C^{1}\xts$ in (\ref{eq:Nbhds-Cr(X,S)}),
$U_{\conf,\bp}$ contains a neighborhood of $\conf$, hence, it is
also a neighborhood.

Next, it is shown that the fact that the values of the derivatives
of elements of $U_{\conf,\bp}$ are close to the injective $D\v{\conf}(\v{\bp})$
everywhere in $\cl W_{\bp}$, implies that these mappings are close
to the linear approximation using $D\v{\conf}(\v{\bp})$, which in
turn, implies injectivity in $\cl W_{\bp}$ of these elements. Specifically,
for and $\v{\bp}_{1},\v{\bp}_{2}\in\cl W_{\bp}$, since 
\begin{equation}
\v{\conf}'(\v{\bp}_{2})-\v{\conf}'(\v{\bp}_{1})=\v{\conf}'(\v{\bp}_{2})-\v{\conf}'(\v{\bp}_{1})-D\v{\conf}(\v{\bp})(\v{\bp}_{2}-\v{\bp}_{1})+D\v{\conf}(\v{\bp})(\v{\bp}_{2}-\v{\bp}_{1}),
\end{equation}
the triangle inequality implies that 
\begin{equation}
\begin{split}\abs{\v{\conf}'(\v{\bp}_{2})-\v{\conf}'(\v{\bp}_{1})} & \ge\abs{D\v{\conf}(\v{\bp})(\v{\bp}_{2}-\v{\bp}_{1})}-\abs{\v{\conf}'(\v{\bp}_{2})-\v{\conf}'(\v{\bp}_{1})-D\v{\conf}(\v{\bp})(\v{\bp}_{2}-\v{\bp}_{1})},\\
 & \ge M\abs{\v{\bp}_{2}-\v{\bp}_{1}}-\abs{\v{\conf}'(\v{\bp}_{2})-\v{\conf}'(\v{\bp}_{1})-D\v{\conf}(\v{\bp})(\v{\bp}_{2}-\v{\bp}_{1})}.
\end{split}
\end{equation}
Using the mean value theorem, there is a point $\v{\bp}_{0}\in\vph(\cl W_{\bp})$
such that 
\begin{equation}
\v{\conf}'(\v{\bp}_{2})-\v{\conf}'(\v{\bp}_{1})=D\v{\conf}'(\v{\bp}_{0})(\v{\bp}_{2}-\v{\bp}_{1}).
\end{equation}
Hence,
\begin{equation}
\begin{split}\abs{\v{\conf}'(\v{\bp}_{2})-\v{\conf}'(\v{\bp}_{1})-D\v{\conf}(\v{\bp})(\v{\bp}_{2}-\v{\bp}_{1})} & =\abs{(D\v{\conf}'(\v{\bp}_{0})-D\v{\conf}(\v{\bp}))(\v{\bp}_{2}-\v{\bp}_{1})},\\
 & \le\abs{D\v{\conf}'(\v{\bp}_{0})-D\v{\conf}(\v{\bp})}\abs{\v{\bp}_{2}-\v{\bp}_{1}},\\
 & <\frac{M}{2}\abs{\v{\bp}_{2}-\v{\bp}_{1}}.
\end{split}
\end{equation}
It follows that 
\begin{equation}
\abs{\v{\conf}'(\v{\bp}_{2})-\v{\conf}'(\v{\bp}_{1})}>\frac{M}{2}\abs{\v{\bp}_{2}-\v{\bp}_{1}},
\end{equation}
which proves the injectivity.

\subsection{Open neighborhoods of embeddings}

Finally, it is shown how every $\conf\in\embds^{1}\xts\subset C^{1}\xts$
has a neighborhood consisting of embeddings only. It will follow that
$\embds^{1}\xts$ is an open subset of $C^{1}\xts$. This has far-reaching
consequences in continuum mechanics and it explains the special role
played by the $C^{1}$-topology in continuum mechanics.

Let $\conf$ be a given embedding. Using the foregoing result, for
each $\bp\in\base$ there is an open neighborhood $W_{\bp}$ of $\bp$
and a neighborhood $U_{\conf,\bp}$ of $\conf$, such that for each
$\conf'\in U_{\conf,\bp}$, $\conf'\resto{\cl W_{\bp}}$ is injective.
The collection of neighborhoods $\{W_{\bp}\}$, $\bp\in\base$, is
an open cover of $\base$ and by compactness, it has a finite sub-cover.
Denote the finite number of open sets of the form $W_{x}$ as above
by $W_{a}$, $a=\oneto A$, so that $K_{a}:=\cl W_{a}$ is a compact
subset of $U_{a}$, $\conf(U_{a})\subset V_{a}$. For each $a$, we
have a neighborhood $U_{\conf,a}$ of $\conf$ such that each $\conf'\in U_{\conf,a}$
satisfies the condition that $\conf'\resto{K_{a}}$ is injective.
Let $\mathcal{N}_{1}=\bigcap_{a=1}^{A}U_{\conf,a}$ so that for each
$\conf'\in\mathcal{N}_{1}$, $\conf'\resto{K_{a}}$ is injective for
all $a$. Let $\mathcal{N}_{2}$ be a neighborhood of $\conf$ which
contains only immersions as in Section \ref{subsec:Immersions}. Thus,
$\mathcal{N}_{0}=\mathcal{N}_{1}\cap\mathcal{N}_{2}$ contains immersions
which are locally injective.

Let $\conf$ be an embedding and $\mathcal{\mathcal{N}}_{0}$ as above.
If there is no neighborhood of $\conf$ that contains only injective
mappings, then, for each $\nu=1,2,\dots$, there is a $\conf_{\nu}\in U_{\conf,\eps_{\nu}}$,
$\eps_{\nu}=\nicefrac{{\displaystyle 1}}{{\displaystyle \nu}}$, and
points $\bp_{\nu},\bp'_{\nu}\in\base$, $\bp_{\nu}\ne\bp'_{\nu}$,
such that $\conf_{\nu}(\bp_{\nu})=\conf_{\nu}(\bp'_{\nu})$. As $\mathcal{N}_{0}$
is a neighborhood of $\conf$, we may assume that $\conf_{\nu}\in\mathcal{N}_{0}$
for all $\nu$. By the compactness of $\base$ and $\base\times\base$,
we can extract a converging subsequence from the sequence $((\bp_{\nu},\bp'_{\nu}))$
in $\base\times\base$ We keep the same notation for the converging
subsequences and let 
\begin{equation}
(\bp_{\nu},\bp'_{\nu})\tto(\bp,\bp'),\qquad\text{as}\quad\nu\tto\infty.
\end{equation}

We first exclude the possibility that $\bp=\bp'$. Assume $\bp=\bp'\in K_{a_{0}}$,
for some $a_{0}=\oneto A$. Then, for any neighborhood $U_{\conf,\eps_{\nu}}$
of $\conf$ and any neighborhood of $\bp=\bp'$, there is a configuration
$\conf_{\nu}$ such that $\conf_{\nu}$ is not injective. This contradicts
the construction of local injectivity above.

Thus, one should consider the situation for which $\bp\ne\bp'$. Assume
$\bp\in K_{a_{0}}$ and $\bp'\in K_{a_{1}}$ for $a_{0},a_{1}=\oneto A$.
By the definition of $U_{\conf,\eps_{\nu}}$, the local representatives
of $\conf_{\nu}\resto{K_{a_{0}}}$ and $\conf_{\nu}\resto{K_{a_{1}}}$
converge uniformly to the local representatives of $\conf\resto{K_{a_{0}}}$
and $\conf\resto{K_{a_{1}}}$, respectively. This implies that 
\begin{equation}
\conf_{\nu}(\bp_{\nu})\tto\conf(\bp),\quad\conf_{\nu}(\bp'_{\nu})\tto\conf(\bp'),\qquad\text{as}\quad\nu\tto\infty.
\end{equation}
However, since for each $\nu$, $\conf_{\nu}(\bp_{\nu})=\conf_{\nu}(\bp'_{\nu})$,
it follows that $\conf(\bp)=\conf(\bp')$, which contradicts the assumption
that $\conf$ is an embedding.

It is finally noted that the set of Lipschitz embeddings equipped
with the Lipschitz topology may be shown to be open in the manifold
of all Lipschitz mappings $\base\to\spc$. See \cite{fukui_topological_2005}
and an application in continuum mechanics in \cite{falach_configuration_2015}.

\section{The General Framework for Global Analytic Stress Theory \label{sec:The-General-Framework}}

The preceding section implied that for the case where the kinematics
of a material body $\base$ is described by its embeddings in a physical
space $\spc$, the collection of configurations\textemdash the \emph{configuration
space} 
\begin{equation}
\csp:=\embds^{\r}\xts
\end{equation}
\textemdash is an open subset of the manifold of mappings $C^{r}\xts$,
for $r\ge1$. As a result, the configuration space is a Banach manifold
in its own right and 
\begin{equation}
T_{\conf}\csp=T_{\conf}C^{r}\xts=T_{\conf}C^{r}(\xi)
\end{equation}
where $\xi:\base\times\spc\to\base$ is the natural projection of
the trivial fiber bundle.

In view (\ref{eq:TanBund_Man_of_Sections}), $T_{\conf}\csp=C^{r}(\conf^{*}V\fb)$,
where now 
\begin{equation}
V\fb=\{v\in T\fb=T\base\times T\spc\mid T\xi(v)=0\in T\base\}.
\end{equation}
Hence, one may make the identifications
\begin{equation}
V\fb=\base\times T\spc
\end{equation}
and 
\begin{equation}
(\conf^{*}V\fb)_{\bp}=(V\fb)_{\conf(\bp)}=T_{\conf(\bp)}\spc.
\end{equation}
A section $\vf$ of $\conf^{*}\tau:\conf^{*}V\fb\to\base$ is of the
form
\begin{equation}
\bp\lmt\vf(\bp)\in T_{\conf(\bp)}\spc
\end{equation}
and may be viewed as a \emph{vector field along} $\conf$, \ie, 
a mapping
\begin{equation}
\vf:\base\tto T\spc,\qquad\text{such that,}\qquad\tau\comp\vf=\conf.
\end{equation}
Thus, a tangent vector to the configuration space at the configuration
$\conf$ may be viewed as a $C^{r}$-vector field along $\conf$.
This is a straightforward generalization of the standard notion of
a virtual velocity field and we summarize these observations by
\begin{equation}
T_{\conf}\csp=C^{r}(\conf^{*}V\fb)=\{\vf\in C^{r}(\base,T\spc)\mid\tau\comp\vf=\conf\}.
\end{equation}

In the case of generalized continua, where $\xi:\fb\to\base$ need
not be a trivial vector bundle, this simplification does not apply
of course. However, the foregoing discussion motivates the definition
of the \emph{configuration space} for a general continuum mechanical
system specified by the fiber bundle $\xi:\fb\to\base$ as
\begin{equation}
\csp=C^{r}(\xi),\qquad r\ge1.
\end{equation}
We note that the condition that configurations are embeddings is meaningless
in the case of generalized continua.

The general framework for global analytic stress theory adopts the
geometric structure for the statics of systems having a finite number
of degrees of freedom. Once a configuration manifold $\csp$ is specified,
\emph{generalized }or\emph{ virtual velocities} are defined to be
elements of the tangent bundle, $T\csp$ and \emph{generalized forces}
are defined to be elements of the cotangent bundle $T^{*}\csp$. The
action of a force $\fc\in T_{\conf}^{*}\csp$ on a virtual velocity
$\vf\in T_{\conf}\csp$ is interpreted as \emph{virtual power} and
as such, the notion of power has a fundamental role in this formulation.

The foregoing discussion, implies that a force at a configuration
$\conf\in C^{r}(\xi)$ is an element of $C^{r}(\conf^{*}V\fb)^{*}$\textemdash a
continuous and linear functional on the Banachable space of $C^{r}$-section
of the vector bundle. Thus, in the following sections we consider
the properties of linear functionals on the space of $C^{r}$-sections
of a vector bundle $\vb$. Of particular interest is the fact that
our base manifold, or body manifold, is a manifold with corners rather
than a manifold without boundary. The relation between such functionals,
on the one hand, and Schwartz distribution and de Rham currents, on
the other hand, is described. In Section \ref{sec:Stresses} we show
that the notions of stresses and hyperstresses emerge from a representation
theorem for such functionals and in Section \ref{sec:Simple-Forces-St}
we study further the properties of stresses.

\section{Duals to Spaces of Differentiable Sections of a Vector Bundle: Localization
of Sections and Functionals\label{sec:Duals-to-Sections}}

As follows from the foregoing discussion, generalized forces are modeled
mathematically as elements of the dual space $C^{r}(\pi)^{*}=C^{r}(\vb)^{*}$
of the space of $C^{r}$-sections of a vector bundle $\pi:\vb\to\base$.
This section reviews the basic notions corresponding to continuous
linear functional in the dual space with particular attention to localization
properties. While we assume that our base manifold $\base$ is compact
with corners, we want to relate the nature of functionals defined
on sections over $\base$ with analogous settings where $\base$ is
a manifold without boundary. Thus, one can make a connection of the
properties of generalized forces and objects like distributions, de
Rham currents and generalized sections on manifolds. (See, in particular,
Section \ref{subsec:Supported-functionals}.) As an additional motivation
for considering sections over manifolds without boundaries, it is
observed that in both the Eulerian formulation of continuum mechanics
and in classical field theories, the base manifold, either space or
space-time, is usually taken as manifold without boundary. We start
with the case where $\base$ is a manifolds without a boundary and
continue with the case where bodies are modeled by compact manifolds
with corners. 

\subsection{Spaces of differentiable sections over a manifold without boundary
and linear functionals}

A comprehensive introduction to the subject considered here is available
in the Ph.D thesis \cite{Steinbauer_PhD} and the corresponding \cite[Chapter 3]{Grosser_et_al_2001}.
See also \cite[Chapter VI]{guillemin_geometric_1977} and \cite{Handbook_Global_Analysis}.

Consider the space of $C^{r}$-sections of a vector bundle $\pi:\vb\to\base$,
for $0\le r\le\infty$. For manifolds without boundary that are not
necessarily compact, the setting of Section \ref{subsec:Cr-Compact}
will not give a norm on the space of sections. Thus, one extends the
settings used for Schwartz distributions and de Rham currents to sections
of a vector bundle (see also \cite[Chapter XVII]{dieudonne_treatise_1972}).
Specifically, we turn our attention to $C_{c}^{r}(\pi)$, the space
of \emph{test sections}\textemdash $C^{r}$-sections of $\pi$ having
compact supports in $\base$.

Let $\{(U_{a},\vph_{a},\Phi_{a})\}_{a\in A}$, be a vector bundle
atlas so that
\begin{equation}
(\vph_{a},\Phi_{a}):\pi^{-1}(U_{a})\tto\reals^{n}\times\reals^{m},\quad v\lmt(\bp^{\ib},v^{\ga}).
\end{equation}
and let $K$ be a compact subset of $\base$. Consider the vector
subspace $C_{c,K}^{r}(\pi)\subset C_{c}^{r}(\pi)$ of sections, the
supports of which are contained in $K$. Let $a_{l}\in A$, indicate
a finite collection of charts such that $\{U_{a_{l}}\}$ cover $K$,
and for each $U_{a_{l}}$ let $K_{a_{l},\mu}$, $\mu=1,2,\dots$,
be a fundamental sequence of compact sets, \ie,  $K_{a_{l},\mu}\subset\inter{K_{a_{l},\mu+1}}$,
covering $\vph_{a_{l}}(U_{a_{l}})\subset\reals^{n}$. Then, for a
section $\vf\in C_{c,K}^{r}(\pi)$, the collection of semi-norms
\begin{equation}
\norm{\vf}_{K,\mu}^{r}=\sup_{a_{l},\ga,\abs{\smi{\mii}}\le r}\braces{\sup_{\vct{\bp}\in K_{a_{l},\mu}}\braces{\abs{(\vf_{a_{l}}^{\ga})_{,\smi{\mii}}(\vct{\bp})}}},
\end{equation}
induce a Fr\'{e}chet space structure for $C_{c,K}^{\infty}(\pi)$.
Since for each compact subset $K$, one has the inclusion mapping
$\incl_{K}:C_{c,K}^{r}(\pi)\to C_{c}^{r}(\pi)$, one may define the
topology on $C_{c}^{r}(\pi)$ as the inductive limit topology generated
by these inclusions, \ie, the strongest topology on $C_{c}^{r}(\pi)$
for which all the inclusions are continuous. A sequence of sections
in $C_{c}^{r}(\pi)$ converges to zero, if there is a compact subset
$K\subset\base$ such that the supports of all sections in the sequence
are contained in $K$ and the $r$-jets of the sections converge uniformly
to zero in $K$.

 A linear functional $T\in C_{c}^{r}(\pi)^{*}$ is continuous when
it satisfies the following condition. Let $(\chi_{j})$ be a sequence
of sections of $\pi$ all of which are supported in a compact subset
$K\subset U_{a}$ for some $a\in A$. In addition, assume that the
local representatives of $\chi_{j}$ and their derivatives of all
orders $k\le r$ converge uniformly to zero in $K$. Then, 
\begin{equation}
\lim_{j\to\infty}T(\chi_{j})=0.
\end{equation}
Functionals in $C_{c}^{r}(\pi)^{*}$ for a finite value of $r$ are
referred to as functionals of order $r$.

For a linear functional $T$, the \emph{support}, $\support T$ is
defined as follows. An open set $U\subset\base$ is termed a null
set of $T$ if $T(\chi)=0$ for any section of $\pi$ with $\support\chi\subset U$.
The union of all null sets, $U_{0}$ is an open set which is a null
set also. Thus, one defines 
\begin{equation}
\support T:=\base\setminus U_{0}.
\end{equation}

\subsection{Localization of sections and linear functionals for manifolds without
boundaries\label{subsec:Local_Functionals_Without-Boudaries}}

Let $\{(U_{a},\chart_{a},\Chart_{a})\}_{a\in A}$ be a locally finite
vector bundle atlas on $\vb$ and consider 
\begin{equation}
E_{U_{a}}:C_{c}^{r}(\pi\resto{U_{a}})\tto C_{c}^{r}(\pi),
\end{equation}
the natural zero extension of sections supported in a compact subsets
of $U_{a}$ to the space of sections that are compactly supported
in $\base$.  This is evidently a linear and continuous injection
of the subspace. On its image, the subspace of sections $\chi$ with
$\support\chi\subset U_{a}$ we have a left inverse, the natural restriction
\begin{equation}
\rho_{U_{a}}:\image E_{U_{a}}\tto C_{c}^{r}(\pi\resto{U_{a}}),
\end{equation}
a surjective mapping. However, it is well known (\eg, \cite[pp. 245--246]{Schwartz-1963,treves_topological_1967})
that the inverse $\rho_{U_{a}}$ is not continuous.

The dual,
\begin{equation}
E_{U_{a}}^{*}:C_{c}^{r}(\pi)^{*}\tto C_{c}^{r}(\pi\resto{U_{a}})^{*},
\end{equation}
is the restriction of functionals on $\base$ to sections supported
on $U_{a}$, and as $\rho_{U_{a}}$ is not continuous, $E_{U_{a}}^{*}$
is not surjective (\emph{loc. cit}.). We will write
\begin{equation}
T\resto{U_{a}}:=\tilde{T}_{a}:=E_{U_{a}}^{*}T.
\end{equation}
We also note that the restrictions $\{\tilde{T}_{a}\}$ satisfy the
condition
\begin{equation}
\tilde{T}_{a}(\chi\resto{U_{a}})=\tilde{T}_{b}(\chi\resto{U_{b}})=T(\chi)\label{eq:compatibility_local_functionals}
\end{equation}
for any section $\chi$ supported in $U_{a}\cap U_{b}$.

Consider the mapping
\begin{equation}
s:\bigoplus_{a\in A}C_{c}^{r}(\pi\resto{U_{a}})\tto C_{c}^{r}(\pi)
\end{equation}
given by 
\begin{equation}
s(\chi_{1},\dots,\chi_{a},\dots):=\sum_{a\in A}E_{U_{a}}(\chi_{a}).
\end{equation}
Due to the overlapping between domains of definition, the mapping
$s$ is not injective. However, $s$ is surjective because using a
partition of unity, $\{u_{a}\}$, which subordinate to this atlas,
for each section, $\chi$, $u_{a}\chi$ is a compactly supported in
$U_{a}$ and $\chi=\sum_{a}u_{a}\chi$. Hence, the dual mapping,
\begin{equation}
s^{*}:C_{c}^{r}(\pi)^{*}\tto\bigoplus_{a\in A}C_{c}^{r}(\pi\resto{U_{a}})^{*},
\end{equation}
given by,
\begin{equation}
(s^{*}T)_{a}:=E_{U_{a}}^{*}T=T\resto{U_{a}},\qquad s^{*}T(\chi_{1},\dots)=T\Bigp{\sum_{a\in A}E_{U_{a}}(\chi_{a})},
\end{equation}
is injective. In other words, a functional is determined uniquely
by the collection of its restrictions. Note that no compatibility
condition is imposed above on the local sections $\{\chi_{a}\}$.

Since $\{\tilde{T}_{a}\}\in\image s^{*}$ satisfy the compatibility
condition (\ref{eq:compatibility_local_functionals}), $s^{*}$ is
not surjective. However, it is easy to see that $\image s^{*}$ is
exactly the subspace of $\bigoplus_{a\in A}C_{c}^{r}(\pi\resto{U_{a}})^{*}$
containing the compatible collections of local functionals. For let
$\{\tilde{T}_{a}\}$ be local functionals that satisfy (\ref{eq:compatibility_local_functionals})
and $\{u_{a}\}$ a partition of unity. Consider the functional $T\in C_{c}^{r}(\pi)^{*}$
given by
\begin{equation}
T(\chi)=\sum_{a\in A}\tilde{T}_{a}(u_{a}\chi).
\end{equation}
If $\chi$ is supported in $U_{b}$ for $b\in A$, then
\begin{equation}
\begin{split}T(\chi) & =\sum_{a\in A}\tilde{T}_{a}(u_{a}\chi),\qquad U_{a}\cap U_{b}\ne\varnothing,\\
 & =\sum_{a\in A}\tilde{T}_{b}(u_{a}\chi),\qquad\text{by (\ref{eq:compatibility_local_functionals}),}\\
 & =\tilde{T}_{b}\Bigp{\sum_{a\in A}u_{a}\chi},\\
 & =\tilde{T}_{b}(\chi).
\end{split}
\end{equation}
Thus, $T$ is a well defined functional on $\pi$ and it is uniquely
determined by the collection $\{\tilde{T}_{a}\}$\textemdash its restrictions,
independently of the partition of unity chosen. 

As mentioned above, a partition of unity induces an injective right
inverse to $s$ in the form
\begin{equation}
p:C_{c}^{r}(\pi)\tto\bigoplus_{a\in A}C_{c}^{r}(\pi\resto{U_{a}}),\qquad p(\chi)=\{(u_{a}\chi)\resto{U_{a}}\},
\end{equation}
that evidently satisfies $s\comp p=\iden$. It is noted that $p$
is not a left inverse. In particular, for a section $\chi_{a}$ supported
in $U_{a}$, with $\chi_{b}=0$ for all $b\ne a$, 
\begin{equation}
(p\comp s\{\chi_{1},\dots\})_{a}=u_{a}\chi_{a}\label{eq:p o s not identity}
\end{equation}
which need not be equal to $\chi_{a}$. Thus, $p$ depends on the
partition of unity.

For the surjective dual mapping
\begin{equation}
p^{*}:\bigoplus_{a\in A}C_{c}^{r}(\pi\resto{U_{a}})^{*}\tto C_{c}^{r}(\pi)^{*},
\end{equation}
\begin{equation}
p^{*}(T_{1},\dots)=\sum_{a\in A}u_{a}(\rho_{U_{a}}^{*}T_{a}),\qquad p^{*}(T_{1},\dots)(\chi)=\sum_{a\in A}T_{a}((u_{a}\chi)\resto{U_{a}}),
\end{equation}
we note that $p^{*}\comp s^{*}=\idnt$ while $s^{*}\comp p^{*}\ne\idnt$,
in general. The surjectivity of $p^{*}$ implies that every functional
$T$ may be represented by a non-unique collection $\{T_{a}\}$ in
the form 
\begin{equation}
T(\chi)=\sum_{a\in A}T_{a}((u_{a}\chi)\resto{U_{a}}),\qquad T=\sum_{a\in A}u_{a}T_{a}.
\end{equation}
which depends on the partition of unity. Here, $u_{a}T$ denotes the
functional defined by $u_{a}T(\chi)=T(u_{a}\chi)$.

Nevertheless, we may restrict $p^{*}$ to the subspace of compatible
local functionals, $\image s^{*}$, \ie,  those satisfying (\ref{eq:compatibility_local_functionals}).
Thus, the restriction
\begin{equation}
p^{*}\resto{\image s^{*}}:\image s^{*}\tto C_{c}^{r}(\pi)^{*},
\end{equation}
 is an isomorphism (which depends on the partition of unity). It follows
that
\begin{equation}
s^{*}\comp p^{*}=\idnt:\image s^{*}\tto\image s^{*}.
\end{equation}
(For additional details, see \cite[p. 244--245]{dieudonne_treatise_1972},
which is restricted to the case of de Rham currents, and \cite[pp. 234--235]{Grosser_et_al_2001}.)

\subsection{Localization of sections and linear functionals for manifolds with
corners\label{subsec:Functionals_Manifolds_with_corners}}

In analogy with Section \ref{subsec:Local_Functionals_Without-Boudaries},
we consider the various aspects of localization relevant to the case
of compact manifolds with corners. Thus, the base manifold for the
vector bundle $\pi:\vb\tto\base$ is assumed to be a manifold with
corners and we are concerned with elements of $C^{r}(\pi)^{*}$ acting
on sections that need not necessarily vanish together with their first
$r$ jets on the boundary of $\base$.

In \cite[pp 10--11]{palais_foundations_1968}, Palais proves what
he refers to as the ``Mayer-Vietoris Theorem''. Adapting the notation
and specializing the theorem to the $C^{r}$-topology, the theorem
may be stated as follows.
\begin{thm}
\label{thm:Mayer-Vietoris-2}Let $\base$ be a compact smooth manifold
and let $K_{1},\dots,K_{A}$ be compact $n$-dimensional submanifolds
of $\base$ whose interiors cover $\base$ $($such as in a precompact
atlas$)$. Given the vector bundle $\pi$, set
\begin{equation}
\tilde{C}^{r}(\pi):=\biggl\{(\chi_{1},\dots,\chi_{A})\in\bigoplus_{a=1}^{A}C^{r}(\pi\resto{K_{a}})\,\Bigl|\,\chi_{a}\resto{K_{b}}=\chi_{b}\resto{K_{a}}\biggr\},\label{eq:Compatibilty_of_Local_Reps-1}
\end{equation}
and define
\begin{equation}
\iota:C^{r}(\pi)\tto\tilde{C}^{r}(\pi),\qquad\text{by}\qquad\iota(\chi)=(\chi\resto{K_{1}},\dots,\chi\resto{K_{A}}).
\end{equation}
Then, $\iota$ is an isomorphism of Banach spaces.
\end{thm}

We will refer to the condition in (\ref{eq:Compatibilty_of_Local_Reps-1})
as the \emph{compatibility condition} for local representatives of
sections. The most significant part of the proof is the construction
of $\iota^{-1}$. Thus, one has to construct a field $\vf$ when a
collection $(\vf_{1},\dots,\vf_{A})$, satisfying the compatibility
condition, is given. This is done using a partition of unity which
is subordinate to the interiors of $K_{1},\dots,K_{A}$.

It is noted that the situation may be viewed as ``dual'' to that
described in Section \ref{subsec:Local_Functionals_Without-Boudaries}.
For functionals on spaces of sections with compact supports defined
on a manifold without boundary, there is a natural restriction of
functionals, $E_{U_{a}}^{*}$, and the images $\{\tilde{T}_{a}\}$
of a functional $T$ under the restrictions satisfy the compatibility
condition (\ref{eq:compatibility_local_functionals}). The collection
of restrictions determine $T$ uniquely. Here, it follows from Theorem
\ref{thm:Mayer-Vietoris-2} that we have a natural restriction of
sections, and the restricted sections satisfy the compatibility condition
(\ref{eq:Compatibilty_of_Local_Reps-1}). The restrictions $\{\chi\resto{K_{a}}\}$
also determine the global section $\chi$, uniquely.

In Section \ref{subsec:Local_Functionals_Without-Boudaries}, we
observed that sections with compact supports on $\base$ cannot be
``restricted'' naturally to sections with compact supports on the
various $U_{a}$. Such restrictions depend on the chosen partition
of unity. The analogous situation for functionals on manifolds with
corners is described below.
\begin{cor}
Let $T\in C^{r}(\pi)^{*}$, then $T$ may be represented (non-uniquely)
by $(T_{1},\dots,T_{A})$, $T_{a}\in C^{r}(\pi\resto{K_{a}})^{*}$,
in the form
\begin{equation}
T(\vf)=\sum_{a=1}^{A}T_{a}(\vf\resto{K_{a}}).
\end{equation}
\end{cor}

Indeed, as $\iota$ in Theorem \ref{thm:Mayer-Vietoris-2} is an embedding
of $C^{r}(\pi)$ into a subspace of $\bigoplus_{a=1}^{A}C^{r}(\pi\resto{K_{a}})$,
one has a surjective
\begin{equation}
\iota^{*}:\bigoplus_{a=1}^{A}C^{r}(\pi\resto{K_{a}})^{*}\tto C^{r}(\pi)^{*},
\end{equation}
given by 
\begin{equation}
\iota^{*}(T_{1},\dots,T_{A})(\vf)=\sum_{a=1}^{A}T_{a}(\vf\resto{K_{a}}).
\end{equation}

\subsection{\label{sub:Indeterminacy_body_forces-1}Supported sections, static
indeterminacy and body forces}

The foregoing observations are indicative of the fundamental problem
of continuum mechanics\textemdash that of static indeterminacy. Given
a force $\fc$ on a body as an element of $C^{r}(\pi)^{*}$ for some
vector bundle $\pi:\vb\to\base$, and a sub-body $\reg\subset\base$,
there is no unique restriction of $\fc$ to a force on $\reg$ in
$C^{r}(\pi\resto{\reg})^{*}$. This problem is evident for standard
continuum mechanics in Euclidean spaces and continues all the way
to continuum mechanics of higher order on differentiable manifolds.

Adopting the notation of \cite{Corners_Melrose}, denote by $\dot{C}^{r}(\pi)$
the space of sections of $\pi$, the $r$-jet extensions of which
vanish on all the components of the boundary $\bdry\base$. Let $\tilde{\base}$
be a manifold without boundary extending $\base$ and let 
\begin{equation}
\tilde{\pi}:\tilde{\vb}\tto\tilde{\base}
\end{equation}
be an extension of $\pi$. Then, we may use zero extension to obtain
an isomorphism
\begin{equation}
\dot{C}^{r}(\pi)\isom\{\chi\in C_{c}^{r}(\tilde{\pi})\mid\support\chi\subset\base).
\end{equation}
If $\reg$ is a sub-body of $\base$, then, one has the inclusion
\begin{equation}
\dot{C}^{r}(\pi\resto{\reg})\lhr\dot{C}^{r}(\pi).\label{eq:Incl_Sect_Comp_Supp}
\end{equation}

The dual $\dot{C}^{r}(\pi)^{*}$ to the space of sections supported
in $\base$ is the space of \emph{extendable functionals.} From \cite[Proposition 3.3.1]{Corners_Melrose}
it follows that the restriction
\begin{equation}
\rho:C^{r}(\pi)^{*}\tto\dot{C}^{r}(\pi)^{*}.
\end{equation}
is surjective and its kernel is the space of functionals on $\tilde{\base}$
supported in $\bdry\base$.

Thus, if we interpret $T\in C^{r}(\pi)^{*}$ as a force, $\rho(T)\in\dot{C}^{r}(\pi)^{*}$
is interpreted as the corresponding \emph{body force}. For a sub-body
$\reg$, using the dual of (\ref{eq:Incl_Sect_Comp_Supp}), one has
\begin{equation}
\rho_{\reg}:\dot{C}^{r}(\pi)^{*}\tto\dot{C}^{r}(\pi\resto{\reg})^{*}.
\end{equation}
We conclude that even in this very general settings, body forces of
any order may be restricted naturally to sub-bodies.

\subsection{Supported functionals\label{subsec:Supported-functionals}}

Distributions on closed subsets of $\reals^{n}$ have been considered
by Glaeser \cite{Glaeser}, Malgrange \cite[Chapter 7]{Malgrange}
and Oksak \cite{Oksak76}. The basic tool in the analysis of distributions
on closed sets is Whitney's extension theorem \cite{Whitney_extension}
(see also \cite{Seeley,Hormander}) which guarantees that a differentiable
function on a closed subset of $\reals^{n}$ may be extended to a
compactly supported smooth function on $\reals^{n}$. The extension
mapping between the corresponding function spaces is continuous. The
extension theorem implies that restriction of functions is surjective
and so, the dual of the restriction mapping associates a unique distribution
in an open subset of $\reals^{n}$ with a linear functional defined
on the given closed set. Distributions and functionals on manifold
with corners have been considered by Melrose \cite[Chapter 3]{Corners_Melrose},
whom we follow below.

Thus, let $T\in C^{r}(\pi)^{*}$ and let $\tilde{\pi}:\tilde{\vb}\to\tilde{\base}$
be an extension of the vector bundle $\pi:\vb\to\base$, where $\tilde{\base}$
is a manifold without a boundary. The Whitney-Seeley extension 
\begin{equation}
E:C^{r}(\pi)\tto C_{c}^{r}(\tilde{\pi})
\end{equation}
is a continuous injection. It follows that the natural restriction
\begin{equation}
\rho_{\base}:C_{c}^{r}(\tilde{\pi})\tto C^{r}(\pi),
\end{equation}
its left inverse satisfying $\rho_{\base}\comp E=\idnt$, is surjective
and the inclusion
\begin{equation}
\rho_{\base}^{*}:C^{r}(\pi)^{*}\tto C_{c}^{r}(\tilde{\pi})^{*}
\end{equation}
is injective. In other words, each functional $T\in C^{r}(\pi)^{*}$,
determines uniquely a functional $\tilde{T}=\rho_{\base}^{*}T$ satisfying
\begin{equation}
\tilde{T}(\tilde{\chi})=T(\tilde{\chi}\resto{\base}).
\end{equation}
The last equation implies also that $\tilde{T}(\tilde{\chi})=0$ for
any section $\tilde{\chi}$ supported in $\tilde{\base}\setminus\base$.
Hence, $\tilde{T}$ is supported in $\base$.

Conversely, every $\tilde{T}\in C_{c}^{r}(\tilde{\pi})$, with $\support\tilde{T}\subset\base$
represents a functional $T\in C^{r}(\pi)^{*}$, \ie, $\tilde{T}=\rho_{\base}^{*}T$.
This may be deduced as follows. For any such $\tilde{T}$, consider
$T=E^{*}\tilde{T}$. One needs to show that $\tilde{T}=\rho_{\base}^{*}\comp E^{*}(\tilde{T})$.
Let $\tilde{\chi}\in C_{c}^{r}(\tilde{\pi})$, then, 
\begin{equation}
\begin{split}(\tilde{T}-\rho_{\base}^{*}\comp E^{*}(\tilde{T}))(\tilde{\chi}) & =\tilde{T}(\tilde{\chi})-\tilde{T}(E\comp\rho_{\base}(\tilde{\chi})),\\
 & =\tilde{T}(\tilde{\chi}-E\comp\rho_{\base}(\tilde{\chi})).
\end{split}
\end{equation}
It is observed that $\tilde{\chi}-E\comp\rho_{\base}(\tilde{\chi})$
vanishes on $\base$ so that $\support(\tilde{\chi}-E\comp\rho_{\base}(\tilde{\chi})\subset\overline{\tilde{\base}\setminus\base}$.
Since $\support\tilde{T}\subset\base$, $\tilde{T}(\chi')=0$ for
any section $\chi'$ supported in $\tilde{\base}\setminus\base$.
However, approximating the section $\tilde{\chi}-E\comp\rho_{\base}(\tilde{\chi})$,
supported in the closure, $\overline{\tilde{\base}\setminus\base}$,
by sections supported in $\tilde{\base}\setminus\base$, one concludes
that $\tilde{T}(\tilde{\chi}-E\comp\rho_{\base}(\tilde{\chi}))=0$
also.

Due to this construction, Melrose \cite[Chapter 3]{Corners_Melrose}
refers to such functionals (distributions) as \emph{supported.} It
is noted that such functionals of compact support are of a finite
order $r$.

\subsection{Density dual and smooth functionals\label{subsec:Density-dual-and-functionals}}

A simple example for functionals on spaces of sections of a vector
bundle $\pi:\vb\to\base$ is provided by \emph{smooth functionals}.
Consider, in analogy with the dual of a vector bundle, the vector
bundle of linear mappings into another one-dimensional vector vector
bundle, that of $n$-alternating tensors. Thus, for a given vector
bundle, $\vb$, we use the notation (see Atiyah and Bott \cite{Atiyah_Bott_I})
\begin{equation}
\ndual{\vb}=\L{\vbts,\ext^{n}T^{*}\base}\isom\vb^{*}\otimes\nfo.\label{eq:def-density_dual}
\end{equation}

Let $A:\vb_{1}\to\vb_{2}$ be a vector bundle morphism over $\base$.
Then, in analogy with the dual mapping, one may consider
\begin{equation}
\ndual A:\ndual{\vb_{2}}\tto\ndual{\vb_{1}},\qquad\text{given by}\qquad f\lmt f\comp A.
\end{equation}
It is also noted that we have 
\begin{equation}
\begin{split}\ndual{(\ndual{\vb})} & =(\vb^{*}\otimes\ext^{n}T^{*}\base)\nd\\
 & =(\vb^{*}\otimes\ext^{n}T^{*}\base)^{*}\otimes\ext^{n}T^{*}\base,\\
 & =\vb\otimes\ext^{n}T\base\otimes\ext^{n}T^{*}\base,
\end{split}
\end{equation}
and as $\ext^{n}T\base\otimes\ext^{n}T^{*}\base$ is isomorphic with
$\reals$, one has a natural isomorphism
\begin{equation}
(\vb\nd)\nd\isom\vb.
\end{equation}
For the vector bundles $\vb$, $\avb$, 
\begin{equation}
(\vb\otimes\avb)'\isom\vb^{*}\otimes\avb^{*}\otimes\nfo\isom\vb^{*}\otimes\avb'.
\end{equation}
We will refer to $\ndual{\vb}$ as the \emph{density}-\emph{dual bundle}
and to $\ndual A$ as the \emph{density}-\emph{dual mapping}.

As an example, for the case $\vb=\ext^{p}T^{*}\base$, we have an
isomorphism (see Section \ref{subsec:Isom-e-1}),
\begin{equation}
e_{\contr}:\ext^{n-p}T^{*}\base\tto\ndual{(\ext^{p}T^{*}\base)},
\end{equation}
given by 
\begin{equation}
e_{\contr}(\go)(\psi)=\go\wedge\psi.\label{eq:Def_e_lrcorner}
\end{equation}

Smooth functionals may be induced by smooth sections of $\vb'$. For
a section $\std$ of $\vb'$, and a section $\chi$ of $\vb$, let
$\std\cdot\chi$ be the $n$-form 
\begin{equation}
\std\cdot\chi(X)=\std(X)(\chi(X)).
\end{equation}
The smooth functional $T_{\std}$ induced by $\std$ is defined by
\begin{equation}
T_{\std}(\chi):=\int_{\base}\std\cdot\chi.
\end{equation}

\subsection{Generalized sections and distributions\label{subsec:Generalized-sections}}

Let $\pi_{0}:\vb_{0}\to\base$ be a vector bundle and consider the
case where the vector bundle $\pi$ above is set to be 
\begin{equation}
\pi:\vb_{0}':=\L{\vb_{0},\nfo}\isom\vb_{0}^{*}\otimes\nfo\tto\base.
\end{equation}
Thus, the corresponding functionals on sections of $\pi$ are elements
of 
\begin{equation}
C^{r}(\pi)^{*}=C^{r}\paren{\L{\vb_{0},\nfo}}^{*}\isom C^{r}\paren{\vb_{0}^{*}\otimes\nfo}^{*}.
\end{equation}
In this case, smooth functionals are represented by smooth sections
of 
\begin{equation}
\begin{split}\L{\vb,\ext^{n}T^{*}\base} & =\L{\vb_{0}^{*}\otimes\nfo,\nfo},\\
 & \isom\paren{\vb_{0}^{*}\otimes\nfo}^{*}\otimes\nfo,\\
 & \isom\vb_{0}\otimes\ext^{n}T\base\otimes\nfo,\\
 & \isom\vb_{0}.
\end{split}
\end{equation}
One concludes that smooth functionals in $C^{r}(\vb_{0}^{*}\otimes\nfo)^{*}$
are represented by sections of $\vb_{0}$ . It is natural therefore
to refer to elements of 
\begin{equation}
C^{-r}(\vb_{0}):=C^{r}(\ndual{\vb_{0}})^{*}\label{eq:def_gen_sect_ndual}
\end{equation}
as \emph{generalized sections} of $\vb_{0}$ (see \cite{Atiyah_Bott_I,guillemin_geometric_1977,Grosser_et_al_2001},
\cite[p. 676]{Handbook_Global_Analysis}).

In the particular case where $\vb_{0}=\base\times\reals$ is the natural
line bundle, smooth functionals are represented by real valued functions
on $\base$. Consequently, elements of 
\begin{equation}
C^{r}\paren{\reals\otimes\nfo}^{*}=C^{r}\paren{\nfo}^{*}
\end{equation}
are referred to as \emph{generalized functions}.

The apparent complication in the definition of generalized sections
using the density dual is justified in the sense that each element
in $C^{-k}(\vb_{0})$ may be approximated by a sequence of smooth
functionals induced by sections of $\vb_{0}'$. (See \cite[p. 241]{Grosser_et_al_2001}.)

In the literature, the term \emph{section distributions} is used in
different ways in this context. For example, in \cite{Grosser_et_al_2001}
and \cite[p. 676]{Handbook_Global_Analysis}, $\vb_{0}$-valued distributions
are defined as elements of $C^{r}\paren{\vb_{0}'}^{*}$, \ie, what
are referred to here as generalized sections of $\vb_{0}$. (In \cite{Atiyah_Bott_I,Atiyah_Singer_I}
they are referred to as \emph{distributional sections.}) On the other
hand, in \cite{guillemin_geometric_1977}, distributions are defined
as generalized sections of $\nfo$\textemdash elements of $C^{r}(\base)^{*}$.
See further comments on this issue and the corresponding terms \emph{section
distributional densities }and \emph{generalized densities} in \cite{guillemin_geometric_1977,Grosser_et_al_2001,Handbook_Global_Analysis}.

\section{de Rham currents\label{sec:de_Rham}}

For a manifold without boundary $\base$, de Rham currents (see \cite{rham_differentiable_1984,schwartz_theorie_1973,federer_geometric_1969})
are functionals corresponding to the case of the vector bundle 
\begin{equation}
\pi:\ext^{p}T^{*}\base\tto\base
\end{equation}
so that test sections are smooth $p$-forms having compact supports.
Thus, a $p$-\emph{current of order} $r$ on $\base$ is a continuous
linear functional on $C_{c}^{r}(\ext^{p}T^{*}\base)$.

A particular type of $p$-currents, smooth currents, are induced by
differential $(n-p)$-forms. Such an $(n-p)$ form, $\go$, induces
the currents $_{\go}T$ and $T_{\go}=(-1)^{p(n-p)}\!\,_{\go}T$ by
\begin{equation}
_{\go}T(\psi)=\int_{\base}\go\wedge\psi,\qquad T_{\go}(\psi)=\int_{\base}\psi\wedge\go.
\end{equation}

Another simple $p$-current, $T_{\mathcal{Z}}$ is induced by an oriented
$p$-dimensional submanifold $\mathcal{Z\subset\base}$. It is naturally
defined by 
\begin{equation}
T_{\mathcal{Z}}(\psi)=\int_{\mathcal{Z}}\psi.
\end{equation}

These two examples illustrate the two points of views on currents.
On the one hand, the example of the current $_{\go}T$ suggests that
a current in $C_{c}^{r}(\ext^{p}T^{*}\base)^{*}$ is viewed as a generalized
$(n-p)$-form. With this point of view in mind, elements of $C_{c}^{r}(\ext^{p}T^{*}\base)$
are referred to as \emph{currents of degree} $n-p$.  Consequently,
the space of $p$-currents on $\base$ is occasionally denoted by
\begin{equation}
C^{-r}(\ext^{n-p}T^{*}\base)=C_{c}^{r}(\ext^{p}T^{*}\base)^{*}.
\end{equation}
On the other hand, the example of the current $T_{\mathcal{Z}}$ induced
by a $p$-dimensional manifolds $\mathcal{Z}$, suggests that currents
be viewed as a geometric object of dimension $p$. Thus, an element
of $C_{c}^{r}(\ext^{p}T^{*}\base)^{*}$ is referred to as a \emph{p-dimensional
current}.

\subsection{Basic operations with currents}

The contraction operations of a $(p+q)$-current $T$ and a $q$-form
$\go$, yields the $p$-currents defined by
\begin{equation}
(T\fcontr\go)(\psi)=T(\psi\wedge\go)\qquad\text{and}\qquad(\go\contr T)(\psi)=T(\go\wedge\psi),\label{eq:Contr_Form_Curr}
\end{equation}
so that 
\begin{equation}
T\fcontr\go=(-1)^{pq}\go\contr T.\label{eq:T cotr omega VS omega fcontr T}
\end{equation}
Note that our notation is different from that of \cite{rham_differentiable_1984}
and different in sign form that of \cite{federer_geometric_1969}.
In particular, given a $p$-current $T$, any $p$-form $\psi$ induces
naturally a zero-current 
\begin{equation}
T\cdot\psi=T\fcontr\psi,\qquad\text{so that}\qquad(T\cdot\psi)(u)=T(u\psi).
\end{equation}
The $p$-current $_{\go}T$ defined above can be expressed using contraction
in the form
\begin{equation}
_{\go}T=\go\contr T_{\base}.
\end{equation}

For a $p$-current $T$ and a $q$-multi-vector field $\xi$, the
$(p+q)$-currents $\xi\wedge T$ and $T\wedge\xi$ are defined by
\begin{equation}
(\xi\wedge T)(\psi):=T(\xi\contr\psi),\qquad(T\wedge\xi)(\psi):=T(\psi\fcontr\xi),\label{eq:eta_wedge_current}
\end{equation}
for a $(p+q)$-form $\psi$. Using, $\xi\contr\psi=(-1)^{qp}\psi\fcontr\xi$,
one has
\begin{equation}
\xi\wedge T=(-1)^{pq}T\wedge\xi,\label{eq:xi wedge T VS T wedge xi}
\end{equation}
in analogy with the corresponding expression for multi-vectors. Thus,
the wedge product of a $p$-current and an $r$-multi-vector is an
$(r+p)$-current. Note that a real valued function $u$ defined on
$\base$ may be viewed both as a zero-form and as a zero-multi-vector.
Hence, we may write $uT$ for any of the four operations defined above
so that $(uT)(\psi)=T(u\psi)$.

The boundary operator
\begin{equation}
\bdry:C^{-r}(\ext^{n-p}T^{*}\base)\tto C^{-(r+1)}(\ext^{n-(p-1)}T^{*}\base),
\end{equation}
defined by 
\begin{equation}
\bdry T(\psi)=T(\dee\psi),
\end{equation}
is a linear and continuous operator. In other words, the boundary
of a $p$-current is a $(p-1)$-current. In particular, for a smooth
current, $_{\go}T$ represented by the $(n-p)$-form $\go$, one has
\begin{equation}
\begin{split}\bdry\,_{\go}T(\psi) & =\int_{\base}\go\wedge\dee\psi,\\
 & =(-1)^{n-p}\int_{\base}\dee(\go\wedge\psi)-(-1)^{n-p}\int_{\base}\dee\go\wedge\psi,\\
 & =(-1)^{n-p+1}T_{\dee\go}(\psi).
\end{split}
\end{equation}
Hence,
\begin{equation}
\bdry\,_{\go}T=(-1)^{n-p+1}T_{\dee\go}.\label{eq:Boundary_of_Smooth_Current}
\end{equation}
Similarly,
\begin{equation}
\bdry T_{\go}=(-1)^{p+1}T_{\dee\go}.\label{eq:Boundary_of_Smooth_Current-1}
\end{equation}

In order to strengthen further the point of view that a $p$-current
is a generalized $(n-p)$-form, the \emph{exterior derivative of a
$p$-current} $\dee T$ is defined by
\begin{equation}
\dee T=(-1)^{n-p+1}\bdry T.\label{eq:Ext_Der_Curr}
\end{equation}
Thus, in the smooth case,
\begin{equation}
\dee\,_{\go}T=T_{\dee\go}.\label{eq:Ext_Der_Smooth_Curr}
\end{equation}

In addition, Stokes's theorem implies that for the $p$-current $T_{\mathcal{Z}}$
induced by the $p$-dimensional submanifold with boundary $\mathcal{Z}$,
the boundary, a $(p-1)$-current, is given by
\begin{equation}
\bdry T_{\mathcal{Z}}=T_{\bdry\mathcal{Z}}.
\end{equation}

It is quite evident, therefore, that the notion of a boundary generalizes
and unites both the exterior derivative of forms and the boundaries
of manifolds.

\subsection{Local representation of currents\label{subsec:Local-represent-Currents}}

We consider next the local representation of de Rham currents in coordinate
neighborhoods.

\subsubsection{Representation by $0$-currents}

Let $\curr=E_{U_{a}}^{*}T$ be the restriction of a $p$-current $T$
to forms supported in a particular coordinate neighborhood\textemdash a
local representative of $T$. Writing
\begin{equation}
\begin{split}R(\psi) & =R(\psi_{\gl}\dx^{\gl}),\qquad\abs{\gl}=p,\\
 & =(\dx^{\gl}\contr R)(\psi_{\gl})
\end{split}
\label{eq:Rep_Curr1a}
\end{equation}
(we could have used $R\fcontr\dx^{\gl}$ just the same as the $\psi_{\gl}$
are real valued functions), one notes that locally 
\begin{equation}
R(\psi)=R^{\gl}(\psi_{\gl}),\qquad\text{where,}\qquad R^{\gl}:=\dx^{\gl}\contr R.\label{eq:Rep_Curr1b}
\end{equation}
Using the exterior product of a multi-vector field $\xi$ and a current
in (\ref{eq:eta_wedge_current}), we may write
\begin{equation}
R(\psi)=R^{\gl}(\bdry_{\gl}\contr\psi)=\bdry_{\gl}\wedge R^{\gl}(\psi),
\end{equation}
and so a current may be represented locally in the form
\begin{equation}
R=\bdry_{\gl}\wedge R^{\gl}.\label{eq:Rep-Curr1c}
\end{equation}

This representation suggests that $T$ be interpreted as a generalized
multi-vector field (\emph{cf.} \cite{whitney_geometric_1957}).

In the sequel, when we refer to local representative of a current
$T$, we will often keep the same notation, $T$, and it will be implied
that we consider the restriction of $T$ to forms (or sections, in
general) supported in a generic coordinate neighborhood.

\subsubsection{Representation by $n$-currents\label{subsec:Representation-curs_by-n--currents}}

Alternatively (\emph{cf.} \cite[p. 36]{rham_differentiable_1984}),
for a $p$-current $R$ defined in a coordinate neighborhood and $\hat{\gl}$
with $|\hat{\gl}|=n-p$, consider the $n$-currents 
\begin{equation}
R_{\hat{\gl}}:=\bdry_{\hat{\gl}}\wedge R,\quad\text{so that}\quad R_{\hat{\gl}}(\theta)=R(\bdry_{\hat{\gl}}\contr\theta).
\end{equation}
Then, for every $p$-form $\go$,
\begin{equation}
\begin{split}(\dx^{\hat{\gl}}\contr R_{\hat{\gl}})(\go) & =R_{\hat{\gl}}(\dx^{\hat{\gl}}\wedge\go),\\
 & =R(\bdry_{\hat{\gl}}\contr(\dx^{\hat{\gl}}\wedge\go)),\\
 & =R_{\hat{\gl}}(\eps^{\hat{\gl}\mu}\go_{\mu}\dx),
\end{split}
\label{eq:Computation_0}
\end{equation}
where we used
\begin{equation}
\dx^{\hat{\gl}}\wedge\go=\eps^{\hat{\gl}\mu}\go_{\mu}\dx.
\end{equation}
Also, 
\begin{equation}
\begin{split}(\bdry_{\hat{\gl}}\contr\dx)(\bdry_{\mu}) & =\dx(\bdry_{\hat{\gl}}\wedge\bdry_{\mu}),\\
 & =\eps_{\hat{\gl}\mu},\\
 & =\eps_{\hat{\gl}\nu}\dx^{\nu}(\bdry_{\mu}),
\end{split}
\end{equation}
implies
\begin{equation}
\bdry_{\hat{\gl}}\contr\dx=\eps_{\hat{\gl}\nu}\dx^{\nu},\label{eq:contr_identity}
\end{equation}
and so, 
\begin{equation}
\bdry_{\hat{\gl}}\contr(\dx^{\hat{\gl}}\wedge\go)=\go_{\gl}\dx^{\gl}=\go,\label{eq:contr_wedge_identity}
\end{equation}
as expected. Hence, 
\begin{equation}
(\dx^{\hat{\gl}}\contr R_{\hat{\gl}})(\go)=R(\go),
\end{equation}
and we conclude that $R$ may be represented by the $n$-currents
\begin{equation}
R_{\hat{\gl}}:=\bdry_{\hat{\gl}}\wedge R,\qquad\text{in the form}\qquad R=\dx^{\hat{\gl}}\contr R_{\hat{\gl}},\label{eq:Rep_Curr2b}
\end{equation}
with 
\begin{equation}
R(\go)=R_{\hat{\gl}}(\dx^{\hat{\gl}}\wedge\go)=\eps^{\hat{\gl}\mu}R_{\hat{\gl}}(\go_{\mu}\dx)=\sum_{\gl}\eps^{\hat{\gl}\gl}R_{\hat{\gl}}(\go_{\gl}\dx).\label{eq:Rep_Curr2c}
\end{equation}
This representation suggests again that a $p$-current $T$ be interpreted
as a generalized $(n-p)$-form. In particular, an $n$-current is
a generalized function and is often referred to as a distribution
on the manifold (\eg, \cite[Chapter 3]{Corners_Melrose}).
\begin{rem}
\label{rem:Change_Order_1}It is noted that one may set 
\begin{equation}
R'_{\hat{\gl}}:=R\wedge\bdry_{\hat{\gl}}.
\end{equation}
Using (\ref{eq:xi wedge T VS T wedge xi}) for the $(n-p)$-multi-vector
$\bdry_{\hat{\gl}}$
\begin{equation}
R'_{\hat{\gl}}=(-1)^{p(n-p)}R_{\hat{\gl}}.\label{eq:R'_gl VS R_gl}
\end{equation}
In addition, by (\ref{eq:Rep_Curr2b}) and (\ref{eq:T cotr omega VS omega fcontr T}),
for the $n$-current $R_{\hat{\gl}}$ and the $(n-p)$-form $\dx^{\hat{\gl}}$,
\begin{equation}
R=R'_{\hat{\gl}}\fcontr\dx^{\hat{\gl}},\label{eq:Rep_Curr2d}
\end{equation}
and
\begin{equation}
R(\go)=R'_{\hat{\gl}}(\go\wedge\dx^{\hat{\gl}})=\sum_{\gl}\eps^{\gl\hat{\gl}}R'_{\hat{\gl}}(\go_{\gl}\dx).\label{eq:Rep_Curr2e}
\end{equation}
\end{rem}

\textbf{}

\section{Vector-valued currents\label{sec:Vector-valued-currents}}

A natural extension of the notions of generalized sections and de
Rham currents yields vector valued vector valued currents that will
be used to model stresses. Vector valued currents and their local
representations will be considered in this section.

\subsection{Vector valued forms}

Let $\pi:\vb\to\base$ be a given vector bundle whose typical fiber
is $m$-dimensional. We will refer to sections of 
\begin{equation}
\L{\vb,\ext^{p}T^{*}\base}\isom\vb^{*}\otimes\ext^{p}T^{*}\base.\label{eq:def_Vec_Val_Form}
\end{equation}
 as \emph{vector valued $p$-forms}, which is short for the more appropriate
\emph{vector bundle valued $p$-form} (\emph{cf.} \cite[p. 340]{schwartz_theorie_1973})\emph{.
}Thus in particular, sections of the density dual, $\vb'=\vb^{*}\otimes\nfo$
are vector valued forms. In the mechanical context, we will also be
concerned with co-vector valued forms, that is, sections of 
\begin{equation}
\L{\vb^{*},\ext^{p}T^{*}\base}\isom\vb\otimes\ext^{p}T^{*}\base.\label{eq:def_covec_val_form}
\end{equation}

The terminology follows from the observation that using the isomorphism
induced by transposition, \ie, $\ext^{p}T\base\otimes\vb\isom\vb\otimes\ext^{p}T\base$,
a co-vector valued form may be viewed as a section of 
\begin{equation}
\ext^{p}T^{*}\base\otimes\vb\isom\ext^{p}(T\base,\vb)\isom L(\ext^{p}T\base,\vb).
\end{equation}

Given a co-vector valued $p$-form, $\chi$, and a vector valued $(n-p)$-form,
$f$, one can define the bilinear action $f\dotwedge\chi$ by setting
\begin{equation}
(g\otimes\go)\dotwedge(\vf\otimes\psi):=g(\vf)\go\wedge\psi,
\end{equation}
for sections $g$, $\vf$, $\go$, $\psi$ of $\vb^{*}$, $\vb$,
$\ext^{n-p}T^{*}\base$, $\ext^{p}T^{*}\base$, respectively. Thus,
$\dotwedge$ induces a bilinear mapping
\begin{equation}
\dotwedge_{\contr}:(\vb^{*}\otimes\ext^{n-p}T^{*}\base)\times(\vb\otimes\ext^{p}T^{*}\base)\tto\ext^{n}T^{*}\base,
\end{equation}
or a linear
\begin{equation}
\dotwedge_{\contr}:\vb^{*}\otimes\ext^{p}T^{*}\base\otimes\vb\otimes\ext^{p}T^{*}\base\tto\ext^{n}T^{*}\base.
\end{equation}
The mapping $\dotwedge_{\contr}$ gives rise to an extension of the
isomorphism $e_{\contr}$ considered above to an isomorphism (we keep
the same notation)
\begin{equation}
e_{\contr}:\vb^{*}\otimes\ext^{n-p}T^{*}\base\tto\ndual{\paren{\vb\otimes\ext^{p}T^{*}\base}},\qquad e_{\contr}(f)(\chi)=f\dotwedge\chi.\label{eq:Def-e_contr}
\end{equation}

Let $\{(U_{a},\chart_{a},\Chart_{a})\}_{a\in A}$ be a vector bundle
trivialization for the vector bundle $\pi:\vb\to\base$ so that 
\begin{equation}
\Chart_{a}:\pi^{-1}(U_{a})\tto U_{a}\times\vs,
\end{equation}
where $\vs$ is the $m$-dimensional typical fiber. Given a basis
in $\vs$, let $\{\sbase_{\ga}\}_{\ga=1}^{m}$ and $\{\sbase^{\ga}\}_{\ga=1}^{m}$
be the local bases and dual bases induced by $\Chart_{a}^{-1}$ on
$\pi^{-1}(U_{a})$. Then, a co-vector valued form $\chi$ and a vector
valued $p$-form $f$ are represented locally in the forms
\begin{equation}
\chi_{\gl}^{\ga}\sbase_{\ga}\otimes\dx^{\gl},\qquad\text{and}\qquad f_{\ga\gl}\sbase^{\ga}\otimes\dx^{\gl},\qquad\abs{\gl}=p,
\end{equation}
respectively.

\subsection{Vector valued currents\label{subsec:Vector-valued-currents}}

We now substitute the vector bundle $\vb^{*}\otimes\ext^{p}T^{*}\base$
for the vector bundle $\vb_{0}$ in definition (\ref{eq:def_gen_sect_ndual})
of generalized sections. Thus,
\begin{equation}
\begin{split}C^{-r}(\vb^{*}\otimes\ext^{p}T^{*}\base) & =C^{r}((\vb^{*}\otimes\ext^{p}T^{*}\base)\nd)^{*}.\end{split}
\label{eq:Def_VVCurr}
\end{equation}
Using the isomorphism $e_{\contr}$ as defined above, it is concluded
that we may make the identifications
\begin{equation}
C^{-r}(\vbts^{*}\otimes\ext^{p}T^{*}\base)=C^{r}\bigp{\vbts\otimes\ext^{n-p}T^{*}\base}^{*}\label{eq:Def_VVcurr-1}
\end{equation}
(see \cite[p. 340]{schwartz_theorie_1973}). Comparing the last equation
to (\ref{eq:def_Vec_Val_Form}) we may refer to elements of these
spaces as \emph{generalized vector valued $p$-forms }or as \emph{vector
valued $(n-p)$-currents}.

A smooth vector valued $(n-p)$-current may be represented by a $\vb^{*}\otimes\ext^{n-p}T\base$
valued $n$-form\textemdash a smooth section $\std$ of $\vb^{*}\otimes\ext^{n-p}T\base\otimes\nfo$
by
\begin{equation}
\chi\lmt\int_{\base}\std\cdot\chi,
\end{equation}
where it is noted that $\std\cdot\chi$ is an $n$-form. Locally,
for $\abs{\mu}=n-p$,
\begin{equation}
\std=\std_{\ga}^{\mu}\sbase^{\ga}\otimes\bdry_{\mu}\otimes\dx,\qquad\std\cdot\chi=\std_{\ga}^{\mu}\chi_{\mu}^{\ga}\dx.\label{eq:Smooth_VVcr_rep_by_n_forms}
\end{equation}

Alternatively, a smooth element of $C^{-r}\paren{\vbts^{*},\ext^{p}T^{*}\base}$
is induced by a section $\curnd{\curd}$ of $\vbts\otimes\ext^{p}T^{*}\base$
in the form
\begin{equation}
\chi\lmt\int_{\base}\curnd{\curd}\dotwedge\chi,\label{eq:Action_Smooth_VVcurrents_rep_by_forms}
\end{equation}
for every $C^{r}$-section $\chi$ of $\vbts\otimes\ext^{n-p}T^{*}\base$.
Locally, for $\abs{\hat{\mu}}=p$, $\abs{\gl}=n-p$,
\begin{equation}
\curnd{\curd}=\curnd{\curd}_{\ga\hat{\mu}}\sbase^{\ga}\otimes\dx^{\hat{\mu}},\qquad\curnd{\curd}\dot{\wedge}\chi=\curnd{\curd}_{\ga\hat{\mu}}\chi_{\gl}^{\ga}\dx^{\hat{\mu}}\wedge\dx^{\gl}=\sum_{\gl}\eps_{\hat{\gl}\gl}\curnd{\curd}_{\ga\hat{\gl}}\chi_{\gl}^{\ga}\dx.\label{eq:Smooth_VVcr_rep_by_n-p_forms}
\end{equation}
Comparing the last two expressions for the resulting densities, one
concludes that
\begin{equation}
\curnd{\curd}_{\ga\hat{\gl}}=\eps_{\hat{\gl}\gl}\std_{\ga}^{\gl}.\label{eq:Relation_between_TSt_rep_and_var_st}
\end{equation}
Globally, it follows that 
\begin{equation}
\curnd{\curd}=\spec C_{\llcorner}(\std),\label{eq:t=00003Dfcontr_S}
\end{equation}
where 
\begin{equation}
\spec C_{\llcorner}:\vb^{*}\otimes\ext^{n-p}T\base\otimes\nfo\tto\vb^{*}\otimes\ext^{p}T^{*}\base
\end{equation}
is induced by the right contraction $\gth\fcontr\eta_{1}(\eta_{2})=\gth(\eta_{2}\wedge\eta_{1})$.
What determined the direction of the contraction is the choice of
action of $\curnd{\curd}$ in (\ref{eq:Def-e_contr}) as in Remark
\ref{rem:Change_Order_1}.

\subsection{Local representation of vector valued currents\label{subsec:Local-represe-VVCurrents}}

We now consider the local representation of the restriction of a vector
valued $p$-current to vector valued forms supported in some given
vector bundle chart. We introduce first some basic operations.

\subsubsection{The inner product of a vector valued current and a vector field\label{subsec:T cdot e_alpha}}

Given a vector valued current $T$ in $C^{r}(\vb\otimes\ext^{p}T^{*}\base)^{*}$
and a $C^{r}$-section $\vf$ of $\vb$, we define the (scalar) $p$-current
$T\cdot\vf$ by
\begin{equation}
T\cdot\vf(\go)=T(\vf\otimes\go).\label{eq:T dot w}
\end{equation}
For local representation, one may consider the $p$-currents 
\begin{equation}
T_{\ga}:=T\cdot\sbase_{\ga}.\label{eq:VVcurr_alpha}
\end{equation}

Thus, in analogy with (\ref{eq:Rep_Curr1a},\ref{eq:Rep_Curr1b})
we have
\begin{equation}
\begin{split}T(\vf\otimes\go) & =T(\vf^{\ga}\sbase_{\ga}\otimes\go),\\
 & =(T\cdot\sbase_{\ga})(\vf^{\ga}\go),\\
 & =T_{\ga}(\vf^{\ga}\go).
\end{split}
\label{eq:Rep_VVcur_By_p-currents}
\end{equation}

\subsubsection{The tensor product of a current and a co-vector field\label{subsec:(e^i) f=00005Cotimes T}}

A (scalar) $p$-current, $T$, and a $C^{r}$-section of $\vb^{*}$,
$g$, induce a vector valued current $g\otimes T\in C^{r}\bigp{\vbts\otimes\ext^{p}T^{*}\base}^{*}$
by setting
\begin{equation}
(g\otimes T)(\vf\otimes\go):=T((g\cdot\vf)\go).
\end{equation}
In particular, locally,
\begin{equation}
(\sbase^{\ga}\otimes T)(\vf\otimes\go):=T(\vf^{\ga}\go).
\end{equation}
Utilizing this definition, one may write for local representatives
\begin{equation}
\begin{split}\sbase^{\ga}\otimes T_{\ga}(\vf\otimes\go) & =T_{\ga}(\vf^{\ga}\go),\\
 & =T(\vf^{\ga}\sbase_{\ga}\otimes\go),
\end{split}
\end{equation}
and so, complementing (\ref{eq:Rep_VVcur_By_p-currents}), one has
\begin{equation}
T=\sbase^{\ga}\otimes T_{\ga}.\label{eq:Rep_VVcur_by_p-currents-2}
\end{equation}

\subsubsection{Representation by $0$-currents\label{subsec:Rep-by-0-currents}}

Proceeding as in Section \ref{subsec:Local-represent-Currents},
the $p$-current $T$ may be represented by the $0$-currents
\begin{equation}
T_{\ga}^{\gl}:=\dx^{\gl}\contr T_{\ga}=\dx^{\gl}\contr(T\cdot\sbase_{\ga}),\qquad\text{in the form}\qquad T(\chi)=T_{\ga}^{\gl}(\chi_{\gl}^{\ga}).\label{eq:Rep_VVcurr_by_0-curr1}
\end{equation}
Using (\ref{eq:eta_wedge_current}) and (\ref{eq:Rep-Curr1c}), we
finally have
\begin{equation}
T=\sbase^{\ga}\otimes(\bdry_{\gl}\wedge T_{\ga}^{\gl}).\label{eq:Rep_VVcurr_by_0-curr2}
\end{equation}

In the case where the $0$-currents $T_{\ga}^{\gl}$ are represented
locally by smooth $n$-forms $\std_{\ga}^{\gl}\dx$, one has
\begin{equation}
T(\chi)=\int_{U}S_{\ga}^{\gl}\chi_{\gl}^{\ga}\dx\label{eq:Rep_by_S^mu_alpha}
\end{equation}
in accordance with (\ref{eq:Smooth_VVcr_rep_by_n_forms}).

\subsubsection{The exterior product of a vector valued current and a multi-vector
field\label{subsec:_eta_wedge_T}}

Next, in analogy with Section \ref{subsec:Local-represent-Currents},
for a vector valued $p$-current $T$ and a $q$-multi-vector $\eta$,
$q\le n-p$, consider the vector valued $(p+q)$-current $\eta\wedge T$
defined by
\begin{equation}
(\eta\wedge T)(\vf\otimes\go):=T(\vf\otimes(\eta\contr\go)).
\end{equation}
In particular, for multi-indices $\hat{\gl}$, $|\hat{\gl}|=n-p$,
we define locally the vector valued $n$-currents
\begin{equation}
T_{\hat{\gl}}:=\bdry_{\hat{\gl}}\wedge T,\qquad T_{\hat{\gl}}(\vf\otimes\gth)=T(\vf\otimes(\bdry_{\hat{\gl}}\contr\gth)),\label{eq:T_lambda}
\end{equation}
so that 
\begin{equation}
T_{\hat{\gl}}(\vf\otimes\dx)=T(\vf\otimes(\bdry_{\hat{\gl}}\contr\dx)).
\end{equation}

\subsubsection{The contraction of a vector valued current and a form\label{subsec:contr_VVcurr_form}}

Also, for a vector valued $p$-current $T$ and a $q$-form $\psi$,
$q\le p$, define the vector valued $(p-q)$-currents $\psi\contr T$
and $T\fcontr\psi$ as
\begin{equation}
(\psi\contr T)(\vf\otimes\go):=T(\vf\otimes(\psi\wedge\go))
\end{equation}
and

\begin{equation}
(T\fcontr\psi)(\vf\otimes\go):=T(\vf\otimes(\go\wedge\psi)),
\end{equation}
so that $T\fcontr\psi=(-1)^{pq}\psi\contr T$. In the case where $q=p$,
we obtain an element $\go\contr T\in C^{r}(\vb)^{*}$, a vector valued
$0$-current, satisfying
\begin{equation}
(\go\contr T)(\vf)=T(\vf\otimes\go).\label{eq:omega_contr_T}
\end{equation}
Locally, one may consider the functionals\textemdash vector valued
$0$-currents,
\begin{equation}
T^{\gl}:=\dx^{\gl}\contr T,\qquad T^{\gl}(\vf)=T(\vf\otimes\dx^{\gl}),\qquad\abs{\gl}=p.
\end{equation}
Hence, 
\begin{equation}
T(\vf\otimes\go)=T^{\gl}(\go_{\gl}\vf).
\end{equation}
is a local representation of the action of $T$ using vector valued
$0$-currents. It is implied by the identity $T^{\gl}(\go_{\gl}\vf)=\bdry_{\gl}\wedge T^{\gl}(\vf\otimes\go)$,
that
\begin{equation}
T=\bdry_{\gl}\wedge T^{\gl}.
\end{equation}

\subsubsection{Representation by $n$-currents\label{subsec:Rep_VVcurr-by-n-currents}}

Next, for a local basis $\dx^{\hat{\gl}}$ of $\ext^{n-p}T^{*}\base$,
using (\ref{eq:T_lambda}) and (\ref{eq:omega_contr_T}),
\begin{equation}
\begin{split}(\dx^{\hat{\gl}}\contr T_{\hat{\gl}})(\vf\otimes\go) & =T_{\hat{\gl}}(\vf\otimes(\dx^{\hat{\gl}}\wedge\go)),\\
 & =T(\vf\otimes(\bdry_{\hat{\gl}}\contr(\dx^{\hat{\gl}}\wedge\go))).
\end{split}
\end{equation}

Following the same procedure as that leading to (\ref{eq:Rep_Curr2b})
and (\ref{eq:Rep_Curr2c}), one concludes that the vector valued $p$-current
$T$ may be represented locally by the vector valued $n$-currents
$T_{\hat{\gl}}$ in the form
\begin{equation}
T=\dx^{\hat{\gl}}\contr T_{\hat{\gl}},\label{eq:eq:Rep_VVCurr2a}
\end{equation}
and
\begin{equation}
T(\vf\otimes\go)=T_{\hat{\gl}}(\vf\otimes(\dx^{\hat{\gl}}\wedge\go))=\sum_{\gl}\eps^{\hat{\gl}\gl}T_{\hat{\gl}}(\go_{\gl}\vf\otimes\dx).\label{eq:eq:Rep_VVCurr2b}
\end{equation}
Using (\ref{eq:T dot w}) and (\ref{eq:Rep_VVcur_by_p-currents-2}),
we may define the (scalar) $n$-currents
\begin{equation}
T_{\ga\hat{\gl}}:=T_{\hat{\gl}}\cdot\sbase_{\ga},\qquad\text{so that}\qquad T_{\hat{\gl}}=\sbase^{\ga}\otimes T_{\ga\hat{\gl}},
\end{equation}
and
\begin{equation}
T_{\ga\hat{\gl}}(\theta)=T_{\hat{\gl}}(\sbase_{\ga}\otimes\theta)=T(\sbase_{\ga}\otimes(\bdry_{\hat{\gl}}\contr\gth)).
\end{equation}
Considering the $p$-currents $T_{\ga}$ in (\ref{eq:VVcurr_alpha}),
the local components $(T_{\ga})_{\hat{\gl}}$ are defined by $(T_{\ga})_{\hat{\gl}}=\bdry_{\hat{\gl}}\wedge(T\cdot\sbase_{\ga})$,
hence,
\begin{equation}
\begin{split}(T_{\ga})_{\hat{\gl}}(\gth) & =\bdry_{\hat{\gl}}\wedge(T\cdot\sbase_{\ga})(\gth),\\
 & =(T\cdot\sbase_{\ga})(\bdry_{\hat{\gl}}\contr\gth),\\
 & =T(\sbase_{\ga}\otimes(\bdry_{\hat{\gl}}\contr\gth)),
\end{split}
\end{equation}
and we conclude that 
\begin{equation}
T_{\ga\hat{\gl}}=(T_{\ga})_{\hat{\gl}}.
\end{equation}
Thus, Equations (\ref{eq:eq:Rep_VVCurr2a}) and (\ref{eq:eq:Rep_VVCurr2b})
may be rewritten as
\begin{equation}
T=\dx^{\hat{\gl}}\contr(\sbase^{\ga}\otimes T_{\ga\hat{\gl}}),\label{eq:Rep_VVcurr_by_n-curr-0}
\end{equation}
and
\begin{equation}
\begin{split}T(\vf\otimes\go) & =T_{\ga\hat{\gl}}(\vf^{\ga}(\dx^{\hat{\gl}}\wedge\go)),\\
 & =\sum_{\gl}\eps^{\hat{\gl}\gl}T_{\ga\hat{\gl}}(\go_{\gl}\vf^{\ga}\dx),\\
 & =\sum_{\gl}\eps^{\hat{\gl}\gl}\dx\contr T_{\ga\hat{\gl}}(\go_{\gl}\vf^{\ga}),
\end{split}
\label{eq:Rep_VVcur_by_n-curr}
\end{equation}
Comparing the last equation with (\ref{eq:Rep_VVcurr_by_0-curr1})
we arrive at
\begin{equation}
\dx\contr T_{\ga\hat{\gl}}=\eps_{\hat{\gl}\gl}T_{\ga}^{\gl},\qquad T_{\ga\hat{\gl}}=\eps_{\hat{\gl}\gl}\bdry_{X}\wedge T_{\ga}^{\gl}.\label{eq:Rel_Represetations by 0 and n currents}
\end{equation}

In the smooth case, the $n$-currents $T_{\ga\hat{\gl}}$ are represented
by functions $\curnd{\curd}_{\ga\hat{\gl}}$ that make up the vector
valued $(n-p)$-form
\begin{equation}
\curnd{\curd}=\curnd{\curd}_{\ga\hat{\mu}}\sbase^{\ga}\otimes\dx^{\hat{\mu}}\label{eq:Rep_VVcurr_by_n-p_forms}
\end{equation}
as in (\ref{eq:Action_Smooth_VVcurrents_rep_by_forms},\ref{eq:Smooth_VVcr_rep_by_n-p_forms}).
\begin{rem}
In summary, the representation by zero currents (\eg, (\ref{eq:Rep_by_S^mu_alpha})
corresponds to viewing the vector valued current as an element of
$C^{r}(\vb\otimes\ext^{p}T^{*}\base)^{*}=:C^{-r}(\vb^{*}\otimes\ext^{p}T\base\otimes\nfo)$.
On the other hand, the representation as in (\ref{eq:Rep_by_S^mu_alpha},\ref{eq:Rep_VVcurr_by_n-p_forms})
corresponds to the point of view that by the isomorphism of $\ext^{p}T\base\otimes\nfo$
with $\ext^{n-p}T^{*}\base$, 
\begin{equation}
C^{r}(\vb\otimes\ext^{p}T^{*}\base)^{*}\isom C^{-r}(\vb^{*}\otimes\ext^{n-p}T^{*}\base)
\end{equation}
\end{rem}

\begin{rem}
\label{rem:Change_Order_VV}In the foregoing discussion we have made
special choices and used, for example, the definitions, $T_{\hat{\gl}}:=\bdry_{\hat{\gl}}\wedge T$
and $T^{\gl}:=\dx^{\gl}\contr T$ rather than $T'_{\hat{\gl}}:=T\wedge\bdry_{\hat{\gl}}$
and $T'^{\gl}:=T\fcontr\dx^{\gl}$, respectively. The correspondence
between the two schemes is a natural extension of Remark \ref{rem:Change_Order_1}.
In particular, $\eps_{\hat{\gl}\gl}$ will be replaced by $\eps_{\gl\hat{\gl}}$.
\end{rem}

\section{The Representation of Forces by Hyper-Stresses and Non-Holonomic
Stresses\label{sec:Stresses}}

\subsection{Stresses and non-holonomic stresses\label{subsec:Rep_Forces_By_Stresses}}

We recall that the tangent space $T_{\conf}C^{r}(\xi)$ to the Banach
manifold of $C^{r}$-sections of the fiber bundle $\xi:\fb\to\base$
at the section $\conf:\base\to\fb$ may be identified with the Banachable
space $C^{r}(\conf^{*}V\fb)$ of sections of the pullback vector bundle
$\conf^{*}\tau_{\fb}:\conf^{*}V\fb\tto\base$. Elements of the tangent
space at $\conf$ to the configuration manifold represent generalized
velocities of the continuous mechanical system. Consequently, a generalized
force is modeled mathematically by and element $\fc\in C^{r}(\conf^{*}V\fb)^{*}$.
The central message of this section is that although such functionals
cannot be restricted naturally to subbodies of $\base$, as discussed
in Section \ref{subsec:Functionals_Manifolds_with_corners}, forces
may be represented, non-uniquely, by stress objects that enable restriction
of forces to sub-bodies. In order to simplify the notation, we will
consider a general vector bundle $\pi:\vb\to\base$, as in Section
\ref{sec:Banachable}, and the notation introduced there will be used
throughout. The construction is analogous to representation theorem
for distributions of finite order (\eg, \cite[p. 91]{schwartz_theorie_1973}
or \cite[p. 259]{treves_topological_1967}).

Consider the jet extension linear mapping 
\begin{equation}
j^{r}:C^{\r}(\pi)\tto C^{0}(\pi^{\r})
\end{equation}
as in Section \ref{subsec:jet-extension}. As noted, $j^{r}$ is an
embedding and under the norm induced by an atlas, it is even isometric.
Evidently, due to the compatibility constraint, $\image j^{r}$ is
a proper subset of $C^{0}(\pi^{r})$ and its complement is open. Hence,
the inverse
\begin{equation}
(j^{r})^{-1}:\image j^{r}\tto C^{r}(\pi)
\end{equation}
is a well defined linear homeomorphism. Given a force $\fc\in C^{r}(\pi)^{*}$,
the linear functional
\begin{equation}
\fc\comp(j^{r})^{-1}:\image j^{r}\tto\reals
\end{equation}
is a continuous and linear functional on $\image j^{r}$. Hence, by
the Hahn-Banach theorem, it may be extended to a linear functional
$\stm\in C^{0}(\pi^{r})^{*}$. In other words, the linear mapping
\begin{equation}
j^{r*}:C^{0}(\pi^{r})^{*}\tto C^{r}(\pi)^{*}
\end{equation}
is surjective.

By the definition of the dual mapping, $\stm$ represents a force
$\fc$, \ie, 
\begin{equation}
j^{r*}\stm=\fc,\label{eq:Equilib}
\end{equation}
if and only if,
\begin{equation}
\fc(\vf)=\stm(j^{r}\vf)\label{eq:PrinVirtWork}
\end{equation}
for all $C^{r}$-virtual velocity fields $\vf$. The object $\stm\in C^{0}(\pi^{r})^{*}$
is interpreted as a generalization of the notion of hyper-stress in
higher-order continuum mechanics and will be so referred to. For $r=1$,
$\stm$ is a generalization of the standard stress tensor. The condition
(\ref{eq:PrinVirtWork}), resulting from the representation theorem,
is a generalization of the principle of virtual work as it states
that the power expended by the force $\fc$ for a virtual velocity
field $\vf$ is equal to the power expended by the hyper-stress for
$j^{r}\vf$\textemdash containing the first $r$ derivatives of the
velocity field. Accordingly, Equation (\ref{eq:Equilib}) is a generalization
of the equilibrium equation of continuum mechanics.

It is noted that $\stm$ is not unique. The non-uniqueness originates
from the fact that the image of the jet extension mapping, containing
the compatible jet fields, is not a dense subset of $C^{0}(\pi^{r})$.
Thus, the static indeterminacy of continuum mechanics follows naturally
from the representation theorem.

In view of (\ref{eq:Isometries}), the same procedure applies if we
use the non-holonomic jet extension $\nj:C^{r}(\pi)\to C^{0}(\hat{\pi}^{r})$.
A force $\fc$ may then be represented by a non-unique, non-holonomic
stress $\hat{\stm}\in C^{0}(\hat{\pi})^{*}$ in the form
\begin{equation}
\fc=\nj^{r*}\hat{\stm}.\label{eq:Gen_Equil_NHS}
\end{equation}
The mapping $C^{0}(\incl^{r})$ of Section \ref{subsec:iterated-jet-ext}
is an embedding. Hence, a hyper-stress $\stm$ may be represented
by some non-unique, non-holonomic stress $\hat{\stm}$ in the form
\begin{equation}
\stm=C^{0}(\incl^{r})^{*}(\hat{\stm}),
\end{equation}
and in the following commutative diagram all mappings are surjective.\begin{equation}
\label{eq:Reps}
\begin{xy}
(40,25)*+{C^0(\nh\pi^\r)} ="TS";
(-10,25)*+{C^\r(\pi)} = "TQ";
(15,25)*+{C^0(\pi^\r)} ="jTS";
{\ar@{<-}@/^{2pc}/^{(\nj^\r)^*} "TQ"; "TS"};
{\ar@{<-}^{j^{\r*}} "TQ"; "jTS"};
{\ar@{<-}^{C^0(\incl^\r)^*} "jTS"; "TS"};
\end{xy}
\end{equation}

\subsection{Smooth stresses\label{subsec:Smooth-stresses}}

In view of the discussion in Section \ref{subsec:Generalized-sections},
hyper-stresses are elements of 
\begin{equation}
C^{0}(\pi^{r})^{*}=C^{-0}((\pi^{r})')=C^{-0}(\pi^{r*}\otimes\ext^{n}T^{*}\base),
\end{equation}
and so they may be approximated by smooth sections of $\pi^{r*}\otimes\nfo$,
\ie, $n$-forms valued in the dual of the $r$-jet bundle.

Similarly, non-holonomic stresses are elements of 
\begin{equation}
C^{0}(\nh{\pi}^{r})^{*}=C^{-0}(\nh{\pi}^{r*}\otimes\ext^{n}T^{*}\base),
\end{equation}
 and smooth non-holonomic stresses are $n$-forms valued in the dual
of the $r$-iterated jet bundle.

\subsection{Stress measures\label{subsec:Stress-measures}}

Analytically, stresses are vector valued zero-currents that are representable
by integration. (See \cite[Section 4.1]{federer_geometric_1969},
for the scalar case).

As noted in Section \ref{subsec:Local_Functionals_Without-Boudaries},
given a vector bundle atlas $\{(U_{a},\vph_{a},\Phi_{a})\}_{a\in A}$,
a linear functional is uniquely determined by its restrictions to
sections supported in the various domains $\vph_{\ga}(U_{\ga})$\textemdash its
local representatives. In particular, for the case of an $m$-dimensional
vector bundle $\pi:\vb\to\base$, and the space of functionals $C^{0}(\pi)^{*}$,
a typical local representative is an element $T_{a}\in\paren{C_{c}^{0}(\vph_{a}(U_{a}))^{*}}^{m}$.
Thus, each component $(T_{a})_{\ga}\in C_{c}^{0}(\vph_{a}(U_{a}))^{*}$
is a Radon measure or a distribution representable by integration.
We will use the same notation for the measure. Consequently, for a
section $\vf_{a}$ compactly supported in $\vph_{a}(U_{a})$, we may
write
\begin{equation}
T_{a}(\vf_{\ga})=\int_{\vph_{a}(U_{a})}\vf_{a}\cdot\dee T_{a}:=\int_{\vph_{a}(U_{a})}\vf_{a}^{\ga}\dee T_{a\ga}.
\end{equation}
Given a partition of unity $\{u_{a}\}$ subordinate to the atlas,
one has
\begin{equation}
T(\vf)=\int_{\base}\vf\cdot\dee T:=\sum_{a\in A}\int_{\vph_{a}(U_{a})}\Phi_{a}(u_{a}\vf)\cdot\dee T_{a}=\sum_{a\in A}\int_{\vph_{a}(U_{a})}u_{a}\vf_{a}^{\ga}\dee T_{a\ga}.
\end{equation}

For the case of stresses, one has to replace $\vb$ by $J^{r}\vb$,
$T$ by $\stm$, and $T_{a\ga}$ by $\stm_{a\ga}^{\mii}$, $\abs{\mii}\le r$.
In addition, as $\base$ is a manifold with corners, representing
measures may be viewed as measures on the extension $\tilde{\base}$
which are supported in $\base$. Thus, for a section $\chi$ of $J^{r}\vb$,
represented locally by $\chi_{a\mii}^{\ga}$,
\begin{equation}
\stm(\chi)=\int_{\base}\chi\cdot\dee\stm:=\sum_{a\in A}\int_{\vph_{a}(U_{a})}\Phi_{a}(u_{a}\chi)\cdot\dee\stm_{a}=\sum_{a\in A}\int_{\vph_{a}(U_{a})}u_{a}\chi_{a\mii}^{\ga}\dee\stm_{a\ga}^{\mii}.
\end{equation}
We note that the components $\stm_{a\ga}^{\mii}$ have the same symmetry
under permutations of $\mii$ as sections of the jet bundle. If $\vf$
is a section of a vector bundle $\vb_{0}$, then,

\begin{equation}
\stm(j^{r}\vf)=\int_{\base}j^{r}\vf\cdot\dee\stm=\sum_{a\in A}\int_{\vph_{a}(U_{a})}u_{a}\vf_{a,\mii}^{\ga}\dee\stm_{a\ga}^{\mii}.
\end{equation}

The same reasoning applies to the representation by non-holonomic
stresses, only here we consider sections $\nh{\chi}$ of the iterated
jet bundle represented locally by $\nh{\chi}_{a\mii_{\rp}}^{\rp\ga_{\rp}}$,
$G_{\rp}\le r$. The local non-holonomic stress measures have components
$\nh{\stm}_{a\ga_{\rp}}^{\rp\mii_{\rp}}$ and 
\begin{equation}
\nh{\stm}(\nh{\chi})=\int_{\base}\nh{\chi}\cdot\dee\nh{\stm}:=\sum_{a\in A}\int_{\vph_{a}(U_{a})}\Phi_{a}(u_{a}\nh{\chi})\cdot\dee\nh{\stm}_{a}=\sum_{a\in A}\int_{\vph_{a}(U_{a})}u_{a}\nh{\chi}_{a\mii_{\rp}}^{\rp\ga_{\rp}}\dee\nh{\stm}_{a\ga_{\rp}}^{\rp\mii_{\rp}},
\end{equation}

\begin{equation}
\nh{\stm}(\nj^{r}\vf)=\int_{\base}\nj^{r}\vf\cdot\dee\nh{\stm}=\sum_{a\in A}\int_{\vph_{a}(U_{a})}u_{a}\vf_{a,\mii_{\rp}}^{\rp\ga_{\rp}}\dee\stm_{a\ga_{\rp}}^{\rp\mii_{\rp}}.
\end{equation}
where summation is implied on all values of $\ga_{\rp}$, $\mii_{\rp}$,
for all values of $\rp$ such that $G_{\rp}\le r$.

It is concluded that for a given force $\fc$, there is some non-unique
vector valued hyper-stress measure $\stm$ and a non-holonomic stress
measure $\hat{\stm}$, such that 
\begin{equation}
\fc(\vf)=\int_{\base}j^{r}\vf\cdot\dee\stm=\int_{\base}\nj^{r}\vf\cdot\dee\nh{\stm}.
\end{equation}

\subsection{Force system induced by stresses\label{subsec:Force-systems}}

It was noted in Section \ref{subsec:Functionals_Manifolds_with_corners}
that given a force on a body $\base$, a manifold with corners, there
is no unique way to restrict it to an $n$-dimensional submanifold
with corners, a sub-body $\reg\subset\base$. We view this as the
fundamental problem of continuum mechanics\textemdash static indeterminacy.

Stress, though not determined uniquely by a force, provide means for
inducing a \emph{force system,} the assignment of a force $\fc_{\reg}$
to each sub-body $\reg$. Indeed, once a stress measure is given,
be it a hyper-stress or a non-holonomic stress, integration theory
makes it possible to consider the force system given by
\begin{equation}
\fc_{\reg}(\vf)=\int_{\reg}j^{r}\vf\cdot\dee\stm=\int_{\reg}\nj^{r}\vf\cdot\dee\nh{\stm}\label{eq:Induc_Force_Sys}
\end{equation}
for any section $\vf$ of $\pi\resto{\reg}$.

Further details on the relation between hyper-stresses and force systems
are available in \cite{segev_consistency_1991}. It is our opinion
that the foregoing line of reasoning captures the essence of stress
theory in continuum mechanics accurately and elegantly.

\section{Simple Forces and Stresses\label{sec:Simple-Forces-St}}

We restrict ourselves now to the most natural setting for continuum
mechanics, the case $r=1$\textemdash the first value for which the
set of $C^{r}$-embeddings is open in the manifold of mappings. (See
\cite{falach_configuration_2015} for consideration of configurations
modeled as Lipschitz mappings.) Evidently, hyper-stresses and non-holonomic
stresses become identical now, and therefore, it is natural in this
case to use the terminology \emph{simple forces} and \emph{stresses}.

\subsection{Simple stresses}

A simple stress $\stm$ on a body $\base$ is an element of 
\begin{equation}
C^{0}(J^{1}\vb)^{*}=:C^{-0}((J^{1}\vb)^{*}\otimes\nfo)
\end{equation}
which implies that smooth stress distributions are sections of 
\begin{equation}
(J^{1}\vb)^{*}\otimes\nfo=\L{J^{1}\vb,\nfo}.
\end{equation}
Following the discussion in Section \ref{subsec:Supported-functionals},
$\stm$ may be viewed as a generalized section of $(J^{1}\tilde{\vb})^{*}\otimes\ext^{n}T^{*}\tilde{\base}$,
which is supported in $\base$, where we use the extension of the
vector bundle to a vector bundle $\tilde{\pi}:\tilde{\vb}\to\tilde{\base}$
over a compact manifold without boundary $\tilde{\base}$.

A typical local representative of a section of the jet bundle is of
the form
\begin{equation}
\chi=\chi^{\ga}\sbase_{\ga}+\chi_{i}^{\ga}\dx^{i}\otimes\sbase_{\ga}
\end{equation}
so that locally,
\begin{equation}
\begin{split}\stm(\chi) & =\stm(\chi^{\ga}\sbase_{\ga}+\chi_{i}^{\ga}\dx^{i}\otimes\sbase_{\ga}),\\
 & =\stm_{\ga}(\chi^{\ga})+\stm_{\ga}^{i}(\chi_{i}^{\ga}).
\end{split}
\label{eq:Loc_Rep_stm}
\end{equation}
Here, $\stm_{\ga}$ and $\stm_{\ga}^{i}$ are $0$-currents defined
by
\begin{gather}
\stm_{\ga}(u):=(\stm\cdot\sbase_{\ga})(u)=\stm(u\sbase_{\ga}),\\
\stm_{\ga}^{i}(u):=(\stm\cdot(\dx^{i}\otimes\sbase_{\ga}))(u)=\stm(u\dx^{i}\otimes\sbase_{\ga}).
\end{gather}

In the smooth case, $\stm$ is represented by a section $\std$ of
$\text{(\ensuremath{J^{1}\vb)^{*}\otimes\nfo}}$ in the form
\begin{equation}
\stm(\chi)=\int_{\base}\std\cdot\chi.
\end{equation}
Locally, such a vector valued form is represented as
\begin{equation}
\std=(\std_{\ga}\sbase^{\ga}+\std_{\ga}^{i}\bdry_{i}\otimes\sbase^{\ga})\otimes\dx\label{eq:smooth simple  stresses}
\end{equation}
so that, for the domain of a chart, $U$, and a section $\chi$ with
$\support\chi\subset U$,
\begin{equation}
\stm(\chi)=\int_{U}(\std_{\ga}\chi^{\ga}+\std_{\ga}^{i}\chi_{i}^{\ga})\dx.\label{eq:smooth_simp_stress_action}
\end{equation}

\subsection{The vertical projection\label{subsec:The-vertical-projection}}

The vertical sub-bundle 
\begin{equation}
V\pi^{1}:VJ^{1}\vb\tto\base
\end{equation}
is kernel of the natural projection
\begin{equation}
\pi_{0}^{1}:J^{1}\vb\tto\vb.
\end{equation}
In other words, elements of the vertical sub-bundle at a point $\bp\in\base$
are jets of sections that vanish at $\bp$. Thus, if a typical element
of $J^{1}\vb$ is represented locally in the form $(\chi^{\ga},\chi_{i}^{\gb}$),
an element of the vertical sub-bundle, $VJ^{1}\vb\subset J^{1}\vb$
has the form $(0,\chi_{i}^{\gb})$ in any adapted coordinate system.
The vertical sub-bundle may be identified with the vector bundle $T^{*}\base\otimes\vb$.
Denoting the natural inclusion by
\begin{equation}
\incl_{V}:VJ^{1}\vb\tto J^{1}\vb,
\end{equation}
one has the induced inclusion
\begin{equation}
C^{0}(\incl_{V}):C^{0}(V\pi^{1})\tto C^{0}(\pi^{1}),\qquad\chi\lmt\incl_{V}\comp\chi.
\end{equation}
Clearly, $C^{0}(\incl_{V})$ is injective and a homeomorphism onto
its image. Hence, its dual
\begin{equation}
C^{0}(\incl_{V})^{*}:C^{0}(\pi^{1})^{*}\tto C^{0}(V\pi^{1})^{*}\isom C^{0}(T^{*}\base\otimes\vb)^{*}=C^{-0}(T\base\otimes\vb^{*}\otimes\nfo),
\end{equation}
is a well defined surjection. Simply put, $C^{0}(\incl_{V})^{*}(\stm)$
is the restriction of the stress $\stm$ to sections of the vertical
sub-bundle. Accordingly, we will refer to an element of $C^{0}(T^{*}\base\otimes\vb)^{*}$
as a \emph{vertical stress} and to $C^{0}(\incl_{V})^{*}$ as the
\emph{vertical projection}.

In case the stress $\stm$ is represented locally by the $0$-currents
$(\stm_{\ga},\stm_{\gb}^{i})$ as in (\ref{eq:Loc_Rep_stm}), then
$C^{0}(\incl_{V})^{*}(\stm)$ is represented by $(\stm_{\gb}^{i})$.

Let $\stm^{+}\in C^{0}(T^{*}\base\otimes\vb)^{*}$ be a vertical stress
and let $\vf\in C^{0}(\pi)$. Then, $\stm^{+}\cdot\vf$ defined by
\begin{equation}
(\stm^{+}\cdot\vf)(\vph):=\stm^{+}(\vph\otimes\vf),\qquad\vph\in C^{0}(T^{*}\base)
\end{equation}
is a 1-current. This is an indication of the fact that $\stm^{+}$
may be viewed as a vector valued $1$-current.

We may use the local representation of currents as in Section \ref{subsec:Local-represe-VVCurrents}
to represent $\stm^{+}$ by the scalar $1$-currents $\stm_{\ga}^{+}=(\stm^{+}\cdot\sbase_{\ga})$
given as
\begin{equation}
\stm_{\ga}^{+}(\vph)=(\stm^{+}\cdot\sbase_{\ga})(\vph)=\stm^{+}(\vph\otimes\sbase_{\ga})
\end{equation}
so that
\begin{equation}
\begin{split}\stm^{+}(\vph\otimes\vf) & =\stm^{+}(\vph^{i}\vf^{\ga}\dx^{i}\otimes\sbase_{\ga}),\\
 & =\stm_{\ga}^{+}(\vph_{i}\vf^{\ga}\dx^{i}),\\
 & =(\vf^{\ga}\contr\stm_{\ga}^{+})(\vph).
\end{split}
\end{equation}
Similarly, the $1$-current $\stm^{+}\cdot\vf$ may be represented
locally as in Section \ref{subsec:Local-represent-Currents} by $0$-currents
$(\stm^{+}\cdot\vf)^{i}$ given by
\begin{equation}
(\stm^{+}\cdot\vf)^{i}(u)=(\dx^{i}\contr(\stm^{+}\cdot\vf))(u)=(\stm^{+}\cdot\vf)(u\dx^{i})
\end{equation}
in the form
\begin{equation}
(\stm^{+}\cdot\vf)(\vph)=(\stm^{+}\cdot\vf)^{i}(\vph_{i}).
\end{equation}
Evidently,
\begin{equation}
\begin{split}(\stm^{+}\cdot\vf)^{i}(u) & =\stm^{+}(u\vf^{\ga}\dx^{i}\otimes\sbase_{\ga}),\\
 & =\stm_{\ga}^{+i}(u\vf^{\ga}),\\
 & =(\vf^{\ga}\contr\stm_{\ga}^{+i})(u),
\end{split}
\end{equation}
and so
\begin{equation}
(\stm^{+}\cdot\vf)^{i}=\vf^{\ga}\contr\stm_{\ga}^{+i}.
\end{equation}
It is concluded that 
\begin{equation}
\stm^{+}=\sbase^{\ga}\otimes(\bdry_{i}\wedge\stm_{\ga}^{+i}),\qquad\stm_{\ga}^{+i}=\dx^{i}\contr(\stm^{+}\cdot\sbase_{\ga}),\label{eq:Rep_Vert_St}
\end{equation}
\begin{equation}
\stm^{+}(\vph\otimes\vf)=\stm_{\ga}^{+i}(\vph_{i}\vf^{\ga})\label{eq:Rep_Vert_St-1}
\end{equation}
where it is recalled that in case $\stm^{+}=C^{0}(\incl_{V})^{*}(\stm)$,
then, $\stm_{\ga}^{+i}=\stm_{\ga}^{i}$.

In the smooth case, the vertical projection of the stress is represented
by a section $\std^{+}$of $T\base\otimes\vb^{*}\otimes\nfo$ so that
\begin{equation}
C^{0}(\incl_{V})^{*}(\stm)(\chi)=\int_{\base}\std^{+}\cdot\chi,
\end{equation}
where
\begin{equation}
(\std^{+}\cdot\chi)(v_{1},\dots,v_{n})(\bp)=\std^{+}(\bp)(\chi(\bp)\otimes(v_{1}(\bp)\wedge\cdots\wedge v_{n}(\bp))).
\end{equation}
 If $\stm$ is represented by a section $\std$ of $(J^{1}\vb)^{*}\otimes\ext^{n}T^{*}\base$
then, $C^{0}(\incl_{V})^{*}(\stm)$ is represented locally by
\begin{equation}
\std^{+}=\std_{\ga}^{i}\bdry_{i}\otimes\sbase^{\ga}\otimes\dx.\label{eq:rep_vert_proj_smooth_stress}
\end{equation}

For a vertical stress $\stm^{+}$which is represented by $S_{\ga}^{i}$
as above and a field $\vf$, the $1$-current $\stm^{+}\cdot\vf$
is given locally by
\begin{equation}
\vph\lmt\int_{U}S_{\ga}^{i}\vf^{\ga}\vph_{i}\dx,
\end{equation}
for a $1$-form $\vph$ supported in $U$. In other words, if the
vertical stress $\stm^{+}$ is represented by the section $\std^{+}$
of $T\base\otimes\vb\otimes\nfo$, the $1$-current $\stm^{+}\cdot\vf$
is represented by the density $\std^{+}\cdot\vf$, a section of $T\base\otimes\nfo$
given by
\begin{equation}
(\std^{+}\cdot\vf)(\vph)=S^{+}(\vph\otimes\vf).
\end{equation}

\subsection{Traction stresses\label{subsec:Traction-stresses}}

Using the transposition $\tr:\vb\otimes T^{*}\base\to T^{*}\base\otimes\vb$,
one has a mapping on the space of vertical stresses
\begin{equation}
C^{0}(\tr)^{*}:C^{0}(T^{*}\base\otimes\vb)^{*}\tto C^{0}(\vb\otimes T^{*}\base)^{*}.
\end{equation}

We define \emph{traction stress} \emph{distributions }to be elements
of

\begin{equation}
C^{0}(\vb\otimes T^{*}\base)^{*}=C^{-0}(\vb^{*}\otimes\ext^{n-1}T^{*}\base).
\end{equation}
Thus, it is noted that a traction stress is not much different than
a vertical stress distribution but transposition enables its representation
as a vector valued current. Using local representation in accordance
with Section \ref{subsec:Rep_VVcurr-by-n-currents}, a traction stress
$\tst$ is represented locally by $n$-currents $\tst_{\ga\hi},$
$|\hi|=n-1$, in the form
\begin{equation}
\tst=\dx^{\hi}\contr\tst_{\hi}=\sbase^{\ga}\otimes\tst_{\ga}=\dx^{\hi}\contr(\sbase^{\ga}\otimes\tst_{\ga\hi}),
\end{equation}
where,
\begin{equation}
\tst_{\hi}:=\bdry_{\hi}\wedge\st,\qquad\tst_{\ga}:=\tst\cdot\sbase_{\ga},\qquad\tst_{\ga\hi}:=\tst_{\hi}\cdot\sbase_{\ga}=(\st_{\ga})_{\hi}=(\bdry_{\hi}\wedge\st)\cdot\sbase_{\ga}.
\end{equation}
Hence, 
\begin{equation}
\begin{split}\tst(\vf\otimes\vph) & =\sum_{i}\eps^{\hi i}\st_{\ga\hi}(\vph_{i}\vf^{\ga}\dx),\\
 & =\sum_{i}(-1)^{n-i}\tst_{\ga\hi}(\vph_{i}\vf^{\ga}\dx),\\
 & =\sum_{i}\tst_{\ga\hi}(\vf^{\ga}\vph\wedge\dx^{\hi}),\\
 & =\sum_{i}\tst_{\ga\hi}\fcontr\dx^{\hi}(\vf^{\ga}\vph).
\end{split}
\label{eq:Rep_Act_Trac_St}
\end{equation}

Introducing the notation 
\begin{equation}
p_{\tst}:=C^{0}(\tr)^{*}\comp C^{0}(\incl_{V})^{*}:C^{0}(\pi^{1})^{*}\tto C^{0}(\vb\otimes T^{*}\base)^{*},\label{eq:Def_p_sigma}
\end{equation}
a simple stress distribution $\stm$ induces a traction stress distribution
$\tst$ by
\begin{equation}
\tst=p_{\tst}(\stm).
\end{equation}

Let $\tst=p_{\tst}(\stm)$, then, comparing the last equation with
(\ref{eq:Rep_Vert_St-1}), it is concluded that, in accordance with
(\ref{eq:Rel_Represetations by 0 and n currents}), locally,
\begin{equation}
(-1)^{n-i}\tst_{\ga\hi}(u\dx)=\stm_{\ga}^{i}(u)\label{eq:Rel_sigma_stress-1}
\end{equation}
for any function $u$, and so
\begin{equation}
\dx\contr\st_{\ga\hi}=(-1)^{n-i}\stm_{\ga}^{i},\qquad\st_{\ga\hi}=(-1)^{n-i}\bdry_{\bp}\wedge\stm_{\ga}^{i}.\label{eq:Rel_sigma_stress-2}
\end{equation}

\begin{rem}
\label{rem:Change_order_trac_st}Continuing Remark \ref{rem:Change_Order_VV},
it is observed that one may consider 
\begin{equation}
\tst'_{\hi}:=\tst\wedge\bdry_{\hi},\qquad\tst'_{\ga\hi}:=\tst'_{\hi}\cdot\sbase_{\ga}=(\st_{\ga})'_{\hi}=(\st\wedge\bdry_{\hi})\cdot\sbase_{\ga},
\end{equation}
so that
\begin{equation}
\tst=\tst'_{\hi}\fcontr\dx^{\hi}=(\sbase^{\ga}\otimes\tst'_{\ga\hi})\fcontr\dx^{\hi}.
\end{equation}
Thus,
\begin{equation}
\begin{split}\tst(\vf\otimes\vph) & =(\sbase^{\ga}\otimes\tst'_{\ga\hi})\fcontr\dx^{\hi}(\vf\otimes\vph),\\
 & =\tst'_{\ga\hi}(\vf^{\ga}\vph\wedge\dx^{\hi}),\\
 & =\sum_{i}(-1)^{i-1}\tst'_{\ga\hi}(\vph_{i}\vf^{\ga}\dx),
\end{split}
\end{equation}
and comparison with (\ref{eq:Rep_Vert_St-1}) implies that 
\begin{equation}
\st'_{\ga\hi}\fcontr\dx=(-1)^{i-1}\stm_{\ga}^{i},\qquad\st'_{\ga\hi}=(-1)^{i-1}\stm_{\ga}^{i}\wedge\bdry_{\bp}=(-1)^{n-1}\tst_{\ga\hi},\label{eq:Change_order_=00005Csigma}
\end{equation}
where it is observed that as $\stm_{\ga}^{i}$ are zero currents,
the order of the contraction and wedge product in the last equation
is immaterial.
\end{rem}

\subsection{Smooth traction stresses\label{subsec:Smooth-traction-stresses}}

In the smooth case, we adapt (\ref{eq:Smooth_VVcr_rep_by_n-p_forms})
to the current context. The traction stress $\tst$ is represented
by a section $\tstd$ of $\vb^{*}\otimes\ext^{n-1}T^{*}\base$ so
that
\begin{equation}
\tst(\chi)=\int_{\base}\tstd\dot{\wedge}\chi.\label{eq:action_tr_st}
\end{equation}
Locally,
\begin{equation}
\tstd=\tstd_{\ga\hi}\sbase^{\ga}\otimes\dx^{\hi},\qquad\tstd\dot{\wedge}\chi=\sum_{i}\eps^{\hi i}\tstd_{\ga\hi}\chi_{i}^{\ga}\dx
\end{equation}
so that the local components of $\tstd$ are the $n$-currents\textemdash functions\textemdash that
represent $\tst$ locally.

Let $\stm$ be a smooth stress represented by the vector valued form
$\std$, a section of $(J^{1}\vb)^{*}\otimes\nfo$ as in (\ref{eq:smooth simple  stresses},
\ref{eq:smooth_simp_stress_action}) and let $\std^{+}$ be its vertical
component as in (\ref{eq:rep_vert_proj_smooth_stress}). In view of
(\ref{eq:t=00003Dfcontr_S}, \ref{eq:Relation_between_TSt_rep_and_var_st}),
$p_{\tst}(S)$ is represented by the section $\tstd$ of $\vb^{*}\otimes\ext^{n-1}T^{*}\base$
with
\begin{equation}
\tstd=\spec C_{\fcontr}(\std^{+}\comp\tr),\qquad\tstd_{\ga\hi}=(-1)^{n-i}\std_{\ga}^{i}.\label{eq:Tr_St_Dens_vs_St_Dens}
\end{equation}

The terminology, traction stress, originates from the fact that a
traction stress $\tst$ represented by a smooth section $\tstd$ of
$\vb^{*}\otimes\ext^{n-1}T^{*}\base$, induces the analog of a traction
field on oriented hyper-surfaces in the body as follows. (See \cite{segev_notes_2013}
for further details.) We consider, for any given section $\tstd$
of $\vb^{*}\otimes\ext^{n-1}T^{*}\base$ and a field $\vf$, the $(n-1)$-form
$\tst\cdot\vf$ given by
\begin{equation}
(\tstd\cdot\vf)(\eta)=\tstd(\vf\otimes\eta)
\end{equation}
for sections $\eta$ of $\ext^{n-1}T\base$. Consider an $(n-1)$-dimensional
oriented smooth submanifold $\srfc\subset\base$. Let
\begin{equation}
\incl_{\srfc}:\srfc\tto\base
\end{equation}
be the natural inclusion and 
\begin{equation}
\incl_{\srfc}^{*}:C^{\infty}(\ext^{n-1}T^{*}\base)\tto C^{\infty}(\ext^{n-1}T^{*}\srfc),
\end{equation}
the corresponding restriction of $(n-1)$-forms. Combining the above,
one may define a linear mapping
\begin{equation}
\rho_{\srfc}:C^{\infty}(\vb^{*}\otimes\ext^{n-1}T^{*}\base)\tto C^{\infty}(\vb^{*}\otimes\ext^{n-1}T^{*}\srfc)
\end{equation}
whereby
\begin{equation}
\rho_{\srfc}(\tstd)\cdot\vf=\incl_{\srfc}^{*}(\tstd\cdot\vf)\in\ext^{n-1}T^{*}\srfc.
\end{equation}

A section $\sfc$ of $\vb^{*}\otimes\ext^{n-1}T^{*}\srfc$ is interpreted
as a \emph{surface traction distribution }on the hyper-surface $\srfc$.
Its action $\sfc\cdot\vf$ is interpreted as the power density of
the corresponding surface force, and may be integrated over $\srfc$.
In particular, the condition
\begin{equation}
\sfc=(-1)^{n-1}\rho_{\srfc}(\tstd)\label{eq:Cauchy_Form_gen}
\end{equation}
is a generalization of Cauchy's formula for the relation between traction
fields and stresses.
\begin{rem}
\label{rem:factor_(-1)^n-1}The factor $(-1)^{n-1}$ that appears
in (\ref{eq:Cauchy_Form_gen}) above, and is absent in \cite{segev_notes_2013},
originates from our use of the choice to use exterior multiplication
on the left as in Remarks \ref{rem:Change_Order_1} and \ref{rem:Change_Order_VV}.
Evidently, if we represented $\tst$ by the smooth vector valued form
$\tstd'$ such that 
\begin{equation}
\tst(\chi)=\int_{\base}\chi\dot{\wedge}\tstd'
\end{equation}
instead of (\ref{eq:action_tr_st}), the factor $(-1)^{n-1}$ would
not appear in the analogous computation and 
\begin{equation}
\sfc=\rho_{\srfc}(\tstd').
\end{equation}
In addition, the second of Equations (\ref{eq:Tr_St_Dens_vs_St_Dens}),
may be rewritten as 
\begin{equation}
\tstd'_{\ga\hi}=(-1)^{n-1}\std_{\ga}^{i}.
\end{equation}

\end{rem}

\subsection{The generalized divergence of the stress}

Let $\stm$ be a stress distribution, $\tst=p_{\tst}(\stm)$, and
$\vf\in C^{1}(\pi)$. We compute using (\ref{eq:Rep_Act_Trac_St})
a local expression for the boundary of the $1$-current $\tst\cdot\vf$
as
\begin{equation}
\begin{split}\bdry(\tst\cdot\vf)(u) & =(\tst\cdot\vf)(\dee u),\\
 & =\tst(\vf\otimes\dee u),\\
 & =\stm_{\ga}^{i}(u_{,i}\vf^{\ga}),\\
 & =\stm_{\ga}^{i}((u\vf^{\ga})_{,i})-\stm_{\ga}^{i}(u\vf_{,i}^{\ga}),\\
 & =\stm_{\ga}^{i}(\bdry_{i}\contr\dee(u\vf^{\ga}))-\stm_{\ga}^{i}(u\vf_{,i}^{\ga})-\stm_{\ga}(u\vf^{\ga})+\stm_{\ga}(u\vf^{\ga}),\\
 & =\bdry_{i}\wedge\stm_{\ga}^{i}(\dee(u\vf^{\ga}))-(\stm\cdot j^{1}\vf)(u)+(\vf^{\ga}\contr\stm_{\ga})(u),\\
 & =\bdry(\bdry_{i}\wedge\stm_{\ga}^{i})(u\vf^{\ga})-(\stm\cdot j^{1}\vf)(u)+(\vf^{\ga}\contr\stm_{\ga})(u),\\
 & =[(\bdry_{i}\stm_{\ga}^{i}\cdot\vf^{\ga})-(\stm\cdot j^{1}\vf)+(\vf^{\ga}\contr\stm_{\ga})](u).
\end{split}
\label{eq:Compute-bdry(sig_cdot_w)}
\end{equation}
Here, for a $0$-current $T=\stm_{\ga}^{i}$, $\bdry_{i}T$ is the
``partial boundary'' operator or the dual to the partial derivative,
a $0$-current defined by 
\begin{equation}
\begin{split}\bdry_{i}T(u) & :=\bdry(\bdry_{i}\wedge T)(u),\\
 & =(\bdry_{i}\wedge T)(du),\\
 & =T(u_{,i}).
\end{split}
\end{equation}
Since $\tst\cdot\vf$ and $\tst_{\ga}$ are $1$-currents, definition
(\ref{eq:Ext_Der_Curr}) implies that the exterior derivatives satisfy
\begin{equation}
\dee(\tst\cdot\vf)=(-1)^{n}\bdry(\tst\cdot\vf),\qquad\dee\st_{\ga}=(-1)^{n}\bdry\tst_{\ga},
\end{equation}
and
\begin{equation}
\dee(\tst\cdot\vf)(u)=(\dee\tst_{\ga}\cdot\vf^{\ga})(u)+(-1)^{n-1}[\stm\cdot j^{1}\vf-\vf^{\ga}\contr\stm_{\ga}](u).
\end{equation}

The computations above imply that there are ``dual'' linear differential
operators 
\begin{equation}
\tilde{\bdry}:C^{-0}(\vb^{*}\otimes\ext^{n-1}T^{*}\base)\tto C^{-1}(\vb^{*}\otimes\nfo)\isom C^{1}(\vb\otimes\reals)^{*},
\end{equation}
and
\begin{equation}
\tilde{\dee}:C^{-0}(\vb^{*}\otimes\ext^{n-1}T^{*}\base)\tto C^{-1}(\vb^{*}\otimes\nfo)\isom C^{1}(\vb\otimes\reals)^{*},
\end{equation}
such that
\begin{equation}
\tilde{\bdry}\tst(\vf\otimes u):=\tst(\vf\otimes\dee u)=\bdry(\tst\cdot\vf)(u)=(\tilde{\bdry}\tst\cdot\vf)(u),\qquad\tilde{\dee}\tst:=(-1)^{n}\tilde{\bdry}\tst.
\end{equation}

Consequently, we define the generalized divergence, a differential
operator
\begin{equation}
\diver:C^{-0}((J^{1}\vb)^{*}\otimes\nfo)\tto C^{-1}(\vb^{*}\otimes\nfo),
\end{equation}
by
\begin{equation}
\diver\stm=-\tilde{\bdry}(p_{\tst}\stm)-j^{1*}\stm.\label{eq:def_divergence}
\end{equation}
Here, we view the various terms as vector valued $0$-currents, elements
of $C^{-1}(\vb^{*}\otimes\nfo)=C^{1}(\vb\otimes\reals)^{*}$, so that
each may be contracted with $\vf$ to give an element of $C^{-1}(\nfo).$
Thus,
\begin{equation}
\begin{split}(\diver\stm\cdot\vf)(u) & =-(\tilde{\bdry}(p_{\tst}\stm)\cdot\vf)(u)-(\stm\cdot j^{1}\vf)(u),\\
 & =-p_{\tst}\stm(w\otimes\dee u)-(\stm\cdot j^{1}\vf)(u).
\end{split}
\label{eq:def_diver_detailed}
\end{equation}
The local expression for the generalized divergence in a coordinate
neighborhood $U$ is obtained using (\ref{eq:Compute-bdry(sig_cdot_w)}).
For a smooth function $u$ having a compact support in $U$,
\begin{equation}
(\diver\stm\cdot\vf)(u)=[-\bdry_{i}\stm_{\ga}^{i}\cdot\vf^{\ga}-\vf^{\ga}\contr\stm_{\ga}](u),
\end{equation}
so that 
\begin{equation}
\diver\stm\cdot\vf=-\bdry_{i}\stm_{\ga}^{i}\cdot\vf^{\ga}-\vf^{\ga}\contr\stm_{\ga}.
\end{equation}

In the smooth case 
\begin{equation}
\begin{split}\bdry_{i}\stm_{\ga}^{i}\cdot\vf^{\ga}(u) & =\bdry_{i}\stm_{\ga}^{i}(\vf^{\ga}u),\\
 & =\stm_{\ga}^{i}((\vf^{\ga}u)_{,i}),\\
 & =\int_{U}\std_{\ga}^{i}(\vf^{\ga}u)_{,i}\dx,\\
 & =\int_{U}(\std_{\ga}^{i}\vf^{\ga}u)_{,i}\dx-\int_{U}\std_{\ga,i}^{i}\vf^{\ga}u\dx,\\
 & =-\int_{U}\std_{\ga,i}^{i}\vf^{\ga}u\dx,
\end{split}
\label{eq:compute_bdry_i}
\end{equation}
where we have used
\begin{equation}
\begin{split}\int_{U}(\std_{\ga}^{i}\vf^{\ga}u)_{,i}\dx & =\int_{U}(\std_{\ga}^{i}\vf^{\ga}u)_{,i}\dx^{i}\wedge\dx^{\hi}\eps_{i\hi},\\
 & =\int_{U}\dee(\std_{\ga}^{i}\vf^{\ga}u)\eps_{i\hi},\\
 & =\int_{\bdry U}\std_{\ga}^{i}\vf^{\ga}u\eps_{i\hi},\\
 & =0,
\end{split}
\end{equation}
as $u$ is compactly supported in $U$. Thus, noting that in the smooth
case $\stm_{\ga}$ are represented by the $n$-forms $S_{\ga}\dx$,
we conclude that locally
\begin{equation}
\diver\stm\cdot\vf=\int_{U}(\std_{\ga,i}^{i}-\std_{\ga})\vf^{\ga}\dx.
\end{equation}
In other words, locally, the vector valued $0$-current is represented
by the vector valued form
\begin{equation}
\diver\std:=(\std_{\ga,i}^{i}-\std_{\ga})\sbase^{\ga}\otimes\dx.\label{eq:def_diver_st-dens}
\end{equation}

\subsection{The balance equation\label{subsec:balance-equation}}

We define now the body force current $\bfcc$ and the boundary force
current $\sfcc$ corresponding to the stress $\stm$, elements of
$C^{-1}(\vb^{*}\otimes\nfo)$, by
\begin{equation}
\bfcc:=-\diver\stm,\qquad\sfcc:=-\tilde{\bdry}(p_{\tst}\stm)=(-1)^{n-1}\tilde{\dee}(p_{\tst}\stm).
\end{equation}
 From the definition of the divergence in (\ref{eq:def_diver_detailed})
we deduce
\begin{equation}
\begin{split}\stm\cdot j^{1}\vf & =\bfcc\cdot\vf+\sfcc\cdot\vf,\end{split}
\label{eq:Principle of VW}
\end{equation}
The last equation is yet another generalization of the principle of
virtual work in continuum mechanics.

For the smooth case, $\tst$ is represented by the smooth vector valued
form $\tstd$ as in (\ref{eq:action_tr_st}) and we can compute, for
any differentiable function $u$ defined on $\base$,
\begin{equation}
\begin{split}\bdry(\tst\cdot\vf)(u) & =\int_{\base}\tstd\dot{\wedge}(\vf\otimes\dee u),\\
 & =\int_{\base}(\tstd\cdot\vf)\wedge\dee u,\\
 & =(-1)^{n-1}\int_{\base}\dee((\tstd\cdot\vf)\wedge u)-(-1)^{n-1}\int_{\base}\dee(\tstd\cdot\vf)\wedge u,\\
 & =(-1)^{n-1}\int_{\bdry\base}(\tstd\cdot\vf)u-(-1)^{n-1}\int_{\base}\dee(\tstd\cdot\vf)u,\\
 & =\int_{\bdry\base}(\sfc\cdot\vf)u-(-1)^{n-1}\int_{\base}\dee(\tstd\cdot\vf)u,
\end{split}
\end{equation}
where Stokes's theorem was utilized in the fourth line and (\ref{eq:Cauchy_Form_gen})
was used in the fifth line. Thus, in the smooth case, $\bdry(\tst\cdot\vf)$
contains, upon appropriate choices of $u$, information regarding
the action of the surface force.

\subsection{Application to non-holonomic stresses}

In spite of numerous attempts (see \cite{segev_forces_1986,segev_geometric_2017,segev_jets_2017,Seg_Snia_JElas_2018}),
for the general geometry of manifolds, we were not able to extend
the foregoing analysis to hyper-stresses, even for the case of stresses
represented by smooth densities. Yet, the introduction of non-holonomic
stresses makes it possible to carry out one step of the reduction.

Let $\vb_{0}$ be a vector bundle over $\base$ and consider forces
in $C^{r}(\vb_{0})^{*}$. Using the representation by non-holonomic
stresses as in (\ref{eq:Gen_Equil_NHS}) in Section \ref{subsec:Rep_Forces_By_Stresses},
let 
\begin{equation}
\vb_{r-1}:=\nh J^{r-1}\vb_{0}.
\end{equation}
Then, a force $F\in C^{r}(\vb_{0})^{*}$ is represented by an element
\begin{equation}
\nh{\stm}\in C^{0}(\nh J^{r}\vb_{0})^{*}=C^{0}(J^{1}\vb_{r-1})^{*}
\end{equation}
in the form
\begin{equation}
\begin{split}\fc(\vf) & =\hj^{r*}(\nh{\stm})(\vf),\\
 & =\nh{\stm}(\hj^{r}\vf),\\
 & =\nh{\stm}(j^{1}(\nj^{r-1}\vf)),\\
 & =j^{1*}\nh{\stm}(\hj^{r-1}\vf).
\end{split}
\end{equation}
Thus, one may apply the foregoing analysis of simple stresses to the
study the action $\nh{\stm}(j^{1}\chi)=j^{1*}\nh{\stm}(\chi)$ for
elements 
\begin{equation}
\chi\in C^{1}(\nh{\pi}^{r-1})^{*}=C^{1}(\vb_{r-1})^{*}.
\end{equation}
In other words, the analysis of simple stresses is used where we substitute
$\vb_{\r-1}$ and $\chi$ for $\vb$ and $\vf$ above respectively.
In particular, the balance equations for this reduction will yield
\begin{equation}
\fc(\vf)=\nh{\stm}(\nj^{r}\vf)=\bfcc(\nj^{r-1}\vf)+\sfcc(\nj^{r-1}\vf),
\end{equation}
where 
\begin{equation}
t,b\in C^{-1}(\vb_{r-1}^{*}\otimes\nfo)
\end{equation}
are interpreted as hyper surface traction and hyper body force distributions,
respectively.

\section{Concluding Remarks\label{sec:Concluding-Remarks}}

The forgoing text is meant to serve as an introduction to global geometric
stress and hyper-stress theory. We used a simple geometric model of
a mechanical system in which forces are modeled as elements of the
cotangent bundle of the configuration space and outlined the necessary
steps needed in order to use it in the infinite dimensional case of
continuum mechanics. The traditional choice of configurations as embeddings
of a body in space, led us to the natural $C^{1}$-topology which
determined the properties of forces as linear functionals. In particular,
the stress object emerges from a representation theorem for force
functionals.

The general stress object we obtain preserves the basic feature of
the stress tensor\textemdash it induces a force system on the body
and its sub-bodies as described in Section \ref{subsec:Force-systems}.
Further details of the relation between hyper-stresses and force systems
are presented in \cite{segev_consistency_1991} for the general case
where stresses are as irregular as measures.

Generalizing continuum mechanics to differentiable manifolds implies
that derivatives can no longer be decomposed invariantly from the
values of vector fields and jets, combining the values of the field
and its derivatives, are used. As a result, simple stresses mix both
components dual to the values of the velocity fields, $\stm_{\ga}$,
and components dual to the derivatives, $\stm_{\ga}^{i}$. This distinction
form the classical stress tensor may be treated if additional mathematical
structure is introduced. It is noted that no conditions of equilibrium,
which are equivalent to invariance of the virtual power under the
action of the Euclidean group, were imposed. In the general case,
one may assume the action of a Lie group on the space manifold and
obtain corresponding balance laws (see \cite{segev_Microstructure_1994}).

Another subject that has been omitted here is that of constitutive
relations. Constitutive relations, in particular the notion of locality
have been considered from the global point of view in \cite{segev_locality_1988}.
Roughly speaking, it is shown in \ref{subsec:Force-systems} that
a local constitutive relation, viewed form the global point of view
as a mapping that assigns a stress distribution to a configuration,
which is continuous relative to the $C^{r}$-topology is a constitutive
relation for a material of grade $r$. Thus, the notion of locality
is tied in with that of continuity.

A framework for the dynamics of a continuous body, for the geometry
of differentiable manifolds, was suggested in \cite{kupferman_continuum_2017}.
The dynamics of the system is specified using a Riemannian metric
on the infinite dimensional configuration space.

\begin{acknowledgement*}
This work has been partially supported by H. Greenhill Chair for Theoretical
and Applied Mechanics and the Pearlstone Center for Aeronautical Engineering
Studies at Ben-Gurion University.
\end{acknowledgement*}
\begin{center}
\par\end{center}

\noindent %

\end{document}